\documentclass[12pt]{article}
\usepackage{amsmath}
\usepackage{amssymb}
\usepackage[mathscr]{eucal}
\usepackage[usenames]{color}
\usepackage[
bookmarks=true,
backref=true,
colorlinks=false,
linkcolor=red,
urlcolor=green]{hyperref}

\newtheorem{lemma}{Lemma}
\newtheorem{prop}{Proposition}
\newtheorem{thm}{Theorem}
\newtheorem{cor}{Corollary}
\newtheorem{rmk}{Remark}
\newtheorem{defn}{Definition}
\numberwithin{equation}{section}
\numberwithin{thm}{section}
\numberwithin{lemma}{section}
\numberwithin{prop}{section}
\numberwithin{cor}{section}
\numberwithin{rmk}{section}
\numberwithin{defn}{section}

\newcommand{\gen}[1]{\partial_{#1}}

\newcommand{\pr}[1]{\rm pr^{(#1)}}

\definecolor{darkolivegreen}{rgb}{0.333333, 0.419608, 0.1843140}

\DeclareMathOperator{\Sl}{sl}

\DeclareMathOperator{\So}{so}

\begin{document}

\title{\Large Symmetry classification of
third-order nonlinear evolution equations.\\ Part I: Semi-simple algebras.
 }

\author{
P.~Basarab-Horwath \\ \small
Department of Mathematics, Link\"oping University,\\ \small S-581 83
Link\"oping, Sweden
\thanks{e-mail: pehor@mai.liu.se}\\ F.~G\"ung\"{o}r\\ \small
Department of Mathematics, Faculty of Arts and Sciences,\\ \small Do\u{g}u\c{s} University, 34722 Istanbul, Turkey \thanks{e-mail: fgungor@dogus.edu.tr}\and
V. Lahno\\
\small Department of Mathematics, Pedagogical University,\\ \small
2 Ostrohradskyj Street, 36000 Poltava, Ukraine\thanks{e-mail:
Lvi@pdpu.poltava.ua} }

\date{}

\maketitle

\abstract{We give a complete point-symmetry classification of all third-order evolution equations of the form $u_t=F(t,x,u,u_x, u_{xx})u_{xxx}+G(t,x,u,u_x, u_{xx})$ which admit semi-simple symmetry algebras and extensions of these semi-simple Lie algebras by solvable Lie algebras. The methods we employ are extensions and refinements of previous techniques which have been used in such classifications.}

\pagebreak

\tableofcontents

\pagebreak

\section{Introduction} This article is the first of two articles in which we give a Lie-algebraic classification of equations of the form

\begin{equation}\label{eveqn}
u_t= F(t,x,u,u_1, u_2)u_3 + G(t,x,u,u_1, u_2)
\end{equation}
which admit non-trivial Lie point symmetries. Here $F$ and $G$ are
arbitrary smooth functions of their  arguments, $F\neq 0$ and we use the notation $u_1=u_x,\; u_2=u_{xx},\; u_3=u_{xxx}$. This
paper continues the application of the methods developed and applied
in \cite{Zhdanov99, Basarab-Horwath01, Zhdanov00, Gungor04-2,
Lahno05, Zhdanov07}, and is a sequel to \cite{Gungor04-2}. Within
the class of equations \eqref{eveqn} are the Korteweg-de Vries (KdV)
equation
\begin{equation}\label{KdV}
u_t=u_3 + uu_1;
\end{equation}
the modified KdV equation
\begin{equation}\label{mKdV}
u_t=u_3 + u^2u_1;
\end{equation}
the cylindrical KdV equation
\begin{equation}\label{cKdV}
u_t=u_3 + uu_1-\frac{1}{2t}u;
\end{equation}
the Harry-Dym equation
\begin{equation}\label{harry-dym}
u_t=u^3u_3;
\end{equation}
as well as the variable coefficient Korteweg-de Vries (vcKdV)
equation
\begin{equation}\label{vcKdV}
u_t=f(t,x)u_3 + g(t,x)uu_1
\end{equation}
with $f\neq 0,\; g\neq 0.$ The symmetries and  integrability
properties of equation (\ref{vcKdV}) were studied in
\cite{Gazeau92, Gungor96}.

In the present article we give a classification of the different types of equation (\ref{eveqn}) for all semi-simple Lie algebras and their semi-direct sum or direct sum extensions. In Part II (the second article) \cite{Basarab-Horwath-10(II)} we classify evolution equations of the form (\ref{eveqn}) which admit solvable Lie algebras up to and including dimension five. At this stage we obtain equations in which the only free parameters are constants (as opposed to functions) and then the calculation of the maximal symmetry algebras  of the equations is possible using the standard Lie-algorithm for point-symmetries.

In \cite{Gungor04-2}, our method of classification was applied to
equations of the form

\begin{equation}\label{eveqn0}
u_t= u_3 +  G(t,x,u,u_1, u_2)
\end{equation}
which is just equation (\ref{eveqn}) with $F=1.$ Here we allow $F$
to be an arbitrary smooth function (other than the zero function).
In particular, in Ref. \cite{Gungor04-2}, among others, it was
shown that how Eqs. \eqref{KdV}, \eqref{mKdV} and \eqref{cKdV}
which is within the class of \eqref{eveqn0}  can be recovered from the
representative equations of the equivalence classes by changes of
point transformations. Eq. \eqref{harry-dym}, which is outside of the class
\eqref{eveqn0}, is known to be integrable and has an infinite
hierarchy of generalized symmetries, and  therefore admits a
recursion operator \cite{Leo83}. Moreover, it allows a Lax and
Hamiltonian formulation. We note that its point symmetry group is
the direct sum of a realization of the Lie algebra $\Sl(2,\mathbb{R})$
with a 2-dimensional non-abelian Lie algebra, and we shall see how it occurs in our classification.

The ideas we exploit and to some extent refine are described in \cite{Zhdanov99} and are given
more fully in \cite{Basarab-Horwath01}. The mechanism behind our
approach is a combination of the usual Lie-algorithm for finding
point symmetries of partial differential equations and the
equivalence group of the class of equations under study. Here we
give a r\'esum\'e of the steps involved, and we refer the reader
to Ref. \cite{Basarab-Horwath01} for details.

The first step is to establish the conditions for a vector field

\[
X=a(t,x,u)\gen t + b(t,x,u)\gen x + c(t,x,u)\gen u
\]
to be a symmetry operator for equation (\ref{eveqn}). This gives
us the determining equations for the coefficients $a(t,x,u),\;
b(t,x,u),\; c(t,x,u)$. In general, these equations are not
solvable explicitly since the functions $F(t,x,u,u_1,u_2)$ and
$G(t,x,u,u_1, u_2)$ are allowed to be arbitrary.

Our other ingredient is the equivalence group of equation
(\ref{eveqn}). This is the group of invertible point
transformations which leave invariant the form of the equation.
The equivalence group is used to find canonical forms for vector
fields which are symmetry operators for the given type of
equation. This is the same as linearizing a vector field using
diffeomorphisms, but with the difference that the diffeomorphisms
allowed belong to a smaller group.

The third part of our calculations involve finding canonical
representations for Lie algebras within the class of symmetry
operators admitted by the equation. This procedure consists in
choosing a canonical representation for one of the operators of
the Lie algebra basis and then invoking the commutation relations
of the Lie algebra to obtain a form for another of the basis
operators. A canonical form for this second operator is found
using those transformations of the equivalence group which
preserve the (canonical form of the) first operator. Here we exploit the so-called {\it residual equivalence group} of a given set of vector fields $\langle Q_1,\dots, Q_k\rangle$: this is just the subgroup of the full equivalence group of equation (\ref{eveqn}) which preserves the form of each of the operators $Q_1, \dots, Q_k$. This
procedure is then continued for all the other basis operators of
the Lie algebra. Having done this, we are able to calculate the
corresponding functions $F$ and $G$ and this gives us canonical
forms for evolution equations of the given type which admit a
given Lie algebra as a symmetry algebra. One may think of this
method as a systematic way of introducing ansatzes for the forms
of the nonlinearities $F$ and $G$ in order to solve the defining
equations for the symmetry operator.

Our article is organized as follows: In sections 2 and 3 we present the basic definitions of equivalence group and symmetry and we give a characterization of symmetries of equation (\ref{eveqn}) using equivalence groups and we give the determining equations for a symmetry vector field as well as calculating
the equivalence group of equation (\ref{eveqn}) (in fact, we show
that the equivalence transformations for a general evolution
equation in $(1+1)$-time-space must be such that the generator of
time translations must be a function of the time $t$ only). Although this result is
known (see \cite{Kingston}, \cite{Kingston-Sophocleous}) we give a self-contained
proof and improve slightly the statement of the result given in  \cite{Kingston},
\cite{Kingston-Sophocleous}. Most of this material is standard knowledge but we present it here from our point of view.

In section 4 we look at admissible abelian symmetry algebras  and we show that any equation of the form (\ref{eveqn}) which admits an abelian symmetry algebra which is of dimension 4, or a rank-one realization of an abelian Lie algebra of dimension 3, is linearizable (equivalent under an equivalence transformation to a linear equation). In fact, we generalize this slightly in Part II to show that if our equation admits a rank-one realization of a three-dimensional solvable Lie algebra then it is linearizable.

In section 5 we give a detailed discussion of our results for the two semi-simple
Lie algebras $\Sl(2, \mathbb{R})$ and $\So(3)$. In particular, we give a detailed
account of the classification of the representations of the algebra $\Sl(2, \mathbb{R})$ (a sketch of this proof was given in \cite{Basarab-Horwath01}), partly just to establish the result and partly to illustrate how our method works.

In section 6 we give a new approach to finding the equations which admit semi-direct sum extensions of semi-simple Lie algebras by solvable Lie algebras. The method used in \cite{Basarab-Horwath01} for classifying equations which admit semi-direct sums of semi-simple and solvable Lie algebras consists in finding the maximal symmetry algebras of the equations which admit semi-simple Lie algebras as symmetries. This is possible because the equations of the form $u_t=F(t,x,u, u_1)u_2+G(t,x,u,u_1)$ which are invariant under semi-simple Lie algebras contain only arbitrary functions of one variable. For our present equation (\ref{eveqn}) this approach is far too complicated to exploit since the arbitrary functions are functions of two variables, and so the usual Lie-algorithm is extremely difficult to apply. Our solution to this problem is to use the representation theory of semi-simple Lie algebras acting on solvable Lie algebras. As a result we are able to identify the maximal symmetry algebras of the equations obtained without using the Lie algorithm directly on the equations. However, it should be pointed out that the Lie-algorithm is built into our calculations since our different realizations are admissible in that they satisfy the conditions given in Proposition \ref{symmcondition}.

Finally, in section 7 we list the equations which admit extensions of $\Sl(2, \mathbb{R})$ by solvable Lie algebras, as symmetry algebras.

\section{Equivalence group and symmetries of (\ref{eveqn}).} By definition, the equivalence group of (\ref{eveqn}) is the group of transformations preserving the form of (\ref{eveqn}). This is given in the following definition.

\begin{defn} The equivalence group of equation (\ref{eveqn}) is the group of smooth transformations $(t,x,u,u_1,\dots, u_n)\to (t',x',u',u'_1,\dots, u'_n)$ preserving the contact structure which map the $n$-th order evolution equation
\[
u_t=F(t,x,u,u_1,\dots, u_n)
\]
to the $n$-th order evolution equation
\[
u'_{t'}=\tilde{F}(t',x',u',u'_1,\dots, u'_n),
\]
where $F$ and $\tilde{F}$ are smooth functions of their arguments satisfying $F_{u_n}\neq 0$ and $\tilde{F}_{u'_n}\neq 0$. Note that we also have $u'_1=u'_{x'},\, u'_2=u'_{x'x'},\dots$.
\end{defn}

\medskip
Transformations preserving the contact structure are known as {\sl tangent transformations} and it is known (see \cite{IbragimovAnderson}) that such transformations are prolongations of point transformations $(t,x,u^1,\dots, u^m)\to (t', x', u'^1, \dots, u'^m)$ (if the number of dependent functions $m$ is greater than one) or prolongations of contact transformations $(t,x,u,p,q)\to (t', x', u', p', q')$ (if there is only one dependent function), where $p=u_t,\; q=u_1$. Here, prolongation refers to extending the transformations to higher derivatives of the dependent functions. Although we use only prolongations of point transformations our classification, we give the general result on contact transformations preserving the form of equation (\ref{eveqn}) for the sake of completeness.

\begin{thm}\label{contactequiv}
Any invertible contact transformation
\[
(t,x,u,p,q)\to (t', x', u', p', q')
\]
with $p=u_t,\;\; q=u_1$ and $p'=u'_{t'},\;\, q'=u'_{x'}$, preserving the form of an evolution equation of order $n$
\[
u_0=F(t,x,u,u_1,\dots, u_n),
\]
with $n\geq 2$, is such that $t'=T(t)$ with $\dot{T}(t)\neq 0$. Then the contact transformation has the form
\[
\begin{split}
t'&=T(t),\; x'=-W_{q}(t,x,u,q),\\
u'&= W(t,x,u,q) - qW_{q}(t,x,u,q),\\
p'&= -p\dot{T}(t) +pW_u(t,x,u,q) +W_t(t,x,u,q),\\
q'&=W_x(t,x,u,q) + qW_u(t,x,u,q)
\end{split}
\]
for some smooth function $W$.
\end{thm}

\smallskip\noindent The proof consists in showing that if we do not have $T_x(t, x, u)=T_u(t,x,u)=0$ then the right-hand side of the equation will contain mixed spatial and temporal derivatives, and we give the proof in the appendix. As a corollary we have the following result:

\begin{cor}\label{pointequiv} The equivalence group $\mathscr{E}$ of point transformations $(t, x, u)\to (t', x', u')$, of equation (\ref{eveqn}) is given by invertible smooth transformations of the form

\begin{equation}\label{eqgrp}
t'=T(t),\quad x'=X(t,x,u),\quad u'=U(t,x,u).
\end{equation}
\end{cor}
An alternative proof of Corollary \ref{pointequiv} can be found in the references (\cite{Kingston}, \cite {Kingston-Sophocleous}).

\medskip\noindent We now recall the definition of symmetries:

\begin{defn}\label{pointsymmdefn} A point symmetry of the evolution equation

\begin{equation}\label{geneveqn}
u_t=F(t,x,u,u_1,\dots, u_n).
\end{equation}
is a transformation $(t, x, u)\to (t', x', u')$ so that
\[
u'_{t'}=F(t', x', u', u'_1,\dots, u'_n).
\]
\end{defn}
One can define contact symmetries in the same way. Then we come to infinitesimal symmetries, defined by vector fields:

\begin{defn}\label{infsymmdefn} The smooth vector field
\[
Q=a(t,x,u)\gen t + b(t,x,u)\gen x + c(t,x,u)\gen u
\]
defines an {\sl infinitesimal point symmetry} of equation (\ref{geneveqn}) if the local flow $\Phi(\epsilon)$ (with $\epsilon$ in some neighbourhood of $0$) of $Q$  defines an invertible (locally) point transformation $(t, x, u)\to (t', x', u')$ which is a symmetry of (\ref{geneveqn}) in the sense of Definition \ref{pointsymmdefn}.
\end{defn}

\begin{thm} A smooth vector field $Q$ which defines an infinitesimal point symmetry of equation (\ref{geneveqn}) for $n\geq 2$, has the form
\[
Q=a(t)\gen t + b(t,x,u)\gen x + c(t,x,u)\gen u,
\]
where $a(t),\; b(t,x,u),\; c(t,x,u)$ are smooth functions of their arguments.
\end{thm}

\smallskip\noindent{\bf Proof:} Note that the local flow $\Phi(\epsilon)$ is  a local one-parameter family of transformations in $\mathscr{E}$ since it maps an evolution equation to an evolution equation. Then, if $(t'(\epsilon), x'(\epsilon), u'(\epsilon))$ denotes the action of $\Phi(\epsilon)$ on $(t,x,u)$, we have for $n\geq 2$ and sufficiently small $\epsilon$
\begin{eqnarray*}
& t'(\epsilon)=t + \epsilon a(t) + o(\epsilon^2)\\
& x'(\epsilon)=x + \epsilon b(t, x, u) + o(\epsilon^2)\\
& u'(\epsilon)=u + \epsilon c(t, x, u) + o(\epsilon^2),
\end{eqnarray*}
for some smooth functions $a(t),\; b(t,x,u),\; c(t,x,u)$, since $\Phi(0)=\text{id}$ and because $t'(\epsilon)$ is independent of $(x,u)$, that is $t'(\epsilon)=T(t, \epsilon)$ for some $T(t,\epsilon)$ depending smoothly on $(t, \epsilon)$, according to Theorem \ref{contactequiv}. From this it follows that
\[
Q=a(t)\gen t + b(t,x,u)\gen x + c(t,x,u)\gen u.
\]
An alternative proof of this result can be found in (\cite{asano90}).

\begin{rmk} It is clear from Theorem \ref{contactequiv} that any infinitesimal contact symmetry of equation (\ref{eveqn}) is of the form
\[
Q=a(t)\gen t + b(t,x,u,q)\gen x + c(t,x,u,q)\gen u + \tau_0(t,x,u,p,q)\gen p + \tau_1(t,x,u,p,q)\gen q,
\]
with $p=u_t,\; q=u_1$ and where
\begin{align*}
& b(t,x,u,q)=-w_q(t,x,u,q),\;\; c(t,x,u,q)=w(t,x,u,q)-qw_q(t,x,u,q),\\ & \tau_0=-p\dot{a}(t)+w_t(t,x,u,q)+pw_u(t,x,u,q),\;\; \tau_1=w_x(t,x,u,q) + qw_q(t,x,u,q)
\end{align*}
for some (smooth) functions $w(t,x,u,q)$ and $a(t)$.
\end{rmk}

\medskip The importance of the equivalence group $\mathscr{E}$ of equation (\ref{eveqn}) lies in the possibility of linearizing vector fields of the form $Q=a(t)\gen t + b(t,x,u)\gen x + c(t,x,u)\gen u$. This is how canonical forms of vector fields (differential operators) are obtained. The fundamental result is the following:

\begin{thm}\label{canonicalform}
A vector field
\[
Q=a(t)\gen t + b(t,x,u)\gen x + c(t,x,u)\gen u
\]
can be transformed by transformations of the form (\ref{eqgrp})
into one of the following canonical forms:

\begin{equation}
Q=\gen t,\qquad Q=\gen x.
\end{equation}
\end{thm}

\smallskip\noindent{\bf Proof:} Any operator
\[
Q= a(t)\gen t + b(t,x,u)\gen x + c(t,x,u)\gen u
\]
is transformed by our allowed transformations into $Q'=Q(T)\gen {t'} + Q(X)\gen {x'} + Q(U)\gen {u'}$. If $a\neq 0$ then we choose $T(t)$ so that $Q(T)=a(t)\dot{T}(t)=1$ (at least locally). We choose $X$ and $U$ to be each of any two independent integrals of the PDE $Q(Y)=0$. This the gives the canonical form $Q=\gen t$ in some coordinate system.

If we now have $a(t)=0$ then $Q$ is transformed into $Q'=Q(X)\gen {x'} + Q(U)\gen {u'}$. We then choose $X, U$ so that $Q(X)=1,\; Q(U)=0$. This gives us the canonical form $Q=\gen x$ in some coordinate system.

\begin{rmk} It is sometimes useful to use the fact that $t'=t,\;\, x'=u,\;\; u'=x$ is an equivalence transformation preserving the form of equation (\ref{eveqn}), and it gives us the following transformations of derivatives:
\[
\gen t\to \gen t,\quad \gen x \to \gen u,\quad \gen u\to \gen x,
\]
so that one can reformulate Theorem \ref{canonicalform} as:
\begin{thm}\label{canonicalformbis}
A vector field
\[
Q=a(t)\gen t + b(t,x,u)\gen x + c(t,x,u)\gen u
\]
can be transformed by transformations of the form (\ref{eqgrp})
into one of the following canonical forms:
\begin{equation}
Q=\gen t,\qquad Q=\gen u.
\end{equation}
\end{thm}
\end{rmk}

Another useful concept is that of {\sl residual equivalence group} of a set of vector fields $\langle Q_1,\dots, Q_k\rangle$.

\begin{defn} The {\sl residual equivalence group} $\mathscr{E}(Q_1,\dots, Q_k)$ of a given set of vector fields $\langle Q_1,\dots, Q_k\rangle$ is the subgroup of $\mathscr{E}$ which leaves invariant in form each of the vector fields $Q_1,\dots, Q_k$.
\end{defn}

As an example of this, consider $Q=\gen t$. Then $\mathscr{E}(\gen t)$ is the set of all transformations $t'=T(t),\; X(t,x,u),\; u'=U(t,x,u)$ of $\mathscr{E}$ which leave $\gen t$ invariant in form, so $Q=\gen t$ is mapped to $Q'=\gen {t'}$. Since $Q'=Q(T)\gen {t'} + Q(X)\gen {x'} + Q(U)\gen {u'}$, then we must have $Q(T)=1,\; Q(X)=Q(U)=0$, so that $X=X(x,u),\; U=U(x,u)$ and $T(t)=t + l$ where $l$ is an arbitrary constant. Thus $\mathscr{E}(\gen t)$ consists of all invertible transformations of $\mathscr{E}$ with $t'=t + l,\; x'=X(x,u),\; u'=U(x,u)$ where $l$ is an arbitrary constant. It is easy to see that $\mathscr{E}(\gen t, \gen x)$ is then the set of all invertible transformations of the form $t'=t+ l,\; x'=x + X(u),\; u'=U(u)$ with $l$ being an arbitrary constant, $X(u)$ being an arbitrary function and $U(u)$ being an arbitrary function satisfying $U'(u)\neq 0$.

\medskip Finally, we come to the {\sl rank of a realization}:

\begin{defn} A {\sl realization} of a Lie algebra $\mathfrak{g}$ with a basis $\langle e_1,\dots, e_n\rangle$ is a Lie algebra isomorphism of $\langle e_1,\dots, e_n\rangle$ to a set of vector fields $\langle Q_1,\dots, Q_n\rangle$ such that $Q_i$ is the image of $e_i$ for $i=1,\dots, n$.

\smallskip\noindent The {\sl rank} of this realization is the smallest positive integer $r=\text{\rm rank}\,\langle Q_1,\dots, Q_n\rangle$ for which all $r+1$-fold products $Q_{i_1}\wedge\dots Q_{i_{r+1}}=0$.
\end{defn}

\smallskip\noindent As an example, both $\langle \gen t, \gen x, \gen u\rangle$ and $\langle \gen x, u\gen x, u^2\gen x\rangle$ are realizations of a three-dimensional abelian Lie algebra $\mathfrak{g}=\langle e_1, e_2, e_3\rangle$. But $\text{rank}\,\langle \gen t, \gen x, \gen u\rangle=3$ while $\text{rank}\,\langle \gen x, u\gen x, u^2\gen x\rangle=1$.

\begin{rmk} Rank is invariant under (local) diffeomorphisms, since for a (local) diffeomorphism $\Phi$ we have $\Phi_{\ast}(Q_1\wedge Q_2\wedge\dots \wedge Q_k)=\Phi_{\ast}Q_1\wedge \Phi_{\ast}Q_2\wedge\dots\wedge\Phi_{\ast}Q_k$ where $\Phi_{\ast}$ is the induced mapping on sections of the tangent bundle.
\end{rmk}

\section{The symmetry condition.} Having given the basic definitions and results, we now come to the conditions on the coefficients of the vector field $\displaystyle Q=a(t)\gen t + b(t,x,u)\gen x + c(t,x,u)\gen u$ in order for it to be a symmetry of equation (\ref{eveqn}).

In order to implement the symmetry algorithm we need to calculate
the third order prolongation $\displaystyle {\pr{3}} Q$ of the vector field $\displaystyle Q=a(t)\gen t + b(t,x,u)\gen x + c(t,x,u)\gen u$ to the third-order jet space $J^3(\mathbb{R}^2, \mathbb{R})$ which has local coordinates $(t,x,u,u_t, u_x, u_{tt}, u_{tx}, u_{xx}, u_{ttt}, u_{ttx}, u_{txx}, u_{xxx})$.

For a general vector field $\displaystyle Q=\xi^0(t,x,u)\gen t + \xi^1(t,x,u)\gen x + \eta(t,x,u)\gen u$, we define the {\sl characteristic function} $\displaystyle \Sigma=\eta - u_t\xi^0-u_1\xi^1$. Then we may express the prolongation ${\pr{3}}Q$ as
\begin{equation}\label{pr}
{\pr{3}} Q = Q + \tau_{\mu} \gen {u_{\mu}}  + \tau_{\mu_1\mu_2} \gen {u_{\mu_1\mu_2}} +\tau_{\mu_1\mu_2\mu_3} \gen {u_{\mu_1\mu_2\mu_3}},
\end{equation}
where $\mu, \mu_i=0, 1$ and $u_0=u_t,\; u_1=u_x,\; u_{00}=u_{tt},\; u_{01}=u_{tx}\dots $, and we sum over repeated indices. The coefficients $\tau_{\mu},\dots$ are given by
\[
\begin{gathered}
  \tau_{\mu}  = D_{\mu}\Sigma  +D_{\mu}(u_{\nu})\xi^{\nu}, \hfill \\
  \tau_{\mu_1\mu_2}  = D_{\mu_1\mu_2}\Sigma  +D_{\mu_1\mu_2}(u_{\nu})\xi^{\nu}, \hfill \\
  \tau_{\mu_1\mu_2\mu_3}  = D_{\mu_1\mu_2\mu_3}\Sigma  +D_{\mu_1\mu_2\mu_3}(u_{\nu})\xi^{\nu},
\hfill \\
\end{gathered}
\]
with $D_{\mu}$ denoting the total derivative with respect to $x^{\mu}$ for $x^0=t,\; x^1=x$, and where we have $D_{\mu_1\mu_2}=D_{\mu_1}D_{\mu_2},\; D_{\mu_1\mu_2\mu_3}=D_{\mu_1}D_{\mu_2}D_{\mu_3}$.
Here $D_x$ and $D_t$ denote the total space and time derivatives. Then the vector field $Q$ is a symmetry of \eqref{eveqn} precisely when its third-order prolongation \eqref{pr} annihilates
equation \eqref{eveqn} on its solution manifold, that is we have
\begin{equation}\label{symcon1}
{\pr{3}}Q(\Delta)\Bigl|_{\Delta=0}=0,\quad \Delta=u_t-Fu_3-G,
\end{equation}
namely
\begin{equation}\label{symcon2}
\tau_0-[Q(F)+\tau_1F_{u_1}+\tau_{11}F_{u_2}]u_3-
\tau_{111}F-Q(G)-\tau_{1}G_{u_1}-\tau_{11}G_{u_2}\Big|_{u_{t}=Fu_3+G}=0.
\end{equation}
We equate coefficients of linearly independent terms in the invariance
condition \eqref{symcon2} to zero and this gives an overdetermined system of
linear PDEs (the {\sl determining equations}). Solving this system we
obtain the following assertion.

\begin{prop}\label{symmcondition}
The symmetry group of the nonlinear equation \eqref{eveqn} for
arbitrary (fixed) functions $F$ and $G$ is generated by the vector
field
\begin{equation}\label{gvf}
Q= a(t)\gen t + b(t,x,u)\gen x + c(t,x,u)\gen u
\end{equation}
where the functions $a$, $b$ and $c$ satisfy the determining
equations
\begin{subequations}\label{deq}
\begin{equation}\label{deq1}
\begin{array}{ll}
&F\,( -a_{t} +
     3\,u_1\,b_{u} +
     3\,b_{x} )  +
  [
  u_1(b_{xx}-2c_{xu})+u_1^2(2b_{xu}-c_{uu})+u_1^3b_{uu}+\\
&
  u_2(2b_x-c_u)+
  3u_1u_2b_u-c_{xx}] \,F_{u_2} +\\
&
  [ u_1^2\,b_{u} +
     u_1(b_x-c_u) - c_{x}]\,F_{u_1} -
  c\,F_{u} - a\,F_{t} -
  b\,F_{x}=0,
\end{array}
\end{equation}
and
\begin{equation}\label{deq2}
\begin{array}{ll}
&G\,\left( -a_{t} -
     u_1\,b_{u} + c_{u}
     \right)  -
  u_1\,b_{t} + c_{t} +\\
&
F\,[u_1(b_{xxx}-3c_{xxu})+3u_1^2(b_{xxu}-c_{xuu})+u_1^3(3b_{xuu}-c_{uuu})
+u_1^4b_{uuu}+\\
&
3u_2(b_{xx}-c_{xu})+3u_1u_2(3b_{xu}-c_{uu})+6u_1^2u_2b_{uu}
+3u_2^2b_u-c_{xxx}]+\\
&
[u_1(b_{xx}-2c_{xu})+u_1^2(2b_{xu}-c_{uu})+u_1^3b_{uu}+u_2(2b_x-c_u)+
3u_1u_2b_u-c_{xx}]\,G_{u_2} +\\
&  [u_1(b_x-c_u)+u_1^2b_u-c_{x}]\,G_{u_1} -
  c\,G_{u} - a\,G_{t} -
  b\,G_{x}=0.
\end{array}
\end{equation}
\end{subequations}
Here the dot over a symbol stands for time derivative.
\end{prop}
We do {\emph not} require that \eqref{deq}
should be satisfied for all possible choices of $F,\; G$ (which is possible only when the
symmetry group is the trivial group consisting of the identity transformation). Our approach is  to find all (locally smooth) functions $F, G$ for which the equation \eqref{eveqn} admits non-trivial symmetry groups. To achieve this, we first find canonical forms for the realizations of given Lie algebras and then find those $F$ and $G$ which make the Lie algebras into symmetry algebras of the equation \eqref{eveqn}. In carrying out this programme we exploit the classification of semi-simple Lie algebras and the results on the classification of low-dimensional solvable Lie
algebras obtained mostly in late sixties \cite{Morozov58,
Mubarakzyanov63-1, Mubarakzyanov63-2, Mubarakzyanov63-3,
Mubarakzyanov66-3, Turkowski88, Turkowski90}.

\section{Abelian algebras as symmetries.} We first deal with abelian algebras $A$ of dimension $\dim A=k$ with $A=\langle Q_1,\dots, Q_k\rangle$, and we give some details of the calculations involved in order to show how our apparatus works. First we note that since each of the vector fields $Q_i$ is of the form $Q=a(t)\gen t + b(t,x,u)\gen x + c(t,x,u)\gen u$, then $\text{rank}\,A\leq 3$. Further, we note that if $A=\langle Q_1,\dots, Q_k\rangle$ and ${\rm rank}\, A=1$, then either $\dim A=1$ or each $Q_i$ must be of the form $Q_i=c_i(t,x)\gen u$ with $Q_1=\gen u$ in canonical form, if $\dim A\geq 2$ (note that $c_i(t,x)$ is independent of $u$ by commutativity). This is so because if we take $Q_1=\gen t$ then $Q_2=a(t)\gen t$, and commutativity then gives $a=\text{constant}$, contradicting $\dim A\geq 2$. We present our results in the form of a theorem:

\begin{thm}\label{abeliansymm} All inequivalent, admissible Lie algebras of vector fields of the form $Q=a(t)\gen t + b(t,x,u)\gen x + c(t,x,u)\gen u$ are given as follows:
\begin{align*}
& A=\langle \gen t\rangle,\quad A=\langle \gen u \rangle\;\; (\dim A=1)\\
& A=\langle \gen t, \gen u \rangle,\quad A=\langle \gen x, \gen u \rangle,\quad A=\langle \gen u, x\gen u\rangle\;\; (\dim A=2)\\
& A=\langle \gen t, \gen x, \gen u\rangle,\quad A=\langle \gen t, \gen u, x\gen u \rangle,\quad A=\langle \gen u, x\gen u, c(t,x)\gen u \rangle,\; c_{xx}\neq 0\;\; (\dim A=3)\\
& A=\langle \gen t, \gen u, x\gen u, c(x)\gen u \rangle,\; c''(x)\neq 0\;\, (\dim A=4)\\
& A=\langle \gen u, x\gen u, c(t,x)\gen u, q_1(t,x)\gen u,\dots, q_k(t,x)\gen u \rangle\;\; c_{xx}\neq 0,\, (q_1)_{xx}\neq 0,\dots, (q_k)_{xx}\neq 0.
\end{align*}
\end{thm}

\smallskip\noindent{\bf Proof:} First, we note that there are only two inequivalent forms for one-dimensional abelian Lie algebras: $A=\langle \gen t\rangle$ and $A=\langle \gen u\rangle$ as noted in Theorem \ref{canonicalformbis}.

Now we take $\dim A=2$. We have $A=\langle Q_1, Q_2\rangle$. If $\text{rank}\, A=1$ then by the above remarks, we take $Q_1=\gen u$ and $Q_2=c(t,x)\gen u$. The residual equivalence group $\mathscr{E}(\gen x)$ is given by invertible transformations of the form $t'=T(t),\, x'=X(t,x),\, u'=u + U(t,x)$ with $\dot{T}(t)\neq 0,\, X_x\neq 0$. Under such a transformation $Q_2$ is mapped to $Q'_2=c(t,x)\gen {u'}$. If $c_x\neq 0$ we may take $X(t,x)=c(t,x)$ so that the $Q'_2=x'\gen {u'}$ and then we have $A=\langle \gen u, x\gen u\rangle$ in canonical form. If $c_x=0$ then $c(t,x)=c(t)$. Substituting $\gen u$ and $c(t)\gen u$ into the equation for $G$ we find $G_u=0,\; \dot{c}(t)=0$ so $c(t)$ is a constant and thus $\dim A=1$ which is a contradiction, so we have no admissible canonical form in this case.

If $\dim A=2,\; \text{rank}\, A=2$ then we may take $Q_1=\gen t$ or $Q_1=\gen x$. If $Q_1=\gen t$ then, by commutativity and the fact that $Q_1, Q_2$ is a basis for $A$, we may take $Q_2=b(x,u)\gen x + c(x,u)\gen u$. The residual equivalence group $\mathscr{E}(\gen t)$ consists of invertible transformations of the form $t'=t+l,\; x'=X(x,u),\; u'=U(x,u)$, with $l=\text{constant}$. Under such a transformation $Q_2$ is mapped to $Q'_2=Q_2(X)\gen {x'} + Q_2(U)\gen {u'}$. We can always choose $X,\, U$ so that $Q_2(U)=1,\; Q_2(X)=0$ (and then $X(x,u),\, U(x,u)$ are functionally independent). Thus we have the canonical form $A=\langle \gen t, \gen u\rangle$. If we take $Q_1=\gen u$, then we have $Q_2=a(t)\gen t + b(t,x)\gen x + c(t,x)\gen u$, and clearly $a(t)\neq 0$ or $b(t,x)\neq 0$ because ${\rm rank}\, A=2$. The residual equivalence group $\mathscr{E}(\gen u)$ is the group of transformations $t'=T(t),\, x'=X(t,x),\, u'=u + U(t,x)$ with $\dot{T}(t)\neq 0,\, X_x\neq 0$. Under such a transformation, $Q_2$ is mapped to $Q'_2=a(t)\dot{T}(t)\gen {t'} + [a(t)X_t(t,u)+b(t,x)X_x]\gen {x'} + [a(t)U_t + b(t,x)u_1 + c(t,x)]\gen {u'}$. If $a(t)\neq 0$ we may choose $T(t),\, X(t,x),\, U(t,x)$ so that $a(t)\dot{T}(t)=1,\, a(t)X_t(t,u)+b(t,x)X_x=0,\, a(t)U_t + b(t,x)u_1 + c(t,x)=0$. Then we may take $Q_1=\gen u,\, Q_2=\gen t$.  If, however, $a(t)=0$, then $Q_2=b(t,x)\gen x + c(t,x)\gen u$ and we then have $b(t,x)\neq 0$ since we have ${\rm rank}\, A=2$. Then $Q'_2=b(t,x)X_x\gen {x'} + [c(t,x)+b(t,x)u_1\gen {u'}$ and because $b(t,x)\neq 0$ we may choose $U(t,x),\; X(t,x)$ so that $b(t,x)u_1 + c(t,x)=0,\; b(t,x)X_x=1$ and this gives $Q'_2=\gen {x'}$, so we have the canonical form $A=\langle\gen u, \gen x\rangle$. Hence there are the following admissible, canonical forms for $\dim A=2$:
\[
A=\langle\gen u, x\gen u \rangle,\quad A=\langle\gen t, \gen u\rangle,\quad A=\langle\gen x, \gen u \rangle.
\]

For $\dim A=3$ we have $A=\langle Q_1, Q_2, Q_3\rangle$. If $\text{rank}\, A=1$ then we may take $Q_1=\gen u,\; Q_2=x\gen u$ by the previous argument for $\dim A=2$. We then have $Q_3=c(t,x)\gen u$ with $c_{xx}\neq 0$. For if $c_{xx}=0$ we have $c(t,x)=\alpha(t)x+\beta(t)$. Then substituting $\gen u, x\gen u, [\alpha(t)x+\beta(t)]\gen u$ into the equation for $G$ we obtain $\dot{\alpha}(t)x + \dot{\beta}(t)=0$, after substituting into the equation for $G$, and this is possible only for constant $\alpha(t),\, \beta(t)$. But this means that $Q_3$ is a constant linear combination of $Q_1,$ and $Q_2$, contradicting $\dim A=3$. So, $c_{xx}(t,x)\neq 0$. For $A=\langle \gen u, x\gen u, c(t, x)\gen u\rangle$ as a symmetry algebra we find the following form for the evolution equation:
\[
u_t=f(t,x)u_3+\frac{c_t(t,x)-c_{xxx}(t,x)f(t,x)}{c_{xx}}u_2+g(t,x)
\]
with $f(t,x)\neq 0$. Thus, the evolution equation is quasi-linear (that is, linear in $u$ and its derivatives)

We now come to $\dim A=3,\; {\rm rank}\, A=2$. If we take $Q_1=\gen t$ then we may take $Q_2=\gen u$ since $\langle Q_1, Q_2\rangle$ is a two-dimensional abelian subalgebra of $A$, and there is only one type of two-dimensional abelian algebra containing $\gen t$ as we have seen. Then we have $Q_3=a\gen t + b(x)\gen x + c(x)\gen u$ with $a=\,{\rm constant}$ and so we may clearly take $Q_3=b(x)\gen x + c(x)\gen u$. The requirement that ${\rm rank}\, A=2$ then gives $b(x)=0$ and we find that $Q_3=c(x)\gen u$. Obviously, $c'(x)\neq 0$ since we must have $\dim A=3$. Now, the residual equivalence group  $\mathscr{E}(\gen t, \gen u)$ is given by transformations of the form $t'=t+l,\; x'=X(x),\; u'=u + U(x)$ with $X'(x)\neq 0$. We see that under such a transformation, $Q_3=c(x)\gen u$ is transformed to $Q'_3=c(x)\gen {u'}$ and we may take $c(x)=X(x)$ giving $Q'_3=x'\gen {u'}$, so we find the canonical form $A=\langle \gen t, \gen u, x\gen u\rangle$.   If $Q_1=\gen u$, then we may take   $Q_2=\gen t$ or $Q_2=\gen x$ if $Q_1\wedge Q_2\neq 0$. If $Q_2=\gen t$ then $Q_3=x\gen u$ as before. If $Q_2=\gen x$ then $Q_3=b(t)\gen x + c(t)\gen u$ because ${\rm rank}\, A=2$. However, putting $Q_1=\gen u,\; Q_2=\gen x,\; Q_3=b(t)\gen x + c(t)\gen u$ into the equation for $G$, we obtain $\dot{c}(t)-\dot{b}(t)u_1=0$, which means that $\dot{c}(t)=\dot{b}(t)=0$ and hence $Q_3$ is a linear combination of $Q_1$ and $Q_2$, which is a contradiction. Thus we cannot have $Q_2=\gen x$ in this case. If $Q_1\wedge Q_2=0$ then $Q_2=c(t,x)\gen u$, and since $\langle Q_1, Q_2\rangle$ is a two-dimensional subalgebra of $A=\langle Q_1, Q_2, Q_3\rangle$ then we know from the above that we may take $Q_2=x\gen u$. This then gives us $Q_3=\gen t$ in canonical form. Hence we have only the case $A=\langle \gen t, \gen u, x\gen u\rangle$ when $\dim A=3,\; {\rm rank}\, A=2$.

Now consider the case $\dim A=3,\, {\rm rank}\, A=3$. We have $A=\langle Q_1, Q_2, Q_3\rangle$ and $Q_i=a_i(t)\gen t + b_i(t,x,u)\gen x + c_i(t,x,u)\gen u$ for $i=1, 2, 3$. Since $Q_1\wedge Q_2\wedge Q_3\neq 0$ at least one of the coefficients $a_i(t)$ must be different from zero. So, assume $a_1(t)\neq 0$. In this case, we may take $Q_1=\gen t$ in canonical form, according to Theorem \ref{canonicalform}. Then we must have $Q_i=a_i\gen t + b_i(x,u)\gen x + c_i(x,u)\gen u$ for $i=2, 3$ and we may take $a_i=0$ in these cases because of linear independence of $Q_1, Q_2, Q_3$. Hence $Q_2=b_2(x,u)\gen x + c_2(x,u)\gen u$ and $Q_3=b_3(x,u)\gen x + c_3(x,u)\gen u$. Consequently we may, by Theorem \ref{canonicalform} take $Q_2=\gen x$. Finally, this means that $Q_3=b_3(u)\gen x + c_3(u)\gen u$ by commutativity, and $c_3(u)\neq 0$ because $Q_1\wedge Q_2\wedge Q_3\neq 0$. The residual equivalence group $\mathscr{E}(\gen t, \gen x)$ is given by transformations of the form $t'=t+l,\; x'=x + Y(u),\; u'=U(u)$ with $U'(u)\neq 0$. Under such a transformation, $Q_3$ is transformed to $Q'_3=[b_3(u)+c_3(u)Y'(u)]\gen {x'} + c_3(u)U'(u)\gen {u'}$. because $c_3(u)\neq 0$ then we may choose $Y(u)$ so that $b_3(u)+c_3(u)Y'(u)=0$ and we may choose $U(u)$ so that $c_3(u)U'(u)=1$ and so we may take $Q_3=\gen u$ in canonical form. Consequently we have only one canonical form $A=\langle \gen t, \gen x, \gen u\rangle$ for an abelian Lie algebra with $\dim A=3,\, \text{rank}\, A=3$.

Finally we come to $\dim A\geq 4$. If ${\rm rank}\, A=3$ then we may take $Q_1=\gen t, Q_2=\gen x, Q_3=\gen u$. For any other symmetry vector field $Q=a(t)\gen t + b(t,x,u)\gen x + c(t,x,u)\gen u$ the coefficients must be constants, so that $Q$ will be a constant linear combination of $Q_1, Q_2, Q_3$, contradicting $\dim A\geq 4$. So there are no admissible abelian Lie algebras $A$ with $\dim A\geq 4,\, {\rm rank}\, A=3$.

For $\dim A= 4,\,{\rm rank}\, A=2$ we may take $Q_1=\gen t,\, Q_2=\gen u,\, Q_3=x\gen u$, and for $Q_4$ we take $Q_4=c(x)\gen u$ because $A$ is abelian and ${\rm rank}\, A=2$. Further $c''(x)\neq 0$, for otherwise $Q_4$ would be a constant linear combination of $Q_2$ and $Q_3$, contradicting $\dim A=4$. Thus we have $A=\langle \gen t, \gen u, x\gen u, c(x)\gen u\rangle$ with $c''(x)\neq 0$. This symmetry algebra gives us the following evolution equation:

\[
u_t=f(x)u_3-\frac{c^{(3)}(x)f(x)}{c^{(2)}(x)}u_2+g(x)
\]
with $f(x)\neq 0$. Again, the evolution equation is quasi-linear.

For $\dim A\geq 5,\; \text{rank}\, A=2$ we find that any additional symmetry operator $Q$ must be of the form $Q=q(x)\gen u$ with $q''(x)\neq 0$ and, substituting the vector fields a symmetries in the equations for $F$ and $G$, we must also have
\[
\frac{c^{(3)}(x)}{c^{(2)}(x)}=\frac{q^{(3)}(x)}{q^{(2)}(x)}.
\]
From this it follows that $\displaystyle \frac{q^{(2)}(x)}{c^{(2)}(x)}=C$ for some constant $C$. Hence $q^{(2)}(x)=Cc^{(2)}(x)$ from which it follows that $q(u)=Cc(x)+C_1x+C_2$ and then $q(x)\gen u$ is a constant linear combination of $\gen x, x\gen u, c(x)\gen u$, and therefore we have no admissible abelian Lie algebras $A$ with $\dim A\geq 5,\; \text{rank}\, A=2$.

For $\dim A\geq 4,\; \text{rank}\, A=1$, arguments similar to those for $\dim A=3,\,\text{rank}\, A=1$ give us abelian Lie algebras $A$ of the form $A=\langle \gen u, x\gen u, c(t,x)\gen u, q_1(t,x)\gen u,\dots, q_k(t,x)\gen u\rangle$ with the sole proviso that each of the $q:$ s satisfy $q_{xx}(t,x)\neq 0$ and
\[
\frac{[c_t-c_{xxx}f(t,x)]}{c_{xx}}=\frac{[q_t-q_{xxx}f(t,x)]}{q_{xx}},
\]
as well as linear independence of the vector fields $\langle \gen u, x\gen u, c(t,x)\gen u, q_1(t,x)\gen u,\dots, q_k(t,x)\gen u\rangle$. We obtain equations of the form
\[
u_t=f(t,x)u_3 + \frac{c_t(t,x)-c_{xxx}(t,x)f(t,x)}{c_{xx}}u_2+g(t,x)
\]
with $f(t,x)\neq 0$.

\bigskip We have the following corollary of the above results:
\begin{cor}\label{linearizableeqns}
The evolution equations of the form (\ref{eveqn}) admitting abelian Lie algebras $\mathsf{A}$ with $\dim\mathsf{A}\geq 4$ or if $\dim\mathsf{A}=3,\;\; {\rm rank}\,\mathsf{A}= 1$ as symmetries are linearizable.
\end{cor}

\smallskip\noindent{\bf Proof:} The proof is just by calculation: for $\dim\mathsf{A}\geq 3,\;{\rm rank}\,\mathsf{A}=1$ we find the evolution equation
\[
u_t=f(t,x)u_3+\frac{[c_t-c_{xxx}f(t,x)]}{c_{xx}}u_2 + g(t, x),
\]
which is linear in $u$ and its derivatives.

Similarly, the rank-two algebra $\mathsf{A}=\langle \gen t, \gen u, x\gen u, c(x)\gen u \rangle,\; c''(x)\neq 0$  gives the equation
\[
u_t=f(x)u_3-\frac{c^{(3)}(x)f(x)}{c^{(2)}(x)}u_2 + g(x).
\]

\bigskip We have the following list of {\bf non-linearized equations} admitting abelian Lie algebras as symmetries:

\begin{enumerate}

\item $\mathsf{A}=\langle \gen t\rangle:$

\[
u_t=F(x,u,u_1, u_2)u_3 + G(x,u,u_1, u_2).
\]

\item $\mathsf{A}=\langle \gen u\rangle:$

\[
u_t=F(t,x,u_1, u_2)u_3 + G(t,x,u_1, u_2).
\]

\item $\mathsf{A}=\langle \gen t, \gen u\rangle:$

\[
u_t=F(x,u_1, u_2)u_3 + G(x,u_1, u_2).
\]

\item $\mathsf{A}=\langle \gen x, \gen u\rangle:$

\[
u_t=F(t,u_1, u_2)u_3 + G(t,u_1, u_2).
\]

\item $\mathsf{A}=\langle \gen u, x\gen u\rangle:$

\[
u_t=F(t,x,u_2)u_3 + G(t,x,u_2).
\]

\item $\mathsf{A}=\langle \gen t, \gen x, \gen u\rangle:$

\[
u_t=F(u_1, u_2)u_3 + G(u_1, u_2).
\]

\item $\mathsf{A}=\langle \gen t, \gen u, x\gen u\rangle:$

\[
u_t=F(x,u_2)u_3 + G(x,u_2).
\]
\end{enumerate}

\section{Semi-simple Lie algebras as symmetries of \eqref{eveqn}.} We now come to the semi-simple Lie algebras, and we recall some basic facts about them.

First, recall that a set $\mathscr{A}$ of linear operators acting non-trivially on a vector space $V$ is said to act {\em semi-simply} if we may decompose $V$ as $V=V_1\oplus\dots\oplus V_m$ where each $V_i, i=1,\dots, m$, is invariant under the action of $\mathscr{A}$ and is irreducible under this action in the sense that no $V_i$ contains a proper subspace invariant under $\mathscr{A}$.

Now, any Lie algebra $\mathfrak{g}$ acts on itself through the adjoint action ${\rm ad}_X(Y)=[X,Y]$ for all $X, Y\in \mathfrak{g}$. Then $\mathfrak{g}$ is a {\em semi-simple} Lie algebra if the set of operators $\mathscr{A}=\{{\rm ad}_X: X\in \mathfrak{g}\}$ acts non-trivially and semi-simply on $\mathfrak{g}$. We say that $\mathfrak{g}$ is {\em simple} if there are no proper invariant subspaces of $\mathfrak{g}$ under the action of $\mathscr{A}=\{{\rm ad}_X: X\in \mathfrak{g}\}$. Thus, every semi-simple Lie algebra $\mathfrak{g}$ can be written as a direct sum of ideals

\[
\mathfrak{g}=\mathfrak{a}_1\oplus\mathfrak{a}\oplus\dots\oplus\mathfrak{a}_m
\]
where the $\mathfrak{a}_i,\; i=1,\dots, m$ are simple Lie algebras. Note that we consider only real Lie algebras.

From these preliminary remarks, it follows that we need only consider the simple Lie algebras as symmetry Lie algebras of our evolution equations and then combine the symmetry operators for the different simple Lie algebras to generate the equations invariant under more general semi-simple Lie algebras.

There are no simple Lie algebras of dimension less than three, and there are two real simple Lie algebras of dimension three: $\So(3, \mathbb{R})$ and  $\Sl(2, \mathbb{R})$, and they can be given the following commutation structure:

\[
\So(3, \mathbb{R})=\langle e_1, e_2, e_3\rangle: [e_1, e_2]=e_3,\;\; [e_2, e_3]=e_1,\,\; [e_3, e_1]=e_2,
\]
and

\[
\Sl(2, \mathbb{R})=\langle e_1, e_2, e_3\rangle: [e_1, e_2]=2e_2,\;\; [e_1, e_3]=-2e_3,\,\; [e_2, e_3]=e_1.
\]

\subsection{Realizations of ${\rm so}(3, \mathbb{R})$:}
\begin{thm}\label{so3} Any realization $\langle Q_1, Q_2, Q_3\rangle$ of the algebra $\So(3, \mathbb{R})$\ by vector fields of the form $Q=a(t)\gen t + b(t,x,u)\gen x + c(t,x,u)\gen u$, as a
symmetry algebra of \eqref{eveqn} is equivalent, under the equivalence group $\mathscr{E}$ of equation \eqref{eveqn}, to the following realization:
\begin{eqnarray}\label{so3realization}
\langle \gen x,\ \tan u\, \sin x\, \gen x + \cos x\,
\gen u,\ \tan u\, \cos x\, \gen x - \sin x\, \gen u
\rangle.
\end{eqnarray}
The functions $F, G$ of equation \eqref{eveqn} are given by

\[
F=\frac{\sec^3 u}{(1+\omega^2)^{\frac{3}{2}}} f(t,\tau),\]

\[
\begin{split}
G=&\left[ 9\omega\tau\tan
u-3\omega\tau^2(1+\omega^2)^{\frac{1}{2}} +
\frac{\omega(1+2\omega^2)}{(\omega^2+1)^{\frac{3}{2}}}
 - \frac{\omega(5+6\omega^2)\tan^2 u}
{(\omega^2+1)^{\frac{3}{2}}}
\right] f(t, \tau)\\
& + (\omega^2+1)^{\frac{1}{2}}h(t,\tau),
\end{split}
\]
where
\[
\omega=u_1\sec u,  \quad \tau =\frac{u_2\sec^2
u+(1+2\omega^2)\tan u}{(1+\omega^2)^{\frac{3}{2}}}.
\]
\end{thm}

\smallskip\noindent{\bf Proof:} First we note that there are no realizations of rank one: for if $Q_1=fX,\, Q_2=gX,\, Q_3=X$ where $X$ is a (locally) smooth vector field and $f, g$ are (locally) smooth functions, then the commutation relations $[Q_1, Q_2]=Q_3,\; [Q_2, Q_3]=Q_1,\; [Q_3, Q_1]=Q_2$ gives $g=Xf,\; f=-Xg,\; fXg-gXf=1$ which in turn gives $f^2+g^2+1=0$ which has no solution for real functions $f,\, g$.

For realizations of rank greater than or equal to two, we must have $Q_1\wedge Q_2\neq 0,\; Q_1\wedge Q_3\neq 0,\; Q_2\wedge Q_3\neq 0$. In fact, we have the Lie derivatives $[Q_2, Q_2\wedge Q_3]= Q_2\wedge Q_1$ and $[Q_2, Q_2\wedge Q_1]=Q_3\wedge Q_2$, from which we see that $Q_2\wedge Q_3=0$ if and only if $Q_1\wedge Q_2=0$. Further $[Q_3, Q_2\wedge Q_3]=Q_3\wedge Q_1$ and $[Q_3, Q_3\wedge Q_1]=Q_3\wedge Q_2$. Hence, if any one of the given exterior products is zero, then all of them are zero. For a realization of rank greater than one, at least one of the exterior products must be different from zero, and therefore all of them must be different from zero.

We may choose $Q_1=\gen t$ or $Q_1=\gen x$. If $Q_3=\gen t$, then with $Q_2=a(t)\gen t + b(t,x,u)\gen x,\; Q_3=A(t)\gen t + B(t,x,u)\gen x + C(t,x,u)\gen u$, we find, using the commutation relations $[Q_1, Q_2]=Q_3$ and $[Q_3, Q_1]=Q_2$ that $\dot{a}=A,\; \dot{A}=-a,\; a\dot{A}-A\dot{a}=1$, which follow from the commutation relations, giving $a(t)^2+A(t)^2+1=0$ and this has no solution for real $a(t),\, A(t)$. Hence we can not have $Q_1=\gen t$.

With $Q_1=\gen x$, the commutation relations give $b_x=B,\; B_x=-b$ and $c_x=C,\; C_x=-c$ as well as $a(t)=A(t)=0$ and so we have only a rank-two realization. We find
\[
b=\alpha(t,u)\cos(x + \phi(t,u)),\; c=\beta(t,u)\cos(x + \psi(t,u)),
\]
and
\[
B=-\alpha(t,u)\sin(x + \phi(t,u)),\; C=-\beta(t,u)\sin(x + \psi(t,u)).
\]
and we must have $b\neq 0,\, B\neq 0,\,c\neq 0,\, C\neq 0$ for a rank-two realization, since all wedge products $Q_1\wedge Q_2,\, Q_1\wedge Q_3,\, Q_2\wedge Q_3$ must be nonzero, as noted above. Thus $\alpha(t,u)\neq 0,\, \beta(t,u)\neq 0$.
So we have
\[
Q_2=\alpha(t,u)\cos(x+\phi(t,u))\gen x + \beta(t,u)\cos(x+\psi(t,u))\gen u.
\]
The proof then proceeds as in \cite{Basarab-Horwath01} and we obtain the canonical forms
\begin{align*}
Q_2&=\tan u\sin x\gen x + \cos x\gen u\\
Q_3&=\tan u\cos x\gen x - \sin x\gen u.
\end{align*}
Substituting these into the symmetry condition for the given evolution equation we find, after some elaborate calculations, that
\[
F=\frac{\sec^3u}{(1+\omega^2)^{3/2}}f(t, \tau).
\]
and
\[
\begin{split}
G=&\left[ 9\omega\tau\tan
u-3\omega\tau^2(1+\omega^2)^{\frac{1}{2}} +
\frac{\omega(1+2\omega^2)}{(\omega^2+1)^{\frac{3}{2}}}
 - \frac{\omega(5+6\omega^2)\tan^2 u}
{(\omega^2+1)^{\frac{3}{2}}}
\right] f(t, \tau)\\
& + (\omega^2+1)^{\frac{1}{2}}h(t,\tau),
\end{split}
\]
where we define
\[
\omega=u_1\sec u,\quad \tau=\frac{u_2\sec^2u + (1+2\omega^2)\tan u}{(1+\omega^2)^{3/2}}.
\]
The equations are awkward to solve, and it is useful to change variable $(t,x,u,u_1, u_2)\to (t,x,u, \omega, \tau)$ in order to make the integration somewhat easier to perform.

\subsection{Realizations of ${\rm sl}(2, \mathbb{R})$:}

\begin{thm}\label{sl2}
Any realization of $\Sl(2, \mathbb{R})$ as vector fields $\langle
Q_1,\, Q_2,\, Q_3\rangle$, of the form $Q=
a(t)\gen t + b(t,x,u)\gen x + c(t,x,u)\gen u$, admissible as a symmetry algebra of equation (\ref{eveqn}) is
equivalent under the equivalence group $\mathscr{E}$ of equation \eqref{eveqn} to one of the following six canonical forms:
\begin{subequations}\label{sl2realiz}
\begin{eqnarray}
& &\langle 2t\gen t + 2x\gen x, -t^2\gen t - (x^2
+2xt)\gen x,
\gen t\rangle,\\
& &\langle 2t\gen t + 2x\gen x, -t^2\gen t
-2xt\gen x -x\gen u,
\gen t\rangle,\\
& &\langle 2x\gen x, -x^2\gen x, \gen x\rangle,\\
& &\langle 2x\gen x + 2u\gen u, -x^2\gen x - 2xu\gen u,
\gen x\rangle,\\
& &\langle 2x\gen x + 2u\gen u, -(x^2-u^2)\gen x - 2xu\gen u, \gen x\rangle,\\
& &\langle 2x\gen x + 2u\gen u, -(x^2 + u^2)\gen x - 2xu\gen u, \gen x\rangle.
\end{eqnarray}
\end{subequations}
\end{thm}

\smallskip\noindent{\bf Proof:} We use the commutation relations

\[
[Q_1, Q_2]= 2Q_2,\quad [Q_3, Q_1] = 2Q_3,\quad [Q_2, Q_3]= Q_1.
\]
We note that $\langle Q_3, Q_1\rangle$ is just a manifestation of the two-dimensional solvable Lie algebra $\mathsf{A}_{2.2}$. Thus we see that $\Sl(2, \mathbb{R})$ is a semi-simple extension of a solvable Lie algebra. We shall return to this point later in this article when we look at extensions of solvable Lie algebras which give extensions of $\Sl(2, \mathbb{R})$ by solvable Lie algebras.

\medskip\noindent The algebra $\mathsf{A}_{2.2}$ is usually given as $\mathsf{A}_{2.2}=\langle e_1, e_2\rangle:\; [e_1, e_2]=e_1$ and on putting $Q_1=2e_2,\; Q_3=e_1$ we obtain $[Q_3, Q_1]=2Q_3$. The inequivalent canonical forms for the admissible realizations of $\mathsf{A}_{2.2}$ are
\begin{align*}
\langle e_1, e_2\rangle&=\langle \gen t, t\gen t + x\gen x\rangle\\
\langle e_1, e_2\rangle&=\langle \gen x, t\gen t + x\gen x\rangle\\
\langle e_1, e_2\rangle&=\langle \gen x, x\gen x + u\gen u\rangle\\
\langle e_1, e_2\rangle&=\langle \gen x, x\gen x\rangle.
\end{align*}
For more details on this, see Part II (\cite{Basarab-Horwath-10(II)})
These then give us the following admissible canonical forms for $Q_1, Q_3$:
\begin{align*}
\langle Q_1, Q_3\rangle&=\langle 2t\gen t + 2x\gen x, \gen t\rangle\\
\langle Q_1, Q_3\rangle&=\langle 2t\gen t + 2x\gen x, \gen x\rangle\\
\langle Q_1, Q_3\rangle&=\langle 2x\gen x + 2u\gen u, \gen x\rangle\\
\langle Q_1, Q_3\rangle&=\langle 2x\gen x, \gen x\rangle.
\end{align*}
We note immediately that if $\langle Q_1, Q_3\rangle=\langle 2t\gen t + 2x\gen x, \gen x\rangle$ then, with $Q_2=a(t)\gen t + b(t,x,u)\gen x + c(t,x,u)\gen u$, the commutation relation $[Q_2, Q_3]=Q_1$ is impossible to implement. Thus we have only the following cases to consider:
\begin{align*}
\langle Q_1, Q_3\rangle&=\langle 2t\gen t + 2x\gen x, \gen t\rangle\\
\langle Q_1, Q_3\rangle&=\langle 2x\gen x + 2u\gen u, \gen x\rangle\\
\langle Q_1, Q_3\rangle&=\langle 2x\gen x, \gen x\rangle.
\end{align*}
If $Q_1=2x\gen x + 2u\gen u$ or $Q_1=2x\gen x$ then $[Q_1, Q_2]=2Q_2$ gives $a(t)=0$ so that we have only a rank-two or rank-one realization for each of these cases. For rank-three realizations we must have $Q_3=\gen t$.

\smallskip\noindent{\bf Rank $\mathbf 1$ realizations:} There is only one possibility for a rank-one realization, and this is $\langle Q_1, Q_3\rangle=\langle 2x\gen x, \gen x\rangle$. In this case the commutation relations give $Q_2=-x^2\gen x$, and we have the rank-one realization $\Sl(2, \mathbb{R})=\langle 2x\gen x, -x^2\gen x, \gen x\rangle$. We can not obtain realization of any other rank in this case.

\smallskip\noindent{\bf Rank $\mathbf 3$ realizations:} As noted above, for a rank-three realization we must have $\langle Q_1, Q_3\rangle=\langle 2t\gen t + 2x\gen x, \gen t\rangle$. the commutation relations and the requirement of a rank-three realization give us $Q_2=-t^2+[-2tx + b(u)x^2]\gen x + c(u)x\gen u$ with $c(u)\neq 0$. The residual equivalence group $\mathcal{E}(Q_1,Q_3)$ is given by invertible transformations of the form $t'=t,\, x'= p(u)x,\; u'= U(u)$ with $U'(u)\neq 0,\; p(u)\neq 0$. Under such a transformation, $Q_2$ is mapped to

\[
Q'_2=-{t'}^2\gen {t'} + [(b(u)p(u)+c(u)p'(u))x^2-2tx']\gen {x'} + c(u)U'(u)x\gen {u'}.
\]
Because $c(u)\neq 0$ we may always choose $p(u)$ so that $b(u)p(u)+c(u)p'(u)=0$, and we may choose $U(u)$ so that $c(u)U'(u)=-p(u)$ from which it follows that we have

\[
Q_2=-{t}^2\gen t - 2tx\gen t - x\gen u
\]
in canonical form. This then gives us $\Sl(2, \mathbb{R})=\langle 2t\gen t + 2x\gen x, -t^2\gen t-2tx\gen x - x\gen u, \gen t\rangle$ in the canonical form of the rank-three realization.

\smallskip\noindent{\bf Rank $\mathbf 2$ realizations:}
We have either $\langle Q_1, Q_3\rangle=\langle 2t\gen t + 2x\gen x, \gen t\rangle$ or $\langle Q_1, Q_3\rangle=\langle 2x\gen x + 2u\gen u, \gen x\rangle$.

\medskip\noindent\underline{$\langle Q_1, Q_3\rangle=\langle 2t\gen t + 2x\gen x, \gen t\rangle$}:
The commutation relations for $\Sl(2, \mathbb{R})$ and the requirement of a rank-two realization give us $Q_2=-t^2\gen t + [b(u)x^2-2tx]\gen x$. If $b(u)=0$ then we have the realization $\langle 2t\gen t + 2x\gen x, -t^2-2tx\gen x, \gen t\rangle$. This realization gives $xu_1=0$ in the equation for $G$ and so we conclude that this is an inadmissible realization and that we must have $b(u)\neq 0$. Under a transformation of the residual equivalence group $\mathcal{E}(Q_1,Q_3): t'=t,\, x'= p(u)x,\; u'= U(u)$ with $U'(u)\neq 0,\; p(u)\neq 0$ the operator $Q_2$ is mapped to
\[
Q'_2=-{t'}^2\gen {t'} + [\frac{b(u)}{p(u)}x'^2-2tx']\gen {x'}
\]
and we choose $p(u)=-b(u)$ so that we obtain $Q_2=-t^2\gen t -[x^2+2tx]\gen x$ in canonical form. So we have the realization $\Sl(2, \mathbb{R})=\langle 2t\gen t+2x\gen x -t^2\gen t -[x^2+2tx]\gen x, \gen t\rangle$.

\medskip\noindent\underline{$\langle Q_1, Q_3\rangle=\langle 2x\gen x + 2u\gen u, \gen t\rangle$}: The commutation relations give $Q_2=(-x^2+\sigma(t)u^2)\gen x + (\tau(t)u^2-2xu)\gen u$. We have the residual equivalence group $\mathcal{E}(Q_1, Q_3): t'=T(t),\; x'=x+p(t)u,\; u'=q(t)u$ with $\dot{T}(t)\neq 0,\; q(t)\neq 0$. Under such a transformation $Q_2$ is mapped to
\[
Q'_2=[-x^2+ (\sigma(t)+p(t)\tau(t))u^2-2p(t)xu]\gen {x'} + [\tau(t)u^2-2xu]q(t)\gen {u'}.
\]
Now we have, after some calculation,
\[
[\tau(t)u^2-2xu]q(t)=\frac{\tau(t)+2p(t)}{q(t)}{u'}^2- 2x'u'.
\]
We choose $p(t)$ so that $\tau(t)+2p(t)=0$. We also have (again after some calculation)
\[
[-x^2+ (\sigma(t)+p(t)\tau(t))u^2-2p(t)xu]=-{x'}^2 + \frac{\sigma(t)+p(t)\tau(t)+p(t)^2}{q(t)^2}{u'}^2,
\]
and with $\tau(t)+2p(t)=0$ we have
\[
[-x^2+ (\sigma(t)+p(t)\tau(t))u^2-2p(t)xu]=-{x'}^2 + \frac{\sigma(t)-p(t)^2}{q(t)^2}{u'}^2.
\]
Hence
\[
Q'_2=\left[-{x'}^2 + \frac{\sigma(t)-p(t)^2}{q(t)^2}{u'}^2\right]\gen {x'} - 2x'u'\gen {u'}.
\]
We have three cases: $\sigma(t)-p(t)^2=0,\;\; \sigma(t)-p(t)^2>0,\;\; \sigma(t)-p(t)^2<0$. If $\sigma(t)-p(t)^2=0$ we have $Q'_2=-{x'}^2\gen {x'} - 2x'u'\gen {u'}$. If $\sigma(t)-p(t)^2>0$ choose $q(t)$ so that $q(t)^2=\sigma(t)-p(t)^2$ and we obtain $Q'_2=-({x'}^2-{u'}^2)\gen {x'} - 2x'u'\gen {u'}$. If $\sigma(t)-p(t)^2<0$ we choose $q(t)$ so that $q(t)^2=p(t)^2-\sigma(t)$ and we obtain $Q'_2=-({x'}^2+{u'}^2)\gen {x'} - 2x'u'\gen {u'}$. This gives three inequivalent realizations of $\Sl(2, \mathbb{R})$: $\langle 2x\gen x + 2u\gen u, -x^2\gen x - 2xu\gen u, \gen x\rangle$, $\langle 2x\gen x + 2u\gen u, -(x^2-u^2)\gen x - 2xu\gen u, \gen x\rangle$ and $\langle 2x\gen x + 2u\gen u, -(x^2+u^2)\gen x - 2xu\gen u, \gen x\rangle$.

\subsection{Equations invariant under ${\rm sl}(2, \mathbb{R})$:}
\begin{enumerate}

\item \underline{$\Sl(2, \mathbb{R})=\langle 2t\gen t + 2x\gen x,
-t^2\gen t-(x^2+2xt)\gen x, \gen t \rangle$:}

\[
u_t=\frac{1}{x^2u^4_1}f(u, \tau)u_3 +\frac{6}{x^3u^2_1}\left(\tau - \frac{1}{xu_1}\right)f(u,\tau) + \frac{1}{x^2u_1}g(u, \tau) + u_1
\]
with $\displaystyle \tau=\frac{u_2}{u_1^2} +
\frac{2}{xu_1}$.

\item\underline{$\Sl(2, \mathbb{R})=\langle 2t\gen t + 2x\gen x,
-t^2\gen t-2tx\gen x - x\gen u, \gen t \rangle$:}

\[
u_t=x^2f(\omega, \tau)u_3+x^{-1}[g(\omega, \tau) - u^2-2u\omega]
\]
with $\displaystyle \omega = xu_1-u,\;\; \tau=x^2u_2$.

\item\underline{$\Sl(2, \mathbb{R})= \langle 2x\gen x, -x^2\gen x,
\gen x \rangle$:}

\[
u_t=f(t,u)\left(\frac{u_3}{u^3_1}-\frac{3}{2}\frac{u_2^2}{u_1^4}\right) +
g(t,u).
\]

\item\underline{$\Sl(2,\mathbb{R})=\langle 2x\gen x + 2u\gen u,
-x^2\gen x - 2xu\gen u, \gen x\rangle$:}

\[
u_t=u^3f(t, \tau)u_3 + ug(t,\tau)
\]
where $\displaystyle \tau= 2uu_2-u^2_1$.

\item\underline{$\Sl(2, \mathbb{R})=\langle 2x\gen x + 2u\gen u,
-(x^2-u^2)\gen x - 2xu\gen u, \gen x\rangle$:}

\begin{equation*}
u_t=u^3f(t, \tau)\left[\frac{u_3}{(1+u^2_1)^{3/2}}-\frac{3u_1u^2_2}{(1+u^2_1)^{5/2}}\right] + u\sqrt{1+u^2_1}h(t, \tau)
\end{equation*}
where $\displaystyle \tau=\frac{1+u^2_1+uu_2}{(1+u^2_1)^{3/2}}$.

\item\underline{$\Sl(2, \mathbb{R})=\langle 2x\gen x + 2u\gen u,
-(x^2+u^2)\gen x - 2xu\gen u, \gen x\rangle$:}

\begin{equation*}
u_t=u^3f(t, \tau)\left[\frac{u_3}{|1-u^2_1|^{3/2}}-\frac{3u_1u^2_2}{|1-u^2_1|^{5/2}}\right] + u\sqrt{|1-u^2_1|}h(t, \tau)
\end{equation*}
where $\displaystyle \tau=\frac{|1-u^2_1|+uu_2}{|1-u^2_1|^{3/2}}$. Note that here we must take the two cases $u^2_1>1$ and $u^2_1<1$ separately.
\end{enumerate}
This disposes of the two simple Lie algebras of lowest dimension. For all other simple or semi-simple Lie algebras we have the following result:

\begin{thm}
The realizations of the algebras\ $\So(3, \mathbb{R})$\ and\
$\Sl(2,{\mathbb{R}})$, given in Theorems \ref{so3} and \ref{sl2},
exhaust the set of all possible realizations of semi-simple Lie
algebras by operators of the form $Q=a(t)\gen t + b(t,x,u)\gen x + c(t,x,u)\gen u$ which are admitted as symmetries of equation \eqref{eveqn}.
\end{thm}

\smallskip\noindent{\bf Proof:} The proof of this theorem is the same as that of the corresponding result in \cite{Basarab-Horwath01}, so that no semi-simple Lie algebra of higher dimension gives an admissible symmetry algebra of equation (\ref{eveqn}).

\section{Equations invariant under semi-direct extensions of semi-simple Lie algebras.} Any Lie algebra $\mathfrak{g}$ has a unique decomposition as the semi-direct sum $\mathfrak{g}=\mathsf{h}\uplus\mathsf{A}$ where $\mathsf{h}$ is a semi-simple Lie algebra and $\mathsf{A}$ is a solvable Lie algebra. This is just the Levi decomposition theorem for Lie algebras (see, for instance, \cite{Varadarajan}). Then $\mathsf{A}$ is a Lie-algebra ideal of $\mathfrak{g}$ under the adjoint action of $\mathfrak{g}$ on itself. If the adjoint action of the semi-simple Lie algebra $\mathsf{h}$ on $\mathsf{A}$ is trivial (that is ${\rm ad}_X(Y)=0$ for all $X\in \mathsf{h},\; Y\in \mathsf{A}$) then we have a Lie-algebra direct sum decomposition $\mathfrak{g}=\mathsf{h}\oplus\mathsf{A}$.

\medskip The adjoint action of $\mathsf{h}$ on $\mathsf{A}$ is the {\em adjoint representation} of $\mathsf{h}$ on $\mathsf{A}$. A {\em finite-dimensional representation} of a Lie algebra $\mathfrak{g}$ is a Lie-algebra homomorphism $\rho: \mathfrak{g}\to \mathsf{gl}(V)$ where $\mathsf{gl}(V)$ is the Lie algebra of all endomorphisms of a finite-dimensional vector space $V$. It is known (see \cite{Varadarajan}) that every finite-dimensional representation of a semi-simple Lie algebra $\mathfrak{g}$ is {\em completely reducible}, that is, if $\rho: \mathfrak{g}\to \mathsf{gl}(V)$ is the representation then the set of operators $\{\rho(X): X\in \mathfrak{g}\}$ acts completely reducibly on $V$ so that if $U\subset V$ is invariant under $\{\rho(X): X\in \mathfrak{g}\}$, there exists a subspace $W\subset V$ with $W$ also invariant under $\{\rho(X): X\in \mathfrak{g}\}$ and with the property that $V=U\dot{+} W$ (the symbol $\dot{+}$ stands for vector-space direct sum). From this it follows that $V$ is given by a direct-sum decomposition $V=V_1\dot{+}\dots\dot{+} V_m$ so that $\{\rho(X): X\in \mathfrak{g}\}$ acts irreducibly (that is,there is no non-trivial invariant subspace) on each of the $V_1,\dots, V_m$.

Applying these remarks to $\mathfrak{g}=\mathsf{h}\uplus \mathsf{A}$, we have then the vector-space direct-sum decomposition $\mathsf{a}=\mathsf{a}_1\dot{+}\dots\dot{+}\mathsf{a}_m$, and each $\mathsf{a}_i,\, i=1,\dots, m$ is invariant under $\{{\rm ad}_X: X\in \mathsf{h}\}$, and is irreducible under this action. Note that this is in general a vector-space direct sum rather than a Lie-algebra direct sum, and that we do not necessarily have $[Y_i, Y_j]=0$ when $Y_i\in \mathsf{a}_i,\, Y_j\in \mathsf{a}_j,\; i\neq j$. Further, it is not obvious that each $\mathsf{a}_i,\, i=1,\dots, m$ is a Lie algebra. However, if each $\mathsf{a}_i,\, i=1,\dots, m$ is a Lie algebra then clearly $\mathsf{a}_i$ is solvable. Further, it is also abelian: $[\mathsf{a}_i, \mathsf{a}_i]$ is then a proper subspace of $\mathsf{a}_i$ and is invariant under the action of $\mathsf{h}$, and because $\mathsf{h}$ acts irreducibly on each $\mathsf{a}_i$, then we must have $[\mathsf{a}_i, \mathsf{a}_i]=0$.

Since $\mathsf{A}$ is solvable, we have the descending chain of ideals

\[
\mathsf{A}\rhd \mathscr{D}^1\mathsf{A}\rhd \mathscr{D}^2\mathsf{A}\rhd\dots \mathscr{D}^k\mathsf{A}\rhd\mathscr{D}^{k+1}\mathsf{A}=\{0\},
\]
where $\mathscr{D}^1\mathsf{A}=[\mathsf{A}, \mathsf{A}],\mathscr{D}^{i+1}\mathsf{A}=[\mathscr{D}^i\mathsf{A}, \mathscr{D}^i\mathsf{A}]$ for $i=1,\dots, k$. In particular $\mathscr{D}^{k+1}\mathsf{A}=\{0\}$. Thus $\mathsf{A}_0=\mathscr{D}^k\mathsf{A}$ is an abelian ideal of $\mathsf{A}$, and it is easy to see that $\mathsf{A}_0$ is invariant under $\mathsf{h}$. Clearly there are two cases to consider: $\mathscr{D}^1\mathsf{A}=\{0\}$ when $\mathsf{A}$ is abelian, and $\mathscr{D}^{k+1}(\mathsf{A})=\{0\}$ for $k\geq 1$. We always have $1\leq \dim \mathsf{A}_0\leq \dim \mathsf{A}$, and we have the decomposition
\[
\mathsf{A}=(\mathsf{a}_1\dot{+}\dots\dot{+}\mathsf{a}_m)\dot{+}\mathsf{A}_0,
\]
where the $\mathsf{a}_1, \dots, \mathsf{a}_m$ are the irreducible subspaces of $\mathsf{A}$ under the action of the semi-simple Lie algebra $\mathsf{h}$ so that $\mathsf{a}_1\dot{+}\dots\dot{+}\mathsf{a}_m$ is the complement of the abelian ideal $\mathsf{A}_0$ in $\mathsf{A}$. Since $\mathsf{A}_0$ is an ideal of $\mathsf{A}$ then it is invariant under the action of $\mathsf{a}_1\dot{+}\dots\dot{+}\mathsf{a}_m$.

Our presentation is as follows: we first consider the representations of $\So(3, \mathbb{R})$ and $\Sl(2, \mathbb{R})$ on abelian Lie algebras $A$, and we show that there are no proper semi-direct sum extensions with abelian Lie algebras $A$ except for one particular case of $\Sl(2, \mathbb{R})$ which admits a two-dimensional semi-direct sum extensions.  Then we construct all the direct-sum extensions by abelian Lie algebras, these algebras $A$ being the direct-sums of one-dimensional representations. We then consider the extension of each of abelian ideal $\mathsf{A}_0$ by irreducible subspaces $\mathsf{k}$ of the solvable algebra $\mathsf{A}$, that is we consider extensions $\mathsf{k}\dot{+}\mathsf{A}_0$ so that our simple Lie algebra acts irreducibly on $\mathsf{k}$ and such that $[\mathsf{k}, \mathsf{A}_0]\subset \mathsf{A}_0$.  We show that these do not exist.

\subsection{Semi-direct sums of  ${\rm so}(3, \mathbb{R})$ with solvable Lie algebras.}

As noted at the beginning of the section, we first take up the representation of our semi-simple Lie algebras on abelian algebras. We have the following results for $\So(3, \mathbb{R})$:

\begin{thm}\label{so3irreps} There is only one (up to equivalence) irreducible representation of
\[
\So(3, \mathbb{R})=\langle\gen x, \tan u\sin x\gen x + \cos x\gen u, \tan u\cos x\gen x - \sin x\gen u\rangle
\]
on an abelian Lie algebra $\mathsf{A}$: $\dim A=1$ and we may take $\mathsf{A}=\langle \gen t\rangle$.
\end{thm}

\smallskip\noindent{\bf Proof:} We look at $\dim\mathsf{A}>1$. We have no representation for $\dim \mathsf{A}=2$ (see Appendix \ref{irreps}). So, let us consider $\dim\mathsf{A}\geq 3$. For any irreducible representation of $\So(3, \mathbb{R})$ on a real Lie algebra $\mathsf{A}$ (whether abelian or not) we have $\dim \mathsf{A}=2J+1$ for some half-integer $J$,  and there exist non-zero vectors $X, Y$ with $[e_1, X]=\alpha Y, \; [e_1, Y]=-\alpha X$ for $\alpha>0$ for any $J>0$ and non-zero vectors $Z$ with $[e_1, Z]=0$ if $J$ is an integer (see Appendix \ref{irreps}).

Now, let $\mathsf{A}$ be abelian and $J\geq 1$ (so that $\dim\mathsf{A}\geq 3$). If $X, Y$ satisfy $[e_1, X]=-JY,\; [e_1, Y]=JX$ then it follows that $X=b\gen x + c\gen u$, $Y=B\gen x + C\gen u$ if $X, Y$ are to be symmetries of our equation, since $e_1, e_2, e_3$ are of this form. In fact, all the vector fields of $\mathsf{A}$ are of this form, since $\mathsf{A}$ is generated from $X$ by applying $\So(3, \mathbb{R})$ to $X$ (this is just a restatement of the fact that $\mathsf{A}$ is irreducible under $\So(3, \mathbb{R})$). Thus $\mathrm{rank}\, \mathsf{A}=1$ or $\mathrm{rank}\, \mathsf{A}=2$.

The equations $[e_1, X]=-JY,\; [e_1, Y]=JX$ have the solution
\begin{align*}
&X=\alpha(t,u)\cos(Jx+\theta(t,u))\gen x + \beta(t,u)\sin(Jx+\phi(t,u))\gen u,\\ &Y=\alpha(t,u)\sin(Jx+\theta(t,u))\gen x - \beta(t,u)\cos(Jx+\phi(t,u))\gen u.
\end{align*}
Further
\begin{multline*}
[e_2, X]=[b_x\tan u\sin x+b_u\cos x-b\tan u\cos x - c(1+\tan^2u)\sin x]\gen x\\ + [c_x\tan u\sin x + c_u\cos x + b\sin x]\gen u.
\end{multline*}

\medskip\noindent\underline{$\mathrm{rank}\, \mathsf{A}=1$:} In this $[e_2, X]\wedge X=0$ and we find that $b=0$ if and only if $c=0$. Thus we must have $b\neq 0,\, c\neq 0$ so that $\alpha(t,u),\, \beta(t,u)\neq 0$. For $\mathrm{rank}\, \mathsf{A}=1$ we also have $X\wedge Y=0$ which then gives, after some elementary calculation, $\alpha\beta\cos(\phi-\theta)=0$ so that $\phi=\theta+ (2n+1)\pi/2$ and then we obtain

\begin{align*}
&X=\alpha(t,u)\cos(Jx+\theta(t,u))\gen x + (-1)^n\beta(t,u)\cos(Jx+\theta(t,u))\gen u,\\ &Y=\alpha(t,u)\sin(Jx+\theta(t,u))\gen x - (-1)^n\beta(t,u)\sin(Jx+\theta(t,u))\gen u.
\end{align*}
However, $X\wedge Y=0$ then gives $\alpha\beta\cos(Jx+\theta(t,u))\sin(Jx+\theta(t,u))=0$ which is a contradiction since $\alpha\beta=0$. Thus we have no rank-one $\mathsf{A}$ for $\dim \mathsf{A}\geq 3$.

\medskip\noindent\underline{$\mathrm{rank}\, \mathsf{A}=2$:} In this case we must have $\dim \mathsf{A}=3$ or $\dim \mathsf{A}=4$ for a symmetry algebra, since the only cases of a rank-two abelian symmetry algebra of our evolution equation are for $\dim \mathsf{A}=2, 3, 4$ and in this case $\dim \mathsf{A}=2$ is inadmissible as an irreducible representation space.

\smallskip\noindent\underline{$\dim\mathsf{A}=3$:} From \cite{Turkowski88} we have only one (up to equivalence) abelian Lie algebra  $\mathsf{A}=\langle Q_1, Q_2, Q_3\rangle$ on which $\So(3, \mathbb{R})=\langle e_1, e_2, e_3\rangle$ is represented irreducibly and we have the following commutation relations:
\begin{align*}
&[e_1, Q_1]=0,\;\; [e_1, Q_2]=Q_3,\;\;  [e_1, Q_3]=-Q_2\\
&[e_2, Q_1]=-Q_3,\;\; [e_2, Q_2]=0,\;\; [e_2, Q_3]=Q_1\\
&[e_3, Q_1]=Q_2,\;\; [e_3, Q_2]=-Q_1,\;\; [e_3, Q_3]=0.
\end{align*}
Now, if an abelian Lie algebra $\mathsf{A}=\langle Q_1, Q_2, Q_3\rangle$ is a rank-two symmetry algebra then one of the $Q_i$ has non-zero wedge product with the other two, and the other two have zero wedge product. Note that $\So(3, \mathbb{R})$ acts irreducibly on the space of bi-vectors $\langle Q_1\wedge Q_2, Q_1\wedge Q_3, Q_2\wedge Q_3\rangle$:
\begin{align*}
&[e_1, Q_1\wedge Q_2]=Q_1\wedge Q_3,\;\; [e_1, Q_1\wedge Q_3]=-Q_1\wedge Q_2,\;\; [e_1, Q_2\wedge Q_3]=0\\
&[e_2, Q_1\wedge Q_2]=Q_2\wedge Q_3,\;\; [e_2, Q_1\wedge Q_3]=0,\;\; [e_2, Q_2\wedge Q_3]=Q_2\wedge Q_1\\
&[e_3, Q_1\wedge Q_2]=0,\;\; [e_3, Q_1\wedge Q_3]=Q_2\wedge Q_3,\;\; [e_3, Q_2\wedge Q_3]=-Q_1\wedge Q_3.
\end{align*}
Thus, if any one of the bivectors $\langle Q_1\wedge Q_2, Q_1\wedge Q_3, Q_2\wedge Q_3\rangle$ is zero, so are all the others. This implies that there are no rank-two realizations of $\So(3, \mathbb{R})$ in this case.

\smallskip\noindent\underline{$\dim \mathsf{A}=4$:} In this case we have, according to \cite{Turkowski88}, one irreducible abelian representation space (up to equivalence) $\mathsf{A}=\langle Q_1, Q_2, Q_3, Q_4\rangle$ with commutation relations
\begin{align*}
&[e_1, Q_1]=\frac{1}{2}Q_4,\;\; [e_1, Q_2]=\frac{1}{2}Q_3,\;\; [e_1, Q_3]=-\frac{1}{2}Q_2,\;\; [e_1, Q_4]=-\frac{1}{2}Q_1\\
&[e_2, Q_1]=\frac{1}{2}Q_2,\;\; [e_2, Q_2]=-\frac{1}{2}Q_1,\;\; [e_2, Q_3]=\frac{1}{2}Q_4,\;\; [e_2, Q_4]=-\frac{1}{2}Q_3,\\
&[e_3, Q_1]=\frac{1}{2}Q_3,\;\; [e_3, Q_2]=-\frac{1}{2}Q_4,\;\; [e_3, Q_3]=-\frac{1}{2}Q_1,\;\; [e_3, Q_4]=\frac{1}{2}Q_2.
\end{align*}
Again, we find that $\So(3, \mathbb{R})$ acts irreducibly on the space of bivectors $\langle Q_1\wedge Q_2, Q_1\wedge Q_3, Q_1\wedge Q_4, Q_2\wedge Q_3,  Q_2\wedge Q_4, Q_3\wedge Q_4\rangle$ and so if any one of them vanishes, so do all the others. Again, if we have $\dim \mathsf{A}=4,\; \mathrm{rank}\, \mathsf{A}=2$ we have to have three wedge products equal to zero, so all must be zero and this means that we have no irreducible representation in this case either.

Consequently we see that we have no admissible irreducible representations of $\So(3, \mathbb{R})$ on abelian Lie algebras $\mathsf{A}$ with $\dim \mathsf{A}\geq 2$. This then implies that we can only have irreducible representations on abelian $\mathsf{A}$ with $\dim \mathsf{A}=1$. In this case, the vector fields of $\mathsf{A}$ must commute with $\langle e_1, e_2, e_3\rangle$. An elementary calculation shows that if $Q$ commutes with $e_1, e_2, e_3$ then $Q=a(t)\gen t$ with $a(t)\neq 0$. The residual equivalence of $\langle e_1, e_2, e_3\rangle$ consists of transformations of the form $t'=T(t),\; x'=x+2n\pi,\; u'=u + n\pi$ with $\dot{T}(t)\neq 0$. Applying such a transformation, we find that $Q$ is mapped to $Q'=a(t)\dot{T}(t)\gen {t'}$ and we choose $T(t)$ so that $a(t)\dot{T}(t)=1$ giving $Q=\gen t$ in canonical form. This proves the theorem.

\begin{thm}\label{so3semidirectsums} There are no semi-direct sum extensions $\So(3, \mathbb{R})\uplus \mathsf{A}$ for any solvable Lie algebra $\mathsf{A}$ other than the extension $\So(3, \mathbb{R})\oplus \mathsf{A}$ with $\dim \mathsf{A}=1$ as given in Theorem \ref{so3irreps}.
\end{thm}

\smallskip\noindent{\bf Proof:} If $\mathsf{A}$ with $\dim\mathsf{A}\geq 2$ is a solvable Lie algebra on which $\So(3, \mathbb{R})$ acts, then, as already remarked on at the beginning of this section, there is an abelian ideal $\mathsf{A}_0$ of $\mathsf{A}$ which is invariant under the action of $\So(3, \mathbb{R})$. Furthermore, $\dim\mathsf{A}_0=1$. In fact, $\mathsf{A}_0$ decomposes into a direct sum of one-dimensional subalgebras, each being invariant under $\So(3, \mathbb{R})$ and if $\mathsf{A}_0=\langle Q_1,\dots, Q_k\rangle$ then $\mathsf{A}_0=\langle Q_1\rangle\oplus\langle Q_2\rangle\oplus\dots\oplus\langle Q_k\rangle$ and each $Q_i$ commutes with each of the generators of $\So(3, \mathbb{R})$. We may then take $Q_1=\gen t$ and then $Q_i=a_i(t)\gen t$ for $i=2,\dots, k$ and since $\mathsf{A}_0$ is abelian, $a_i(t)=\,{\rm constant}$, so that $\dim{A}_0=1$.  Thus we may take $\mathsf{A}_0=\langle\gen t\rangle$. Now suppose that $\mathsf{k}$ is the complement of $\mathsf{A}_0$ in $\mathsf{A}$: $\mathsf{A}=\mathsf{k}\dot{+}\mathsf{A}_0$ with $\mathsf{k}$ invariant under $\So(3, \mathbb{R})$. Put $\mathsf{k}=\{X_1,\dots, X_k\}$. Now, $\mathsf{k}$ decomposes into a direct vector sum $\mathsf{k}=\mathsf{k}_1\dot{+}\dots\dot{+}\mathsf{k}_m$ for some $m\in \mathbb{N}$. From the proof of Theorem \ref{so3irreps} we know that any vector field $X\in \mathsf{k}_i$ is of the form $X=b(t,x,u)\gen x + c(t,x,u)\gen u$ and hence we have $X_i=b_i(t,x,u)\gen x + c_i(t,x,u)\gen u$. Then since $\mathsf{A}_0=\langle \gen t\rangle$ is an ideal in the algebra, we must have $[X_i, \gen t]=0$. This implies that $b_i=b_i(x,u),\, c_i=c_i(x,u)$, and this implies that $[X_i, X_j]\in \mathsf{k}$. Thus $\mathsf{k}$ is a Lie algebra, and it is solvable because it is a subalgebra of the solvable Lie algebra $\mathsf{A}$, and therefore it contains an abelian ideal invariant under $\So(3, \mathbb{R})$. As we have seen, such an ideal must be one-dimensional, and its generator must then commute with $\So(3, \mathbb{R})$. However, the only operator of the form $X=b(x,u)\gen x + c(x,u)\gen u$ which commutes with $\So(3, \mathbb{R})$ is $X=0$. Thus, we cannot have an extension $\mathsf{k}\dot{+}\mathsf{A}_0$ of $\mathsf{A}_0$. This proves the result.

We now have the following result:

\begin{cor}\label{so3DirectExtension} The Lie algebra\ $\So(3, \mathbb{R})$\  given by
\[
\langle \gen x,\ \tan u\, \sin x\, \gen x + \cos x\,
\gen u,\ \tan u\, \cos x\, \gen x - \sin x\, \gen u
\rangle.
\]
admits only one (up to equivalence) direct sum extension $\So(3, \mathbb{R})\oplus\mathsf{A}$ :
\[
\langle \gen x,\ \tan u\, \sin x\, \gen x + \cos x\,
\gen u,\ \tan u\, \cos x\, \gen x - \sin x\, \gen u
\rangle\oplus\langle \gen t\rangle.
\]
The corresponding non-linearities are
\[
F=\frac{\sec^3 u}{(1+\omega^2)^{\frac{3}{2}}} f(\tau)
\]
and
\[
\begin{split}
G=&\left[ 9\omega\tau\tan
u-3\omega\tau^2(1+\omega^2)^{\frac{1}{2}} +
\frac{\omega(1+2\omega^2)}{(\omega^2+1)^{\frac{3}{2}}}
 - \frac{\omega(5+6\omega^2)\tan^2 u}
{(\omega^2+1)^{\frac{3}{2}}}
\right] f(\tau)\\
& + (\omega^2+1)^{\frac{1}{2}}h(
\tau),
\end{split}
\]
where
\[
\omega=u_1\sec u,  \quad \tau =\frac{u_2\sec^2
u+(1+2\omega^2)\tan u}{(1+\omega^2)^{\frac{3}{2}}}.
\]
\end{cor}

\subsection{Direct-sum extensions of ${\rm sl}(2, \mathbb{R})$ by solvable Lie algebras.} Here we look at direct sum extensions $\Sl(2, \mathbb{R})\oplus \mathsf{A}$ where $\mathsf{A}$ is a solvable Lie algebra: these are the reducible Lie algebras and  $\mathsf{A}=\mathsf{k}_0$ which is a direct vector sum of irreducible $J=0$ representations of $\Sl(2, \mathbb{R})$ (spin zero representations). The elements of $\mathsf{A}$ must then commute with those of $\Sl(2, \mathbb{R})$. In our classification we use the fact that any solvable Lie algebra possesses a descending chain of ideals
\[
\mathsf{A}=\mathsf{a}_n\rhd\mathsf{a}_{n-1}\rhd\dots\rhd\mathsf{a}_1\rhd\mathsf{a}_0=\{0\}
\]
where each $\mathsf{a}_i$ is an ideal in $\mathsf{a}_{i+1}$ of codimension one. In fact, this property characterizes solvable Lie algebras, as is well known. We record our result as the following theorem:

\begin{thm}\label{directsumsthm} We have the following inequivalent admissible direct-sum extensions of $\Sl(2, \mathbb{R})$ by solvable Lie algebras $\mathsf{A}$:

\smallskip\noindent $\dim\mathsf{A}=1:$
\begin{align*}
&\langle 2t\gen t + 2x\gen x,
-t^2\gen t-(x^2+2xt)\gen x, \gen t \rangle\oplus\langle \gen u\rangle\\
&\langle 2t\gen t + 2x\gen x,
-t^2\gen t-2xt\gen x - x\gen u, \gen t \rangle\oplus\langle \gen u\rangle\\
&\langle 2t\gen t + 2x\gen x,
-t^2\gen t-2xt\gen x - x\gen u, \gen t \rangle\oplus\langle x\gen x +  u\gen u\rangle\\
&\langle 2x\gen x ,
-x^2\gen x, \gen x \rangle\oplus\langle \gen t\rangle\\
&\langle 2x\gen x ,
-x^2\gen x, \gen x \rangle\oplus\langle \gen u\rangle\\
&\langle 2x\gen x + 2u\gen u,
-x^2\gen x-2xu\gen u, \gen x \rangle\oplus\langle \gen t\rangle\\
&\langle 2x\gen x + 2u\gen u,
-x^2\gen x-2xu\gen u, \gen x \rangle\oplus\langle u\gen u\rangle\\
&\langle 2x\gen x + 2u\gen u,
-x^2\gen x-2xu\gen u, \gen x \rangle\oplus\langle tu\gen u\rangle\\
&\langle 2x\gen x + 2u\gen u,
-(x^2-u^2)\gen x-2xu\gen u, \gen x \rangle\oplus\langle \gen t\rangle\\
&\langle 2x\gen x + 2u\gen u,
-(x^2+u^2)\gen x-2xu\gen u, \gen x \rangle\oplus\langle \gen t\rangle.
\end{align*}

\noindent $\dim\mathsf{A}=2:$

\begin{align*}
&\langle 2t\gen t + 2x\gen x,
-t^2\gen t-(x^2+2xt)\gen x, \gen t \rangle\oplus\langle \gen u, u\gen u\rangle\\
&\langle 2t\gen t + 2x\gen x,
-t^2\gen t-2xt\gen x - x\gen u, \gen t \rangle\oplus\langle \gen u,  x\gen x +  u\gen u\rangle\\
&\langle 2x\gen x ,
-x^2\gen x, \gen x \rangle\oplus\langle \gen t, \gen u\rangle\\
&\langle 2x\gen x ,
-x^2\gen x, \gen x \rangle\oplus\langle \gen t, t\gen t + u\gen u\rangle\\
&\langle 2x\gen x ,
-x^2\gen x, \gen x \rangle\oplus\langle \gen u, t\gen t + u\gen u\rangle\\
&\langle 2x\gen x + 2u\gen u,
-x^2\gen x-2xu\gen u, \gen x \rangle\oplus\langle \gen t, u\gen u\rangle\\
&\langle 2x\gen x + 2u\gen u,
-x^2\gen x-2xu\gen u, \gen x \rangle\oplus\langle \gen t, t\gen t + qu\gen u\rangle,\;\; q\in \mathbb{R},\; q\neq 0.
\end{align*}

\smallskip\noindent $\dim\mathsf{A}=3:$

\[
\langle 2x\gen x, -x^2\gen x, \gen x\rangle\oplus\langle \gen t, \gen u, t\gen t + \frac{u}{3}\gen u\rangle, \quad q\in \mathbb{R}.
\]
\end{thm}

\smallskip\noindent{\bf Proof:} We give the details of the calculations for the different admissible realizations of $\Sl(2, \mathbb{R})$.

\medskip\noindent{\underline{$\Sl(2, \mathbb{R})=\langle 2t\gen t + 2x\gen x,
-t^2\gen t-(x^2+2xt)\gen x, \gen t \rangle.$}} If a vector field $$Q=a(t)\gen t + b(t,x,u)\gen x + c(t,x,u)\gen u$$ commutes with $e_1, e_2, e_3$ then $Q=c(u)\gen u$, as is easily computed. The residual equivalence group of $\langle e_1, e_2, e_3\rangle$ is given by transformations $t'=t,\quad x'= x, \quad u'=U(u)$ with $U'(u)\neq 0$. Under such a transformation, $Q=c(u)\gen u$ is transformed to $Q'=c(u)U'(u)\gen {u'}$ and we can then choose $U(u)$ so that $c(u)U'(u)=1$. So we have only one admissible one-dimensional extension
\[
\langle 2t\gen t + 2x\gen x,
-t^2\gen t-(x^2+2xt)\gen x, \gen t \rangle\oplus\langle \gen u\rangle.
\]

\smallskip\noindent If $\dim\mathsf{A}=2$ then we have $\mathsf{A}=\langle Q_1, Q_2\rangle$ where  $[Q_1,Q_2]=0$ or $[Q_1, Q_2]=Q_1$ (this is the canonical form of two-dimensional non-abelian Lie algebras). We take $Q_1=\gen u$ in canonical form and put $Q_2=c(u)\gen u$, and then $[Q_1, Q_2]=0$ is impossible since this means that $c'(u)=0$ and then $Q_2=c\gen u$ which contradicts $\dim\mathsf{A}=2$. If we now take $[Q_1, Q_2]=Q_1$ we find that $c'(u)=1$ so that $c(u)=u + \beta$. The residual equivalence group of $\langle e_1, e_2, e_3, Q_1\rangle$ is given by transformations $t'=t, x'=x, u'=u + k$. Under such a transformation $Q_2=(u +\beta)\gen u$ is mapped to $Q'_2=(u +\beta)\gen {u'}=(u'+\beta - k)\gen {u'}$ and we choose $k=\beta$ so that $Q_2=u\gen u$ in canonical form. Thus we have the two-dimensional extension
\[
\langle 2t\gen t + 2x\gen x,
-t^2\gen t-(x^2+2xt)\gen x, \gen t \rangle\oplus\langle \gen u, u\gen u\rangle.
\]
If $\dim\mathsf{A}=3$ we have $A=\langle Q_1, Q_2, Q_3\rangle$ and $\langle Q_1, Q_2\rangle$ is an ideal of $\langle Q_1, Q_2, Q_3\rangle$. We take $Q_1=\gen u, Q_2=u\gen u$ in canonical form and we put $Q_3=c(u)\gen u$. Then $[Q_1, Q_3]=\alpha Q_1+\beta Q_2$ gives $c(u)=\frac{1}{2}\beta u^2 + \alpha u +\gamma$. Further, $[Q_2, Q_3]=\delta Q_1 + \epsilon Q_2$ gives $uc'(u)-c(u)=\epsilon u + \delta$. Now, this gives us $\beta=0$ and so we find that $Q_3=\alpha Q_2+\gamma Q_1$ so that $\dim\mathsf{A}\geq 3$ is impossible.

\medskip\noindent{\underline{$\Sl(2, \mathbb{R})=\langle 2t\gen t + 2x\gen x,
-t^2\gen t-2xt\gen x-x\gen u, \gen t \rangle.$}} We first note that if $Q=a(t)\gen t + b(t,x,u)\gen x + c(t,x,u)\gen u$ commutes with $e_1, e_2, e_3$ then we find that $Q=\alpha(x\gen x + u\gen u) + \beta \gen u$, and from this it follows that we have only one- and two-dimensional direct sum extensions:
\begin{align*}
&\langle 2t\gen t + 2x\gen x,
-t^2\gen t-2xt\gen x-x\gen u, \gen t \rangle\oplus\langle \gen u\rangle\\
&\langle 2t\gen t + 2x\gen x,
-t^2\gen t-2xt\gen x-x\gen u, \gen t \rangle\oplus\langle x\gen x+u\gen u\rangle\\
&\langle 2t\gen t + 2x\gen x,
-t^2\gen t-2xt\gen x-x\gen u, \gen t \rangle\oplus\langle \gen u, x\gen x + u\gen u\rangle.
\end{align*}

\medskip\noindent{\underline{$\Sl(2, \mathbb{R})=\langle 2x\gen x,
-x^2\gen x, \gen x\rangle.$}} In this case, if $Q=a(t)\gen t + b(t,x,u)\gen x + c(t,x,u)\gen u$ commutes with $e_1, e_2, e_3$ then $Q=a(t)\gen t + c(t,u)\gen u$. The residual equivalence group of $\langle e_1, e_2, e_3\rangle$ is given by transformations of the form $t'=T(t),\; x'=x,\; u'=U(t,u)$ with $\dot{T}(t)\neq 0,\; U_u\neq 0$. Under such a transformation $Q$ is mapped to $Q'=a(t)\dot{T}(t)\gen {t'} + [a(t)U_t + c(t,u)U_u]\gen {u'}$ and if $a(t)\neq 0$ then we choose $T(t)$ so that $a(t)\dot{T}(t)=1$ and we choose $U(t,u)$ so that $a(t)U_t + c(t,u)U_u=0$, giving $Q=\gen t$ in canonical form. If $a(t)=0$ then $Q'=c(t,u)U_t,u)\gen {u'}$ and we choose $U(t,u)$ so that $c(t,u)U_u=1$, giving $Q=\gen u$ in canonical form. Thus we have two one-dimensional direct sum extensions:
\begin{align*}
&\langle 2x\gen x,-x^2\gen x, \gen x\rangle\oplus\langle \gen t\rangle\\
&\langle 2x\gen x,-x^2\gen x, \gen x\rangle\oplus\langle \gen u\rangle.
\end{align*}

\smallskip\noindent For $\dim\mathsf{A}=2$ we have $\mathsf{A}=\langle Q_1, Q_2\rangle$ and $[Q_1, Q_2]=0$ or $[Q_1, Q_2]=Q_1$. First, consider $[Q_1, Q_2]=0$. Putting $Q_1=\gen t$ and $Q_2=a(t)\gen t + c(t,u)\gen u$ we find that $\dot{a}(t)=0,\; c_t=0$. Thus we have $\mathsf{A}=\langle Q_1, Q_2\rangle=\langle \gen t, \gen c(u)\gen u\rangle$ with $c(u)\neq 0$. The residual equivalence group of $e_1, e_2, e_3, \gen t$ is given by transformations of the form $t'=t+k,\, x'=x,\, u'=U(u)$ with $U'(u)\neq 0$. Under such a transformation, $Q_2= c(u)\gen u$ is mapped to $Q'_2=c(u)U'(u)\gen {u'}$ and we may choose $U(u)$ so that $c(u)U'(u)=1$ so that $Q_2=\gen u$ in canonical form. This gives the canonical form $\mathsf{A}=\langle \gen t, \gen u\rangle$. If we take $Q_1=\gen u$ then $[Q_1, Q_2]=0$ gives $Q_2=a(t)\gen t + c(t)\gen u$. If $a(t)=0$ then $Q_2=c(t)\gen u$ and putting $\gen u$ and $c(t)\gen u$ into the symmetry equation for $G$ gives $\dot{c}(t)=0$, so that we have $Q_2=cQ_1$ and this contradicts the requirement $\dim\mathsf{A}=2$. Thus $a(t)\neq 0$. The residual equivalence group of $e_1, e_2, e_3, \gen u$ is given by transformations of the form $t'=T(t),\, x'=x,\, u'=u+p(t)$. Under such a transformation, $Q_2$ is mapped to $Q'_2=a(t)\dot{T}(t)\gen {t'} + [a(t)\dot{p}(t)+ c(t)]\gen {u'}$ and we choose $T(t)$ so that $a(t)\dot{T}(t)=1$ and $p(t)$ so that $a(t)\dot{p}(t)+ c(t)=0$, giving $Q_2=\gen t$ in canonical form. Thus we find that $\mathsf{A}=\langle \gen t, \gen u\rangle$ is the canonical form of two-dimensional (admissible) abelian Lie algebras.

\smallskip\noindent For $\mathsf{A}=\langle Q_1, Q_2\rangle$ with $[Q_1, Q_2]=Q_1$ we take $Q_1=\gen t$ and $Q_2=a(t)\gen t + c(t,u)\gen u$, and then the commutation relation gives us $a(t)=t +\beta$ and $c_t=0$ so that $Q_2=(t + \beta)\gen t + c(u)\gen u$. The residual equivalence group of $e_1, e_2, e_3, Q_1$ is given by transformations of the form $t'=t+k,\; x'=x,\; u'=U(u)$ with $U'(u)\neq 0$. Under such a transformation $Q_2$ is mapped to $Q'_2= (t + \beta)\gen {t'} + c(u)U'(u)\gen {u'}$. If $c(u)=0$ then we have $Q'_2=(t' + \beta-k)\gen {t'}$. We then choose $k$ so that $\beta-\alpha k=0$, giving $A=\langle \gen t, t\gen t\rangle$. This is however inadmissible as a symmetry since this implies that $F=0$ which is a contradiction. Thus we must have $c(u)\neq 0$. In this case we may choose $U(u)$ so that $c(u)U'(u)=U(u)$ if $\alpha=0$, giving $Q'_2= (t' + \beta-k)\gen {t'} + c(u)U'(u)\gen {u'}=t'\gen {t'} + u'\gen {u'}$, so that $A=\langle \gen t, t\gen t + u\gen u\rangle$ in canonical form. If we take $Q_1=\gen u$ then $[Q_1, Q_2]=Q_1$ gives $c_u(t,u)=1$ so that $Q_2=a(t)\gen t + [u + \beta(t)]\gen u$. Applying a transformation of the residual equivalence group of $e_1, e_2, e_3, \gen u$, $Q_2$ is mapped to
\[
Q'_2=a(t)\dot{T}(t)\gen {t'} + [u'+\beta(t)-p(t)+a(t)\dot{p}(t)]\gen {u'}.
\]
If $a(t)\neq 0$ then we may choose $T(t)$ so that $a(t)\dot{T}(t)=T(t)$ and $p(t)$ so that $\beta(t)-p(t)+a(t)\dot{p}(t)=0$, giving $Q_2=t\gen t + u\gen u$ in canonical form. If $a(t)=0$ then we choose $p(t)$ so that $\beta(t)-p(t)=0$, giving $Q_2=u\gen u$ in canonical form and we obtain the canonical two-dimensional (admissible) solvable Lie algebras $A=\langle \gen u, t\gen t + u\gen u\rangle$ and $A=\langle \gen u, u\gen u\rangle$. However, if $\gen u,\, u\gen u$ are symmetries then we find that $F=0$ since $F=f(t)/u_1^3$. Thus we have the following two-dimensional extensions:
\begin{align*}
&\langle 2x\gen x,-x^2\gen x, \gen x\rangle\oplus\langle \gen t, \gen u\rangle\\
&\langle 2x\gen x,-x^2\gen x, \gen x\rangle\oplus\langle \gen t, t\gen t + u\gen u\rangle\\
&\langle 2x\gen x,-x^2\gen x, \gen x\rangle\oplus\langle \gen u, t\gen t + u\gen u\rangle.
\end{align*}

\smallskip\noindent $\dim\mathsf{A}=3$. The admissible three-dimensional solvable Lie algebras are extensions of the admissible solvable two-dimensional solvable Lie algebras, containing them as ideals of codimension one.

\smallskip\noindent $\mathsf{A}=\langle \gen t, \gen u, Q_3\rangle$. We require that $Q_3=a(t)\gen t + c(t,u)\gen u$ satisfy $[\gen t, Q_3]\in \langle \gen t, \gen u\rangle$ and $[\gen u, Q_3]=\langle \gen t, \gen u\rangle$. An elementary calculation gives $Q_3=(\alpha t + \beta)\gen t + (\gamma t + \delta u + \epsilon)\gen u$. Obviously, we may assume $\beta=0,\, \epsilon=0$, so we take $Q_3=\alpha t\gen t + (\gamma t + \delta u)\gen u$. If $\delta=0$ then putting $Q_3$ as a symmetry in the equation for $F$ gives $\alpha F=0$, so we must have $\alpha =0$ to avoid the contradiction $F=0$. However, then $\gamma \neq 0$ if $Q_3\neq 0$ and $\gen u, t\gen u$ in the equation for $G$ gives us $1=0$, a contradiction. Thus, $\delta\neq 0$. Then put $Q_3$ into the equation for $F$ and we obtain (noting that $F_t=F_x=F_u=0$)
\[
\delta(u_1F_{u_1} + u_2F_{u_2})+ \alpha F=0.
\]
From the defining equations for $F$ we find that $F_{u_2}=0$ and $u_1F_{u_1} + 2u_2F_{u_2}+ 3F=0$, so we conclude that $\alpha =3\delta$ and we may write
\[
Q_3=t\gen t + \left[\frac{u}{3}+pt\right]\gen u,
\]
where $p\in \mathbb{R}$ is arbitrary. Thus we have the extension
\[
\langle 2x\gen x,-x^2\gen x, \gen x\rangle\oplus\langle \gen t, \gen u, t\gen t + \left[\frac{u}{3}+pt\right]\gen u\rangle.
\]
Now, if we make the equivalence transformation $\displaystyle t'=t,\; x'=x,\; u'=u-\frac{3p}{2}t$ then $\displaystyle \gen t\to \gen {t'}-\frac{3p}{2}\gen {u'},\; \gen x\to \gen {x'},\; \gen u\to \gen {u'}$ and $\displaystyle t\gen t + \left[\frac{u}{3}+pt\right]\gen u\to t'\gen {t'} + \frac{u'}{3}\gen {u'}$. Thus the algebra $\displaystyle \langle \gen t, \gen u, t\gen t + \left[\frac{u}{3}+pt\right]\gen u\rangle$ is equivalent to the algebra $\displaystyle \langle \gen t-\frac{3p}{2}\gen u, \gen u, t\gen t + \frac{u}{3}\gen u\rangle=\langle \gen t, \gen u, t\gen t + \frac{u}{3}\gen u\rangle$. We recognize $\displaystyle \langle \gen t, \gen u, t\gen t + \frac{u}{3}\gen u\rangle$ as the solvable Lie algebra $\mathsf{A}_{3.7}$ with $q=1/3$, in Mubarakzyanov's classification. Hence we may write our direct-sum extension as
\[
\langle 2x\gen x,-x^2\gen x, \gen x\rangle\oplus\langle \gen t, \gen u, t\gen t + \frac{u}{3}\gen u\rangle.
\]
Put this into the equations for $F$ and $G$ and we obtain the evolution equation
\[
u_t=K\left(\frac{u_3}{u_1^3}-\frac{3}{2}\frac{u_2^2}{u_1^4}\right).
\]
Note also that if we make the equivalence transformation $t'=t,\; x'=x,\; u'=x$ then this direct-sum extension is mapped to
\[
\langle 2u\gen u,-u^2\gen u, \gen u\rangle\oplus\langle \gen t, \gen x, t\gen t + \frac{x}{3}\gen x\rangle,
\]
and the corresponding evolution equation is
\[
u_t=Ku_3-\frac{3K}{2}\frac{u_2^2}{u_1}.
\]
The maximal symmetry of this equation is the six-dimensional direct-sum extension
\[
\langle 2u\gen u,-u^2\gen u, \gen u\rangle\oplus\langle \gen t, \gen x, t\gen t + \frac{x}{3}\gen x\rangle.
\]
This is the six-dimensional extension of the five dimensional solvable Lie algebra
\[
\mathsf{A}_{3.7}\oplus\mathsf{A}_{2.2}=\langle \gen t, \gen x, t\gen t + \frac{x}{3}\gen x, \gen u, u\gen u\rangle,
\]
which gives the evolution equation
\[
u_t=Ku_3+M\frac{u_2^2}{u_1}.
\]
For a non-linear equation we must have $M\neq 0$. This equation has maximal symmetry algebra
\[
\langle 2u\gen u,-u^2\gen u, \gen u\rangle\oplus\langle \gen t, \gen x, t\gen t + \frac{x}{3}\gen x\rangle
\]
when $M=-3K/2$, otherwise the maximal symmetry algebra is just
\[
\mathsf{A}_{3.7}\oplus\mathsf{A}_{2.2}=\langle \gen t, \gen x, t\gen t + \frac{x}{3}\gen x, \gen u, u\gen u\rangle.
\]

\smallskip\noindent $\mathsf{A}=\langle \gen t, t\gen t + u\gen u, Q_3\rangle$. In this case we require $[\gen t, Q_3]\in \langle \gen t, t\gen t + u\gen u\rangle$ and $[t\gen t + u\gen u, Q_3]\in \langle \gen t, t\gen t + u\gen u\rangle$. Standard calculations yield $Q_3=(\alpha t + \beta)\gen t+qu\gen u$. This can be written as $Q_3=(\alpha-q)\gen t + \beta \gen t + q(t\gen t + u\gen u)$. If $\alpha\neq q$ then we have $\mathsf{A}=\langle \gen t, t\gen t + u\gen u, t\gen t\rangle$, and we see that this is inadmissible since this requires $\gen t$ and $t\gen t$ to be symmetries, which leads to $F=0$. So we must have $\alpha=q$ and then $Q_3=\beta Q_1 + qQ_2$ so that $\dim\mathsf{A}=2$. Hence, we have no three-dimensional extension in this case.

\smallskip\noindent $\mathsf{A}=\langle \gen u, t\gen t + u\gen u, Q_3\rangle$. Again we require $[\gen t, Q_3]\in \langle \gen t, t\gen t + u\gen u\rangle$ and $[t\gen t + u\gen u, Q_3]\in \langle \gen t, t\gen t + u\gen u\rangle$. This gives us, after standard calculations, $Q_3=\alpha \gen t + (\beta u + \gamma t + \delta)\gen u$. Again, we have $Q_3=\delta Q_1 + \alpha Q_2 + (\beta-\alpha)u\gen u + \gamma t\gen u$, so we may assume that $Q_3=(au+qt)\gen u$. Putting this form of $Q_3$ in the equation for $F$ we obtain $a(u_1F_{u_1}+u_{2}F_{u_2})=0$. Since we have $F=Kt^2/u_1^3$ (being invariant under $\Sl(2, \mathbb{R})$ and under $\gen u,\, t\gen t + u\gen u$) we must have $a=0$. Thus $Q_3=bt\gen u$. However, putting $\gen u$ and $bt\gen u$ as symmetries in the equation for $G$ gives $b=0$. Thus we have $a=b=0$ and so $Q_3\in \langle \gen u, t\gen t + u\gen u\rangle$, and we do not have a three-dimensional extension in this case.

\smallskip\noindent We see then that we have only one three-dimensional extension for this case of $\Sl(2, \mathbb{R})$:
\[
\langle 2x\gen x,-x^2\gen x, \gen x\rangle\oplus\langle \gen t, \gen u, t\gen t + \frac{u}{3}\gen u\rangle
\]
with corresponding evolution equation
\[
u_t=K\left(\frac{u_3}{u_1^3}-\frac{3}{2}\frac{u_2^2}{u_1^4}\right).
\]

\smallskip\noindent $\dim\mathsf{A}=4$: In this case we only have to consider $\mathsf{A}=\langle \gen t, \gen u, t\gen t + \frac{u}{3}\gen u, Q_4\rangle$. Then $[\gen t, Q_4]\in \langle \gen t, \gen u, t\gen t + \frac{u}{3}\gen u\rangle$ and $[\gen u, Q_4]\in \langle \gen t, \gen u, t\gen t + \frac{u}{3}\gen u\rangle$ give us $Q_4=(\alpha t + \beta)\gen t + (\lambda u +\mu t + \nu)\gen u$. Then we find that $[Q_4, Q_3]=\beta Q_1+\nu Q_2 + \frac{2}{3}\mu t\gen u$. Since we require $Q_4$ to be a symmetry, then $[Q_4, Q_3]$ is also a symmetry, and in the equation for $G$ we obtain $\mu =0$, so that  $Q_4=\alpha Q_3 +\beta Q_1 + \nu Q_2 + bu\gen u$ with $b=\lambda-\alpha/3$. It is easy to verify that $u\gen u$ is not a symmetry of the equation, so we must have $b=0$ and thus $Q_4$ is a linear combination of $Q_1,\, Q_2,\; Q_3$ so that we have no extension of this direct-sum.

\smallskip\noindent{\underline{$\Sl(2,\mathbb{R})=\langle 2x\gen x + 2u\gen u,
-x^2\gen x - 2xu\gen u, \gen x\rangle:$}}
From $[e_1, Q]=[e_2, Q]=[e_3, Q]=0$ we find that $Q=a(t)\gen t + c(t)u\gen u$. The equivalence transformations are given  by $t'=T(t),\; x'=x,\; u'= p(t)u$ with $\dot{T}\neq 0, p(t)\neq 0$. Then $Q=a(t)\gen t + c(t)u$ is transformed to $Q'=a(t)\dot{T}\gen {t'} + (a(t)\dot{p}(t)u + c(t)p(t)u)gen {u'}$. If $a(t)\neq 0$ we choose $T$ with $a(t)\dot{T}=1$ and $a(t)\dot{p}(t) + c(t)p(t)=0$, and this gives the canonical form $Q=\gen t$. If $a(t)=0$ then we have $Q'=c(t)p(t)u\gen {u'}=c(t)u'\gen {u'}$. In this case, if $\dot{c}=0$ we have the canonical form $Q=u\gen u$; but if $\dot{c}\neq 0$ we choose $c(t)=T(t)$ and this gives us the canonical form $Q=tu\gen u$.

Thus we have the three canonical one-dimensional extensions:
\begin{align*}
&\langle 2x\gen x + 2u\gen u,
-x^2\gen x - 2xu\gen u, \gen x\rangle\oplus\langle \gen t \rangle,\\
&\langle 2x\gen x + 2u\gen u,
-x^2\gen x - 2xu\gen u, \gen x\rangle\oplus\langle u\gen u \rangle,\\
&\langle 2x\gen x + 2u\gen u,
-x^2\gen x - 2xu\gen u, \gen x\rangle\oplus\langle tu\gen u \rangle.
\end{align*}

\smallskip\noindent $\dim\mathsf{A}=2:$ We have two types of algebra here: $\mathsf{A}=\langle Q_1, Q_2\rangle$ with $[Q_1, Q_2]=0$ or $[Q_1, Q_2]=Q_1$. First, consider $[Q_1, Q_2]=0$. If $Q_1=\gen t$ then $[Q_1, Q_2]=0$ gives $Q_2=a\gen t + cu\gen u$, so that $\mathsf{A}=\langle \gen t, u\gen u\rangle$. If $Q_1=u\gen u$, then $Q_2=a(t)\gen t + c(t)u\gen u$. The residual equivalence group consists of transformations of the form $t'=T(t),\, x'=x,\, u'=p(t)u$ with $\dot{T}(t)\neq 0,\, p(t)\neq 0$. Under such a transformation, $Q_2$ is transformed to $Q'_2=a(t)\dot{T}(t)\gen {t'} + [a(t)\frac{\dot{p}(t)}{p(t)}+c(t)]u'\gen {u'}$. If $a(t)\neq 0$ then we choose $p(t)$ such that $a(t)\frac{\dot{p}(t)}{p(t)}+c(t)=0$ and $T(t)$ such that $a(t) \dot{T}(t)=1$, giving $\mathsf{A}=\langle \gen t, u\gen u\rangle$. If $a(t)=0$ then putting $u\gen u,\, c(t)u\gen u$ into the equation for $G$ gives $\dot{c}(t)=0$ so that we find that $\dim\mathsf{A}=1$, contradicting the requirement that $\dim\mathsf{A}=2$. If $Q_1=tu\gen u$ then again $[Q_1, Q_2]=0$ gives $a(t)=0$ and so $Q_2=c(t)u\gen u$. Putting $tu\gen u,\, c(t)u\gen u$ into the equation for $G$ we find that $t\dot{c}(t)=c(t)$ so that $Q_2=\alpha Q_1$ so that $\dim\mathsf{A}=1$. Thus, we only find $\mathsf{A}=\langle \gen t, u\gen u\rangle$ as the canonical form of two-dimensional admissible abelian Lie algebras.

\smallskip\noindent Next we look at $\mathsf{A}=\langle Q_1, Q_2\rangle$ with $[Q_1, Q_2]=Q_1$. If $Q_1=\gen t$ then we obtain $Q_2=(t+k)\gen t + qu\gen u$ where $q\in \mathbb{R}$, and we may write $Q_2=t\gen t + qu\gen u$. The case $q=0$ is inadmissible since if $\gen t,\, t\gen t$ are symmetries, then the equation for $F$ gives $F=0$, a contradiction. Thus we must have $q\neq 0$. This case gives the extension
\[
\langle 2x\gen x + 2u\gen u,
-x^2\gen x - 2xu\gen u, \gen x\rangle\oplus\langle \gen t, t\gen t + qu\gen u \rangle,
\]
which gives the evolution equation
\[
u_t=K\frac{u^3u_3}{\tau^{3/2+1/2q}} + Lu\tau^{-1/2q},
\]
with $\tau=2uu_2-u_1^2$. We recover the Harry Dym equation when $K=1, L=0, q=-1/3$.

\smallskip\noindent If we now put $Q_1=u\gen u$ or $Q_1=tu\gen u$ we find that $[Q_1, Q_2]=0$ so we have no realization of $[Q_1, Q_2]=0$ in these cases. So we have the following two two-dimensional extensions:
\begin{align*}
&\langle 2x\gen x + 2u\gen u,
-x^2\gen x - 2xu\gen u, \gen x\rangle\oplus\langle \gen t, u\gen u \rangle,\\
&\langle 2x\gen x + 2u\gen u,
-x^2\gen x - 2xu\gen u, \gen x\rangle\oplus\langle \gen t, t\gen t + qu\gen u \rangle.
\end{align*}

\smallskip\noindent $\dim\mathsf{A}=3:$ Here we have two cases. $\mathsf{A}=\langle \gen t, u\gen u, Q_3\rangle$ and $\mathsf{A}=\langle \gen t, t\gen t + qu\gen u, Q_3\rangle$ with $\langle Q_1, Q_2\rangle$ contained as an ideal. First, take $\mathsf{A}=\langle \gen t, u\gen u, Q_3\rangle$. Then $[\gen t, Q_3]\in \langle \gen t, u\gen u\rangle$ and $[u\gen u, Q_3]\in \langle \gen t, u\gen u\rangle$ give $Q_3=(at+b)\gen t + cu\gen u$. Putting $\gen t, u\gen u$ and $Q_3=(at+b)\gen t + cu\gen u$ as symmetries in the equation for $F$ gives $aF=0$ so that we must have $a=0$ in order to have $F\neq 0$. Thus $Q_3=bQ_1+cQ_2$ and this means that $\dim\mathsf{A}=2$, contradicting the requirement that  $\dim\mathsf{A}=3$. Hence we have no realization in this case. Next, take $\mathsf{A}=\langle \gen t, t\gen t + qu\gen u, Q_3\rangle$. Then $[\gen t, Q_3]\in \langle \gen t, t\gen t + qu\gen u\rangle$ and $[t\gen t  + qu\gen u, Q_3]\in \langle \gen t, t\gen t + qu\gen u\rangle$ gives, after elementary calculations, $Q_3=(at+b)\gen t + cu\gen u$. We can rewrite this as $Q_3=bQ_1+aQ_2 + (c-a)u\gen u$. If $c\neq a$ then $u\gen u$ is also a symmetry, so that $t\gen t$ will also be a symmetry. Since we know that $\gen t$ is a symmetry, we find that $F=0$, a contradiction. Thus we must have $a=c$ and thus $Q_3=aQ_2+bQ_1$ so that $\dim\mathsf{A}=2$, and we have no extension in this case. Thus, we have only the two two-dimensional extensions given above.

\smallskip\noindent{\underline{$\langle 2x\gen x + 2u\gen u,
-(x^2 - u^2)\gen x - 2xu\gen u, \gen x\rangle.$}} In this case $[e_1, Q]=[e_2, Q]=[e_3, Q]=0$ gives $Q=a(t)\gen t$. The residual equivalence group is given by transformations of the form $t'=T(t),\, x'=x,\, u'=\pm u$ with $\dot{T}(t)\neq 0$. Under such a transformation, $Q$ is mapped to $Q'=a(t)\dot{T}(t)\gen {t'}$ and we choose $T(t)$ so that $a(t)\dot{T}(t)=1$. Thus we have the canonical one-dimensional algebra $\mathsf{A}=\langle \gen t\rangle$. It is clear that if $\mathsf{A}=\langle Q_1, Q_2\rangle$ with $Q_1=\gen t$ then we cannot have $[Q_1, Q_2]=0$ and $\dim\mathsf{A}=2$. Further, if we require $[Q_1, Q_2]=Q_1$ with $Q_1=\gen t$ then $Q_2=(t + k)\gen t$, which is inadmissible since $\gen t,\, t\gen t$ as symmetries in the equation for$F$ gives $F=0$, a contradiction. Thus we obtain only a one-dimensional extension:
\[
\langle 2x\gen x + 2u\gen u,
-(x^2 - u^2)\gen x - 2xu\gen u, \gen x\rangle\oplus\langle \gen t\rangle.
\]

\smallskip\noindent{\underline{$\langle 2x\gen x + 2u\gen u,
-(x^2 + u^2)\gen x - 2xu\gen u, \gen x\rangle.$}} As in the previous case (with exactly the same argument) we have only one extension, the one-dimensional extension

\[
\langle 2x\gen x + 2u\gen u,
-(x^2 + u^2)\gen x - 2xu\gen u, \gen x\rangle\oplus\langle \gen t\rangle.
\]

\subsection{Semi-direct sums of ${\rm sl}(2, \mathbb{R})$ with solvable Lie algebras.}

We now consider the case of semi-direct extensions of $\Sl(2, \mathbb{R})$, and we proceed case by case. We use the following notation in the proofs:

\begin{enumerate}

\item $\mathsf{A}_0$ denotes the abelian ideal of the solvable Lie algebra $\mathsf{A}$

\item $\mathsf{k}$ is the invariant (under the action of $\Sl(2, \mathbb{R})$) subspace of $\mathsf{A}$ which is the complement of $\mathsf{A}_0$, so that $\mathsf{A}=\mathsf{k}\dot{+}\mathsf{A}_0$

\item $\mathsf{k}_J=\langle Q_1,\dots, Q_{2J+1}\rangle$ is an irreducible representation space for $\Sl(2, \mathbb{R})$, with $\mathsf{k}_J\subset \mathsf{k}$ for $J=\frac{1}{2},\, 1, \, \frac{3}{2},\dots$

\item $\mathsf{k}_0\subset \mathsf{k}$ is the representation space for $\Sl(2, \mathbb{R})$, with spin $J=0$. It is the direct sum of irreducible one-dimensional subspaces and $\Sl(2, \mathbb{R})$ commutes with all the elements of $\mathsf{k}_0$.

\item $\mathsf{k}_{\star}$ is the sum of all $\mathsf{k}_J$, for $J\geq \frac{1}{2}$, which are subspaces of $\mathsf{k}$
\end{enumerate}

Thus we have the decomposition $\mathsf{A}=\mathsf{k}_{\star}\dot{+}\mathsf{k}_0\dot{+}\mathsf{A}_0$. First we have the following useful result:

\begin{lemma}\label{usefullemma} Suppose that $\mathsf{A}_0$ commutes with $\Sl(2, \mathbb{R})$. Then  $[\mathsf{k}_{\star}, \mathsf{A}_0]=\{0\}$ as well as $[\mathsf{k}_{\star}, \mathsf{k}_0]\subset \mathsf{k}_{\star}$.

\end{lemma}
\smallskip\noindent{\bf Proof:} First we note that $[\mathsf{k}_{\star}, \mathsf{A}_0]\subset \mathsf{A}_0$ since $\mathsf{A}_0$ is an ideal. Now for any $\mathsf{k}_J$ with $J>0$ we also have $[\mathsf{k}_J, \mathsf{A}_0]\subset \mathsf{k}_{\star}$. In fact, if $X\in \mathsf{A}_0$ then put $Q'_i=[Q_i, X],\; i=1, \dots 2J+1$. Then it is straightforward to show that $\mathsf{k}'_J=\langle Q'_1,\dots, Q'_{2J+1}\rangle$ is an irreducible representation space for $\Sl(2, \mathbb{R})$ if $\mathsf{k}'_J\neq \{0\}$ (in fact it is clear that if any $Q'_i=0$ then $\mathsf{k}'_J=\{0\}$) and since $\mathsf{k}'_J\subset \mathsf{A}$ we have $\mathsf{k}'_J\subset \mathsf{k}_{\star}$. Hence $[\mathsf{k}_{\star}, \mathsf{A}_0]\subset \mathsf{k}_{\star}\cap\mathsf{A}_0=\{0\}$. The same reasoning shows that $[\mathsf{k}_{\star}, \mathsf{k}_0]\subset \mathsf{k}_{\star}$.

\begin{lemma} $ \displaystyle \Sl(2, \mathbb{R})=\langle 2x\gen x, -x^2\gen x, \gen x\rangle$ admits only direct-sum extensions\newline  $\langle 2x\gen x, -x^2\gen x, \gen x\rangle\oplus\mathsf{A}$ by solvable Lie algebras $\mathsf{A}$.
\end{lemma}

\smallskip\noindent{\bf Proof:} Let $\langle Q_1,\dots, Q_{2J+1}\rangle$ where $J>0$ be an irreducible representation subspace in $\mathsf{A}$. Put $Q_{2J+1}=a(t)\gen t + b(t,x,u)\gen x + c(t,x,u)\gen u$. Since $\mathsf{k}_J$ is irreducible under the action of $\Sl(2, \mathbb{R})$ it follows that $a(t)=0$. The conditions $[e_3, Q_{2J+1}]=0$ and $[e_1, Q_{2J+1}]=-2JQ_{2J+1}$ then give $c(t,x,u)=0,\; b_x=0$ and $b=Jb$. For $J\neq 1$ we obviously have $b=0$, giving $Q_{2J+1}=0$, and hence $A=\{0\}$. For $J=1$ we then have $2J+1=3$ and $Q_3=b(t,u)\gen x$. Then $Q_2=[e_2, Q_3]=2xb(t,u)\gen x,\; Q_1=\frac{1}{2}[e_2, Q_3]=-x^2b(t,u)\gen x$ so that $\langle Q_1, Q_2, Q_3\rangle$ is just a realization of $\Sl(2, \mathbb{R})$, which is impossible for a subspace of a solvable Lie algebra $\mathsf{A}$. Hence, any solvable Lie algebra which is a representation space for this choice of $\Sl(2, \mathbb{R})$ must be a direct sum of irreducible one-dimensional ($J=0$) subspaces. $\Sl(2, \mathbb{R})$ acts trivially on irreducible one-dimensional spaces, so that each $Q\in \mathsf{A}$ must commute with $e_1, e_2, e_3$, and therefore the extension must be a direct-sum extension.

\begin{lemma} $\displaystyle \Sl(2, \mathbb{R})=\langle 2x\gen x + 2u\gen u, -(x^2-u^2)\gen x-2xu\gen u, \gen x\rangle$ admits only direct-sum extensions $\displaystyle \langle 2x\gen x + 2u\gen u, -(x^2-u^2)\gen x-2xu\gen u, \gen x\rangle\oplus\mathsf{A}$ by solvable Lie algebras $\mathsf{A}$.
\end{lemma}

\smallskip\noindent{\bf Proof:} Let $\langle Q_1,\dots, Q_{2J+1}\rangle$ where $J>0$ be an irreducible representation subspace in $\mathsf{A}$. Put $Q_{2J+1}=a(t)\gen t + b(t,x,u)\gen x + c(t,x,u)\gen u$. Since $\mathsf{k}_J$ is irreducible under the action of $\Sl(2, \mathbb{R})$ it follows that $a(t)=0$ and we also have that any element of $\mathsf{k}_J$ is of the form $X=b(t,x,u)\gen x + c(t,x,u)\gen u$, so that either ${\rm rank}\,\mathsf{k}_J=1$ or ${\rm rank}\,\mathsf{k}_J=2$. The conditions $[e_3, Q_{2J+1}]=0$ and $[e_1, Q_{2J+1}]=-2JQ_{2J+1}$ give $Q_{2J+1}=b(t)u^{-J+1}\gen x + c(t)u^{-J+1}\gen u$ and then $Q_{2J}=[e_2, Q_{2J+1}]$ gives
\[
Q_{2J}=2JxQ_{2J+1}-2c(t)u^{-J+2}\gen x + 2b(t)u^{-J+2}\gen u.
\]
We have ${\rm rank}\,\mathsf{k}_J=1$ if and only if $Q_{2J+1}\wedge Q_{2J}=0$ which gives $b(t)^2+c(t)^2=0$, which is impossible if $Q_{2J+1}\neq 0$, so we must have ${\rm rank}\,\mathsf{k}_J=2$ and $b(t)^2+c(t)^2\neq 0$.

Now assume that $\mathsf{k}_J$ is abelian. Then $[Q_{2J}, Q_{2J+1}]=0$ and we find that
\[
(2J-1)b(t)^2 + (J-2)c(t)^2=0,\quad b(t)c(t)=0.
\]
If $c(t)\neq 0$ then $b(t)=0$ since we have $b(t)c(t)=0$. This gives $(J-2)c(t)^2=0$ so we have either $c(t)=0$ or $J=2$. However, $J=2$ corresponds to $\dim\,\langle Q_1, \dots, Q_{2J+1}\rangle=5$, and admissible abelian Lie algebras of dimension five are all rank-one realizations, which is a contradiction. Thus, $c(t)=0$ and we have $b(t)\neq 0$. This gives $(2J-1)b(t)^2=0$ so that $Q_{2J+1}\neq 0$ only for $J=1/2$. In this case we have $Q_2=b(t)u^{1/2}\gen x,\;\; Q_1=xb(t)u^{1/2}\gen x + 2b(t)u^{3/2}\gen u$. Putting these into the equation for $F$ and noting that $F_x=0$ we have the two equations
\begin{align*}
&4u^2F_u+2uu_1F_{u_1}+[2uu_2-u_1^2]F_{u_2}=0\\
&4u^2F_u+uu_1F_{u_1}+[2uu_2+u_1^2]F_{u_2}=6uF.
\end{align*}
We also have the defining equations
\begin{align*}
&uF_u=u_2F_{u_2} + 3F\\
&u(1+u_1^2)F_{u_1}+[u_1(1+u_1^2)+ 3uu_1u_2]F_{u_2} + 3uu_1F=0.
\end{align*}
Writing these equations in the form of the matrix equation $Mv=0$ where $v=(F, F_u, F_{u_1}, F_{u_2})^T$ we find that $\det M\neq 0$ except on some subsets of dimension less than five in the space parametrized locally by $(t,x,u,u_1, u_2)$ and so $F=0$ because $F$ is smooth. This is a contradiction and so we have no irreducible representations of $\Sl(2, \mathbb{R})$ on abelian algebras other than those for which $J=0$, that is one-dimensional representation spaces. Consequently, the abelian ideal $\mathsf{A}_0$ of $\mathsf{A}$ is a direct sum of one-dimensional irreducible subspaces for $\Sl(2, \mathbb{R})$, and so if $Q\in \mathsf{A}_0$ then $Q$ commutes with $e_1, e_2, e_3$. Now note that by definition $\mathsf{k}_0$ commutes with all of $\Sl(2, \mathbb{R})$, so that $\mathsf{k}_0\dot{+}\mathsf{A}_0$ must commute with $\Sl(2, \mathbb{R})$. From Theorem \ref{directsumsthm} it follows that $\dim\,\mathsf{k}_0\dot{+}\mathsf{A}_0=1$, and we may assume that $\mathsf{k}_0\dot{+}\mathsf{A}_0=\langle \gen t\rangle$ so we have $\mathsf{A}=\mathsf{k}_{\star}\dot{+}\langle \gen t\rangle$ and $\mathsf{k}_{\star}$ is the direct sum of irreducible subspaces $\mathsf{k}_J=\langle Q_1,\dots, Q_{2J+1}\rangle$ with $J>0$. We know that $Q_{2J+1}=b(t)u^{-J+1}\gen x + c(t)u^{-J+1}\gen u$. Since $\langle \gen t\rangle$ is an ideal in $\mathsf{A}$, we must have $[Q_{2J+1}, \gen t]=0$ so that $Q_{2J+1}=bu^{-J+1}\gen x + cu^{-J+1}\gen u$ with $b, c$ constants.  It now follows that $[\mathsf{k}_{\star}, \mathsf{k}_{\star}]\subset \mathsf{k}_{\star}$, so that $\mathsf{k}_{\star}$ is a subalgebra of $\mathsf{A}$, and as such it is solvable. Assume that $\mathsf{k}_{\star}\neq \{0\}$. Then, being solvable, $\mathsf{k}_{\star}$ contains a non-trivial abelian ideal which is invariant under $\Sl(2, \mathbb{R})$, and this abelian ideal must be a direct sum of irreducible one-dimensional representation spaces by our reasoning above, so that $\mathsf{k}_{\star}$ contains the representation $J=0$, which contradicts the definition of $\mathsf{k}_{\star}$. Hence $\mathsf{k}_{\star}=\{0\}$ and so the only extension of $\Sl(2, \mathbb{R})$ by a solvable Lie algebra $\mathsf{A}$ is a direct-sum extension (so each of the operators of $\mathsf{A}$ commutes with $e_1, e_2, e_3$).

\begin{lemma} $\displaystyle \Sl(2, \mathbb{R})=\langle 2x\gen x + 2u\gen u, -(x^2+u^2)\gen x-2xu\gen u, \gen x\rangle$ admits only direct-sum extensions $\displaystyle \langle 2x\gen x + 2u\gen u, -(x^2+u^2)\gen x-2xu\gen u, \gen x\rangle\oplus\mathsf{A}$ by solvable Lie algebras $\mathsf{A}$.
\end{lemma}

\smallskip\noindent{\bf Proof:} The proof is essentially the same as for $\Sl(2, \mathbb{R})=\langle 2x\gen x + 2u\gen u, -(x^2-u^2)\gen x-2xu\gen u, \gen x\rangle$. For an irreducible representation subspace $\mathsf{k}_J=\langle Q_1,\dots, Q_{2J+1}\rangle$ of $\mathsf{A}$ with $J>0$ we find $Q_{2J+1}=b(t)u^{-J+1}\gen x + c(t)u^{-J+1}\gen u$ and
\[
Q_{2J}=2JxQ_{2J+1}+2c(t)u^{-J+2}\gen x + 2b(t)u^{-J+2}\gen u.
\]
Then $[Q_{2J}, Q_{2J+1}]=0$ gives
\[
(2J-1)b(t)^2 - (J-2)c(t)^2=0,\quad b(t)c(t)=0.
\]
Then the same type of argument as for $\Sl(2, \mathbb{R})=\langle 2x\gen x + 2u\gen u, -(x^2-u^2)\gen x-2xu\gen u, \gen x\rangle$ shows that there are no irreducible abelian representation spaces $\mathsf{k}_J$ for $J>0$. The final part of the argument for $\Sl(2, \mathbb{R})=\langle 2x\gen x + 2u\gen u, -(x^2+u^2)\gen x-2xu\gen u, \gen x\rangle$ goes through without modification, and we conclude that only direct-sum extensions by solvable Lie algebras are allowed.

\begin{lemma} $\displaystyle \Sl(2, \mathbb{R})=\langle 2x\gen x + 2u\gen u, -x^2\gen x-2xu\gen u, \gen x\rangle$ admits only direct-sum extensions $\displaystyle \langle 2x\gen x + 2u\gen u, -x^2\gen x-2xu\gen u, \gen x\rangle\oplus\mathsf{A}$ by solvable Lie algebras $\mathsf{A}$.
\end{lemma}

\smallskip\noindent{\bf Proof:} As in the previous calculations, we have ${\rm rank}\,\mathsf{k}_J\leq 2$ for the irreducible representation space $\mathsf{k}_J=\langle Q_1, \dots, Q_{2J+1}\rangle$ with $J>0$.The elements of $\mathsf{k}_J$ are of the form $X=b(t,x,u)\gen x + c(t,x,u)\gen u$ and we obtain from the conditions $[e_3, Q_{2J+1}]=0$ and $[e_1, Q_{2J+1}]=-2JQ_{2J+1}$ that
\[
Q_{2J+1}=b(t)u^{-J+1}\gen x + c(t)u^{-J+1}\gen u,
\]
and then $Q_{2J}=[e_2, Q_{2J+1}]$ yields
\[
Q_{2J}=2JxQ_{2J+1} +2b(t)u^{-J+2}\gen u.
\]
If ${\rm rank}\,\mathsf{k}_J=1$ then $Q_{2J+1}\wedge Q_{2J}=0$ and this gives $b(t)=0$ so that $c(t)\neq 0$ and $Q_{2J+1}=c(t)u^{-J+1}\gen u,\; Q_{2J}=2JxQ_{2J+1}$ and we find that $\mathsf{k}_J$ is abelian. Putting these as symmetries into the equation for $F$ gives
\begin{align*}
&u^2F_u-(J-1)uu_1F_{u_1}+(J-1)[Ju^2_1-uu_2]F_{u_2}=0\\
&uF_{u_1}-2(J-1)u_1F_{u_2}=0.
\end{align*}
We also have the defining equations for $F$:
\[
F_x=0,\quad uF_u=u_2F_{u_2}+3F,\quad uF_{u_1}+u_1F_{u_2}=0.
\]
These last three equations give $F=u^3f(t, \tau)$ where $\tau=uu_2-u^2_1$ and substituting this into the first equation above gives
\[
3f(t, \tau)=(J-2)\tau f_{\tau}(t, \tau) - J^2u_1^2f_{\tau}(t, \tau).
\]
In the new coordinate system $(t,x,u,u_1, \tau)$ we note that $u_1$ and $\tau$ are independent, so that we must have $f_{\tau}=0$ for $J>0$ and consequently $F=0$. Thus we have no admissible realization for $J>0$ in this case.

If now ${\rm rank}\; \mathsf{k}_J=2$ then $b(t)\neq 0$. Assume that $\mathsf{k}_J$ is abelian. Then commutativity gives us the equations
\[
(2J-1)b^2=0,\quad bc=0.
\]
Since $b(t)\neq 0$ then $c(t)=0$ and $J=1/2$ and we obtain $Q_2=b(t)u^{1/2}\gen x,\;\; Q_1=2JxQ_2+b(t)u^{3/2}\gen u$. Putting these as symmetries into the equation for $F$ gives $F=0$, which is a contradiction. Thus we have no irreducible abelian representation space for $J>0$, and so all the  elements of the abelian ideal $\mathsf{A}_0$ of $\mathsf{A}$ commute with $\Sl(2, \mathbb{R})$, so that $\mathsf{A}_0$ is a direct sum of one-dimensional invariant subspaces.

We have the decomposition $\mathsf{A}=\mathsf{k}_{\star}\dot{+}\mathsf{k}_0\dot{+}\mathsf{A}_0$ where $\mathsf{k}_{\star}$ is a direct sum of irreducible representation spaces $\mathsf{k}_J$ with $J>0$, and from the above we conclude that $\mathsf{k}_0\dot{+}\mathsf{A}_0$ is a solvable Lie subalgebra of $\mathsf{A}$: in fact $[\mathsf{k}_0, \mathsf{A}_0]\subset \mathsf{A}_0$ since $\mathsf{A}_0$ is an ideal, and $[\mathsf{k}_0, \mathsf{k}_0]$ is a subspace which commutes with $\Sl(2, \mathbb{R})$ so that we must have $[\mathsf{k}_0, \mathsf{k}_0]\subset \mathsf{k}_0\dot{+}\mathsf{A}_0$. Furthermore, the action of $\Sl(2, \mathbb{R})$ on $\mathsf{k}_0\dot{+}\mathsf{A}_0$ is a direct-sum action. Thus $\mathsf{k}_0\dot{+}\mathsf{A}_0$ provides a direct-sum extension of $\Sl(2, \mathbb{R})$, and from Theorem \ref{directsumsthm} we have the following possibilities. $\mathsf{k}_0\dot{+}\mathsf{A}_0=\langle \gen t\rangle$, $\mathsf{k}_0\dot{+}\mathsf{A}_0=\langle u\gen u\rangle$, $\mathsf{k}_0\dot{+}\mathsf{A}_0=\langle tu\gen u\rangle$, for which $\mathsf{k}_0=\{0\}$ as well as $\mathsf{k}_0\dot{+}\mathsf{A}_0=\langle \gen t, u\gen u\rangle$ and $\mathsf{k}_0\dot{+}\mathsf{A}_0=\langle \gen t, t\gen t + qu\gen u\rangle$ with $q\neq 0$.

\medskip\noindent\underline{$\mathsf{A}_0=\langle \gen t\rangle,\; \mathsf{k}_0=\{0\}$:} in this case we note that $\mathsf{k}_{\star}$ consists of operators of the form $b(x,u)\gen x + c(x,u)\gen u$ since we have $[\mathsf{k}_J, \gen t]=0$ by lemma \ref{usefullemma}, and thus we can only have $[\mathsf{k}_{\star}, \mathsf{k}_{\star}]\subset \mathsf{k}_{\star}$ since $\mathsf{k}_0=\{0\}$. If $\mathsf{k}_{\star}\neq \{0\}$ then it is a solvable subalgebra of $\mathsf{A}$ and so contains an abelian ideal which commutes with $\Sl(2, \mathbb{R})$ because the representation of $\Sl(2, \mathbb{R})$ on abelian Lie algebras is a direct sum of one-dimensional ($J=0$) representations, as we have seen above. However, $\mathsf{k}_{\star}$ contains only representations for $J>0$ so we have a contradiction. Hence $\mathsf{k}_{\star}=\{0\}$.

\medskip\noindent\underline{$\mathsf{A}_0=\langle u\gen u\rangle,\; \mathsf{k}_0=\{0\}$:} in this case we have $[\mathsf{k}_J, u\gen u]=\{0\}$ for $J>0$, by Lemma \ref{usefullemma}. So with $Q_{2J+1}=b(t)u^{-J+1}\gen x + c(t)u^{-J+1}\gen u$ the condition $[Q_{2J+1}, u\gen u]=0$ gives $(J-1)b(t)u^{-J+1}\gen x + Jc(t)u^{-J+1}\gen u=0$. Since $J>0$ we must have $c(t)=0$ and we find that $b(t)=0$ unless $J=1$. For $J=1$ we have $Q_3=b(t)\gen x$ and then $Q_2=2b(t)[x\gen x + u\gen u]$ and $Q_1=-b(t)[x^2\gen x + 2xu\gen u]$ so that $\mathsf{k}:J$ is just a copy of $\Sl(2, \mathbf{R})$ if $b(t)\neq 0$, which is impossible since $\mathsf{k}_J$ is a subspace of a solvable Lie algebra. Hence we have $\mathsf{k}_{\star}=\{0\}$. The same reasoning holds for $\mathsf{A}_0=\langle tu\gen u\rangle$ and for $\mathsf{A}_0=\langle u\gen u\rangle,\; \mathsf{k}_0=\langle \gen t\rangle$ as well as $\mathsf{A}_0=\langle \gen t, u\gen u\rangle$.

\medskip\noindent\underline{$\mathsf{A}_0=\langle \gen t\rangle,\; \mathsf{k}_0=\{u\gen u\}$:} in this case we have  $[\mathsf{k}_{\star}\dot{+}\mathsf{k}_0, \mathsf{k}_{\star}\dot{+}\mathsf{k}_0]\subset \mathsf{k}_{\star}\dot{+}\mathsf{k}_0$ by Lemma \ref{usefullemma} and from the structure of the operators of $\mathsf{k}_{\star}$, so that $\mathsf{k}_{\star}\dot{+}\mathsf{k}_0$ is a solvable subalgebra of $\mathsf{A}$ and so it contains an abelian ideal invariant under $\Sl(2, \mathbb{R})$. By the above, this abelian ideal must be $\langle u\gen u\rangle$, so that $\langle \gen t, u\gen u\rangle$ is an abelian ideal in $\mathsf{A}$, so we have $\mathsf{k}_{\star}=\{0\}$ as above.

\medskip\noindent\underline{$\mathsf{A}_0=\langle \gen t\rangle,\; \mathsf{k}_0=\{t\gen t +qu\gen u\},\; q\neq 0$:} in this case the operators of $\mathsf{k}_{\star}$ are of the form $b(x,u)\gen x + c(x,u)\gen u$ since $\gen t$ commutes with $\mathsf{k}_{\star}$. It is clearly impossible to have $[\mathsf{k}_{\star}, \mathsf{k}_{\star}]\cap \mathsf{k}_0\neq \{0\}$ or $[\mathsf{k}_{\star}, \mathsf{k}_{\star}]\cap \mathsf{A}_0\neq \{0\}$ so we have $[\mathsf{k}_{\star}, \mathsf{k}_{\star}]\subset \mathsf{k}_{\star}$, and thus $\mathsf{k}_{\star}$ is a solvable Lie algebra, containing an abelian ideal if $\mathsf{k}_{\star}\neq \{0\}$. This is impossible as we have seen above, so $\mathsf{k}_{\star}=\{0\}$ in this case as well.

\begin{lemma} $\displaystyle \Sl(2, \mathbb{R})=\langle 2t\gen t + 2x\gen x, -t^2\gen t-(x^2+2tx)\gen x, \gen t\rangle$ admits only direct-sum extensions $\displaystyle \langle 2t\gen t + 2x\gen x, -t^2\gen t-(x^2+2tx)\gen x, \gen t\rangle\oplus\mathsf{A}$ by solvable Lie algebras $\mathsf{A}$.
\end{lemma}

\smallskip\noindent{\bf Proof:} Denote an irreducible representation space of $\mathsf{A}$ by $\mathsf{k}_J=\langle Q_1, \dots, Q_{2J+1}\rangle$ where $J>0$. We first assume that $\mathsf{k}_J$ is an abelian algebra. Putting $Q_{2J+1}=a(t)\gen t + b(t,x,u)\gen x + c(t,x,u)\gen u$, the conditions $[e_3, Q_{2J+1}]=0$ and $[e_1, Q_{2J+1}]=-2JQ_{2J+1}$ give
\[
Q_{2J+1}=a\gen t + b(u)x^{-J+1}\gen x + c(u)x^{-J}\gen u.
\]
with $a=0$ for $J\neq 1$. Now $Q_{2J}=[e_2, Q_{2J+1}]$ gives us
\[
Q_{2J}=2JtQ_{2J+1}+(J+1)b(u)x^{-J+2}\gen x + c(u)x^{-J+1}\gen u.
\]
The condition that $[Q_{2J+1}, Q_{2J}]=0$ then yields $a=0$ and the conditions
\[
(J+1)b(u)^2+c(u)b'(u)=0,\quad b(u)c(u)=0.
\]
Note that $b(u)c(u)=0$ is the same as requiring $Q_{2J+1}\wedge Q_{2J}=0$, and this is the same as having a rank-one realization space $A$. In fact, $e_2$ acts nilpotently on $A$ and we have $k!Q_{2J-k+1}=({\rm ad}_{e_2})^k(Q_{2J+1})$ so that $Q_{2J+1}\wedge Q_{2J}=0$ implies that  $Q_i\wedge Q_k=0$ for all $Q_i,\, Q_k\in A$.

Obviously $c(u)\neq 0$ since otherwise $b(u)=0$ by $(J+1)b(u)^2+c(u)b'(u)=0$. Hence, $b(u)=0$ and $Q_{2J+1}=c(u)x^{-J}\gen u$ with $c(u)\neq 0$. Now, the residual equivalence group of $\langle e_1, e_2, e_3\rangle$ is given by invertible transformations of the form $t'=t,\; x'=x,\; u'=U(u)$. Under such a transformation, $Q_{2J+1}$ is mapped to $Q'_{2J+1}=c(u)U'(u)x^{-J}\gen {u'}$ and we choose $U(u)$ so that $c(u)U'(u)=1$ so that we may take $Q_{2J+1}=x^{-J}\gen u$. This gives
\[
Q_{2J}=2JtQ_{2J+1} + x^{-J+1}\gen u.
\]
Putting $Q_{2J+1}$ and $Q_{2J}$ as symmetries in the equation for $F$ we obtain the following two equations:
\begin{align*}
& x^2F_u-JxF_{u_1}+J(J-1)F_{u_2}=0\\
& x^2F_u-(J-1)xF_{u_1}+J(J-1)F_{u_2}=0,
\end{align*}
and this immediately gives $F_{u_1}=0$. The defining equations for $F$ (obtained from $\langle e_1, e_2, e_3\rangle$ as symmetries) are
\begin{align*}
& xF_x=u_1F_{u_1}+2u_2F_{u_2}+2F\\
& x^2F_x=2xu_1F_{u_1}+(4xu_2+2u_1)F_{u_2}+6xF.
\end{align*}
The second equation $x^2F_x=2xu_1F_{u_1}+(4xu_2+2u_1)F_{u_2}+6xF$ then gives $F_{u_2}=0$, and thus $F_u=0$. This gives us the reduced system $xF_x=2F,\; xF_x=6F$, so $F=0$, which is a contradiction. So we can only have irreducible representations of $\langle e_1, e_2, e_3\rangle$ on abelian Lie algebras for $J=0$.

We have the decomposition  $\mathsf{A}=\mathsf{k}_{\star}\dot{+}\mathsf{k}_0\dot{+}\mathsf{A}_0$, and $\mathsf{k}_0\dot{+}\mathsf{A}_0$ is a solvable Lie algebra whose operators commute with $e_1, e_2, e_3$, so the action of $\Sl(2, \mathbb{R})$ on $\mathsf{k}_0\dot{+}\mathsf{A}_0$ is a direct sum action. From Theorem \ref{directsumsthm} we know that this realization of $\Sl(2, \mathbb{R})$ admits only a one-dimensional direct-sum extension by solvable Lie algebras and we may put $\mathsf{k}_0\dot{+}\mathsf{A}_0=\mathsf{A}_0=\langle \gen u\rangle$. Hence for any $\mathsf{k}_J,\; J>0$ we have $[Q_{2J+1}, \gen u]=0$ so that the coefficients of $Q_{2J+1}$ are independent of $u$. Thus we may write

\[
Q_{2J+1}=a\gen t + bx^{-J+1}\gen x + cx^{-J}\gen u
\]
where $b,c$ are constants and $b^2+c^2\neq 0$ for otherwise $Q_{2J+1}=a\gen t$ giving $\mathsf{k}_J=\langle e_1, e_2, e_3\rangle$ which is impossible for a subspace $\mathsf{k}_J$ of a solvable Lie algebra. Since $\gen t$ is a symmetry of our equation (we have $e_3=\gen t$) we put $Q'_{2J+1}=bx^{-J+1}\gen x + cx^{-J}\gen u$ and then $Q'_{2J}=[e_2, Q'_{2J+1}]=2JtQ'_{2J+1} + (J+1)b(u)x^{-J+2}\gen x + c(u)x^{-J+1}\gen u$. It is now straightforward to show that $Q'_{2J+1}$ and $Q'_{2J}$ as symmetries in the equation for $F$ give $F=0$, which is a contradiction. Thus we have no $\mathsf{k}_J$ in $\mathsf{A}$ for $J>0$. Consequently we have only direct-sum extensions by solvable Lie algebras $\mathsf{A}$.

\medskip\noindent We now come to the final case of $\Sl(2, \mathbb{R})$, and we obtain the following result which we prove in a series of Lemmas:

\begin{thm}\label{specialsl2} The algebra

\[
\Sl(2, \mathbb{R})=\langle 2t\gen t + 2x\gen x,
-t^2\gen t-2tx\gen x - x\gen u, \gen t \rangle
\]
has no other extensions by solvable Lie algebras other than the following:
\begin{align*}
&\langle 2t\gen t + 2x\gen x,
-t^2\gen t-2tx\gen x - x\gen u, \gen t \rangle\oplus\langle \gen u \rangle\\
&\langle 2t\gen t + 2x\gen x,
-t^2\gen t-2tx\gen x - x\gen u, \gen t \rangle\oplus\langle x\gen x + u\gen u\rangle\\
&\langle 2t\gen t + 2x\gen x,
-t^2\gen t-2tx\gen x - x\gen u, \gen t \rangle\oplus\langle \gen u, x\gen x + u\gen u\rangle\\
&\langle 2t\gen t + 2x\gen x,
-t^2\gen t-2tx\gen x - x\gen u, \gen t \rangle\uplus D_{1/2}
\end{align*}
where $\displaystyle D_{1/2}=\langle x^{1/2}\gen x + \frac{1}{2}x^{-1/2}u\gen u, tx^{1/2}\gen x + \frac{1}{2}(tx^{-1/2}u+x^{1/2})\gen u \rangle$
\end{thm}
The first step in the proof is the following result on the representation of this realization of $\Sl(2, \mathbb{R})$ on abelian Lie algebras:

\begin{lemma} The only irreducible representations of the admissible realizations of

\[
\Sl(2, \mathbb{R})=\langle 2t\gen t + 2x\gen x,
-t^2\gen t-2xt\gen x - x\gen u, \gen t \rangle,
\]
on abelian Lie algebras are one dimensional, except for the irreducible representation space

\[
D_{1/2}=\langle x^{1/2}\gen x + \frac{1}{2}x^{-1/2}u\gen u, tx^{1/2}\gen x + \frac{1}{2}(tx^{-1/2}u+x^{1/2})\gen u \rangle.
\]
\end{lemma}

\smallskip\noindent{\bf Proof:} As before, $\mathsf{k}_J=\langle Q_1,\dots, Q_{2J+1}\rangle$ denotes an irreducible representation space for $J>0$. We also assume that $\mathsf{k}_J$ is abelian. Note that $e_1$ acts semi-simply on $\mathsf{k}_J=\langle Q_1,\dots, Q_{2J+1}\rangle$ and that $e_2$ and $e_3$ act nilpotently on $A$. We have in particular $[e_1, Q_k]=2(J-k+1)Q_k$ and
\[
[e_3, Q_{2J+1}]=0,\quad Q_k=\frac{({\rm ad}\, e_3)^{k-1}}{(k-1)!}(Q_1)
\]
as well as
\[
[e_2, Q_1]=0,\quad Q_{2J-k+1}=\frac{({\rm ad}\, e_2)^{k}}{k!}(Q_{2J+1}).
\]
Thus, if $Q_1=A(t)\gen t + B(t,x,u)\gen x + C(t,x,u)\gen u$ then $A(t), B(t,x,u), C(t,x,u)$ are polynomials of degree no greater than $2J$ in $t$.

\medskip\noindent The commutator $[e_1, Q_1]=2JQ_1$ gives $t\dot{A}(t)=(J+1)A(t)$ so that $A(t)=aT^{J+1}$. Further, $[e_2, Q_1]=0$ gives $(J-1)a=0$ so $A(t)=0$ if $J\neq 1$. But we also require $\mathsf{k}_J$ to be abelian and this gives us $a=0$ (by considering $[Q_2, Q_3]=0$ in the case of $J=1$). Thus we have $Q_1=B(t,x,u)\gen x + C(t,x,u)\gen u$ for all half-integers $J>0$ and consequently all the $Q_i$ have the structure $Q_i=a_i(t,x,u)\gen x + b_i(t,x,u)\gen u$ so that ${\rm rank}\,\mathsf{k}_J\leq 2$. We also note that ${\rm rank}\, \mathsf{k}_J=2$ is impossible for $\dim \mathsf{k}_J>2$: any admissible rank-two abelian Lie algebra $\mathsf{A}$ has dimension $\dim \mathsf{A}=2, 3, 4$ and for $\dim \mathsf{A}=3, 4$ there are wedge products which are zero, as is clear from the list of admissible abelian Lie algebras. For $\dim \mathsf{A}=3, 4$, we know that $\langle e_1, e_2, e_3\rangle$ acts irreducibly on the spaces of bivectors $\langle Q_1\wedge Q_2, Q_1\wedge Q_3, Q_2\wedge Q_3\rangle$ and $\langle Q_1\wedge Q_2, Q_1\wedge Q_3, Q_1\wedge Q_4, Q_2\wedge Q_3, Q_2\wedge Q_4, Q_3\wedge Q_4\rangle$, so that if any wedge product is zero, so are all the others. Thus we have to investigate only the cases ${\rm rank}\,\mathsf{k}_J=1$ for $\dim \mathsf{k}_J\geq 2$ and ${\rm rank}\, \mathsf{k}_J=2, \dim \mathsf{k}_J=2$.

The commutation relations $[e_1, Q_1]=2JQ_1$ and $[e_2, Q_1]=0$ give us the system of equations
\begin{align*}
& tB_t+xB_x=(J+1)B\\
& tC_t+xC_x=JC\\
& t^2B_t+2txB_x+xB_u=2tB\\
& t^2C_t+2txC_x+xC_u=B,
\end{align*}
whose solution is
\[
B(t,x,u)=x^{-J+1}t^{2J}b(\tau),\quad C(t,x,u)=x^{-J}t^{2J}[c(\tau)+xt^{-1}b(\tau)]
\]
where $\tau=u-x/t$. Now $B,C$ have to be polynomials of degree not greater than $2J$ in $t$ and we find that $b(\tau), c(\tau)$ are polynomials of degree $2J$ and $2J+1$ respectively. Performing a Taylor expansion in $x/t$ about $u$ we obtain
\[
B(t,x,u)=x^{-J+1}\sum_{k=0}^{2J}(-1)^k\frac{b^{(k)}(u)}{k!}x^kt^{2J-k}
\]
and
\[
C(t,x,u)=x^{-J}\left[c(u)t^{2J} + \sum_{k=1}^{2J}(-1)^k\frac{(-1)^k}{k!}[c^{(k)}(u)-kb^{(k-1)}(u)]x^kt^{2J-k}\right]
\]
with $c^{(2J+1)}(u)=(2J+1)b^{(2J)}(u)$. From these coefficients we find that
\begin{align*}
& Q_{2J+1}=b(u)x^{-J+1}\gen x + c(u)x^{-J}\gen u,\\
& Q_{2J}=2JtQ_{2J+1} -b'(u)x^{-J+2}\gen x - (c'(u)-b(u))x^{-J+1}\gen u,
\end{align*}
where $b(u),\; c(u)$ are polynomials of degree at most $2J$ and $2J+1$ respectively, and $c^{(2J+1)}(u)=(2J+1)b^{(2J)}(u)$. Note that these calculations are true without assuming that $\mathsf{k}_J$ is abelian.

\smallskip\noindent{\underline{$\dim \mathsf{k}_J\geq 2,\; {\rm rank}\,\mathsf{k}_J=1$}:} We have ${\rm rank}\, \mathsf{k}_J=1$ if and only if $Q_{2J+1}\wedge Q_{2J}=0$. This gives us
\begin{equation}\label{rank1}
b^2=bc'-b'c.
\end{equation}
Further, $[Q_{2J+1}, Q_{2J}]=0$ gives us
\begin{align}
& 2bb'=b'c'-cb'' \label{abelian1}\\
& (J-1)[b^2+b'c] - (J-2)bc'+cc''-(c')^2=0 \label{abelian2}.
\end{align}
If $c(u)=0$ then equation (\ref{rank1}) gives $b(u)=0$ so that $\mathsf{k}_J=\{0\}$, so we must have $c(u)\neq 0$. If now $b(u)=0$ then equation (\ref{abelian2}) gives $cc''-(c')^2=0$ which is impossible  for a polynomial $c(u)$ unless $c(u)={\rm constant}$. In this case we may take $Q_{2J+1}=x^{-J}\gen u$ in canonical form, and this gives $Q_{2J}=2JtQ_{2J+1}$. Putting $Q_{2J+1}$ and $Q_{2J}$ as symmetries into the equation for $G$ and using the fact that $G_t=0$ we obtain $2Jx^{-J}=0$, which is impossible for $J>0$.

\medskip\noindent We are left with $b\neq 0,\, c\neq 0$ and then equation (\ref{rank1}) may be written as
\[
\frac{b'}{c}-\frac{bc'}{c^2}+\left(\frac{b}{c}\right)^2=0,
\]
which is just
\[
\frac{d}{du}\left(\frac{b}{c}\right)=-\left(\frac{b}{c}\right)^2,
\]
from which it follows that $c(u)=(u+l)b(u)$ and we obtain
\[
Q_{2J+1}=b(u)x^{-J+1}\gen x + (u+l)b(u)x^{-J}\gen u.
\]
The residual equivalence group of $\langle e_1, e_2, e_3\rangle$ is given by invertible transformations of the form $t'=t,\, x'=\alpha x,\, u'=\alpha u+k$ with $\alpha\neq 0$. Choosing $k=l$ we see that $Q_{2J+1}$ is mapped to
\[
Q'_{2J+1}=\tilde{b}(u')(x')^{-J+1}\gen {x'} + u'\tilde{b}(u')(x')^{-J}\gen {u'},
\]
and we see that we may take
\[
Q_{2J+1}=b(u)x^{-J+1}\gen x + ub(u)x^{-J}\gen u
\]
in canonical form, with $b(u)$ a polynomial of degree $2J$. Now, putting $c(u)=ub(u)$ into equation (\ref{abelian1}) we obtain $bb'+u[bb''-(b')^2]=0$, and so we obtain $\displaystyle u\frac{d}{du}\left(\frac{b'}{b}\right)+\frac{b'}{b}=0$ so that we have $b(u)=\lambda u^m$ for some constant $\lambda\neq 0$ and some integer $m\geq 0$. It is clear that we may assume $\lambda=1$ and this gives
\[
Q_{2J+1}=u^mx^{-J+1}\gen x + u^{m+1}x^{-J}\gen u.
\]
This gives $Q_{2J}=2JtQ_{2J+1}-mu^{m-1}x^{-J+2}\gen x - mu^mx^{-J+1}\gen u$.

\medskip\noindent For $m=0$ we have $Q_{2J+1}=x^{-J+1}\gen x + ux^{-J}\gen u$ and $Q_{2J}=2JtQ_{2J+1}$. Putting $Q_{2J+1}$ together with $Q_{2J}$ as symmetries into the equation for $G$ and noting that $G_t=0$, we obtain
\[
-2Jx^{-J+1}u_1+2Jx^{-J}=0,
\]
which is a contradiction for $J>0$. Thus, we have no irreducible representation of $\langle e_1, e_2, e_3\rangle$ in this case. For $m\geq 1$, putting $Q_{2J+1}=u^mx^{-J+1}\gen x + u^{m+1}x^{-J}\gen u$ and $Q_{2J}=2JtQ_{2J+1}-mu^{m-1}x^{-J+2}\gen x - mu^mx^{-J+1}\gen u$ into the equation for $F$ gives (after a standard calculation) $F=0$, which is a contradiction. Hence we have no realization of $\Sl(2, \mathbb{R})$ on  rank-one abelian Lie algebras $\mathsf{A}$ with $\dim \mathsf{A}\geq 2$.

\smallskip\noindent{\underline{$\dim \mathsf{A}=2,\; {\rm rank}\, \mathsf{A}=2$}.} In this case $\mathsf{A}=\langle Q_1, Q_2\rangle$ with $Q_2=x^{1/2}b(u)\gen x + x^{-1/2}c(u)\gen u$ and $Q_1=tQ_2 + x^{3/2}b'(u)\gen x + x^{1/2}[c'(u)-b(u)]\gen u$. Further $Q_2\wedge Q_1\neq 0$ for a rank-two algebra. We know that $\deg b(u)\leq 1$ and that $c''(u)=2b'(u)$, so $b(u)=\alpha u+\beta$ and $c''(u)=2\alpha$, so that $c(u)=au^2+\gamma u + \delta$ for some constants $\alpha, \beta, \gamma, \delta$. Putting this into equation (\ref{abelian1}) we find that $\alpha(2\beta-\gamma)=0$. Then equation (\ref{abelian2}) gives $\alpha=0$. We note that $b(u)=l\neq 0$ since otherwise we would have a rank-one realization, so we may assume that $b(u)=1$ and with $c(u)=\gamma u + \delta$ we obtain $\gamma=1$ or $\gamma=1/2$, giving $Q_2=x^{1/2}\gen x + x^{-1/2}[u+\delta]\gen u$ and $\displaystyle Q_2=x^{1/2}\gen x + x^{-1/2}[\frac{u}{2}+\delta]\gen u$ respectively. For $Q_2=x^{1/2}\gen x + x^{-1/2}[u+\delta]\gen u$ we find that $Q_1=tQ_2$ and this gives us a rank-one realization, which is a contradiction. Thus we have only $\displaystyle Q_2=x^{1/2}\gen x + x^{-1/2}[\frac{u}{2}+\delta]\gen u$. Using an equivalence transformation $t'=t, x'=x, u'= u+2\delta$ which is in the residual equivalence group of $\langle e_1, e_2, e_3\rangle$, we find that we may put $Q_2$ in the canonical form $\displaystyle Q_2=x^{1/2}\gen x + x^{-1/2}\frac{u}{2}\gen u$. We take $\displaystyle Q_2=2x^{1/2}\gen x + x^{-1/2}u\gen u$ for convenience, and this gives $Q_1=tQ_2+x^{1/2}\gen u$.

\medskip\noindent With these as symmetries in the equations for $F$ and $G$ already obtained for $\langle e_1, e_2, e_3\rangle$, we find
\[
F=Kx^2\tau^{-1/3},\quad G=x^{-1}\left[\frac{3}{4}K(xu_1-u)\tau^{-1/3}+L(xu_1-u)^{3/2} + u^2 - 2xuu_1\right]
\]
where $\tau=2x^2u_2-xu_1+u$ and $K\neq 0, L$ are constants.

\begin{lemma} If $\displaystyle \langle 2t\gen t + 2x\gen x, -t^2\gen t - 2tx\gen x -x\gen u\rangle$ is represented on a solvable Lie algebra $\mathsf{A}$ such that $\displaystyle D_{1/2}\subset\mathsf{A}$, then $\displaystyle \mathsf{A}=D_{1/2}$.
\end{lemma}

\smallskip\noindent{\bf Proof:} Suppose that $\mathsf{A}$ also contains $\mathsf{k}_J$ with $J\geq 1/2$. Then since $\dim\mathsf{A}< \infty$ there is a maximal value $j$ of $J$ for which $\mathsf{k}_J\subset \mathsf{A}$. Now put $\displaystyle X=x^{1/2}\gen x + \frac{u}{2}x^{-1/2}\gen u\in D_{1/2}$. Then, if $Q=[Q_{2j+1}, X]\neq 0,$ we have $[e_3, Q]=0$ and $\displaystyle [e_1, Q]=-2(j+\frac{1}{2})Q$. Further
\[
({\rm ad}\, e_2)^{2j+2}(Q)=\sum_{N=0}^{2j+2}[({\rm ad}\, e_2)^{N}(Q_{2j+1}), ({\rm ad}\, e_2)^{2j+2-N}(X)]=0
\]
since $({\rm ad}\, e_2)^{2j+1}(Q)_{2j+1}=0,\; ({\rm ad}\, e_2)^2(X)=0$. Then put $Q'_{2j+2}=Q$ we can define an irreducible space $k_{j+\frac{1}{2}}=\langle Q'_1,\dots, Q'_{2j+\frac{1}{2}}\rangle$ by defining $\displaystyle Q'_k=\frac{({\rm ad}\, e_2)^k}{k!}(Q'_{2j+2})$. Clearly, $k_{j+\frac{1}{2}}\subset\mathsf{A}$ and corresponds to spin $j+1/2$ if $k_{j+\frac{1}{2}}\neq \{0\}$, contradicting the fact that $j$ was the largest value of $J$ for which $\mathsf{k}_J\subset\mathsf{A}$. Thus we must have $[Q_{2j+1}, X]=0$. Writing $Q_{2j+1}=b(u)x^{-j+1}\gen x + c(u)x^{-j}\gen u$, and using the fact that $c^{(2j+1)}(u)=(2j+1)b^{(2j)}(u)$, the condition $[Q_{2j+1}, X]=0$ gives us
\[
b(u)=qu^{2j-1},\quad c(u)=\frac{q}{2}u^{2j}.
\]
We assume $q\neq 0$ so that we may take
\[
Q_{2j+1}=u^{2j-1}x^{-j+1}\gen x + \frac{u^{2j}}{2}x^{-j}\gen u.
\]
At this point, we note that in fact we have, for the case $J=1$, that $Q_3=a\gen t + b(u)\gen x + c(u)x^{-1}$. If $a\neq 0$ then $Q_3, Q_2, Q_1$ will contain the terms $ae_3, ae_2, ae_1$ which form $\Sl(2, \mathbb{R})$, and this is not permissible if $Q_1, Q_2, Q_3$ belong to a solvable Lie algebra. Thus we must have $a=0$ in this case and then our calculations cover all cases of $J\geq 1/2$.

It is a straightforward (but painstaking) calculation to show that putting this operator into the equation for $F$ gives $F=0$ when $2j-1\neq 0$. Hence we conclude that there are no irreducible subspaces $\mathsf{k}_J$ with $J>1/2$ contained in $\mathsf{A}$. It also follows from this (and using the same argument) that there are no other irreducible subspaces $\mathsf{k}_{1/2}$ of spin $J=1/2$ other than $D_{1/2}$. Then we have $\mathsf{A}=D_{1/2}\dot{+}\mathsf{k}_0$ where $\mathsf{k}_0$ is the space of operators commuting with $e_1, e_2, e_3$. We also have the abelian ideal $\mathsf{A}_0\subset\mathsf{k}_0$. The same argument as given previously shows that $[D_{1/2}, \mathsf{k}_0]\subset D_{1/2}$. Now any operator $Q\in \mathsf{k}_0$ is of the form $Q=\alpha\gen u + \beta(x\gen x+u\gen u)$, as follows from Theorem \ref{directsumsthm}. Then we have
\[
[Q, X]=\alpha x^{-1/2}\gen u -\frac{\beta}{2}X,
\]
and it is clear that $[Q, X]\notin D_{1/2}$ unless $\alpha=0$. Thus we must have $\mathsf{k}_{0}=\langle x\gen x + u\gen u\rangle$ or $\mathsf{k}_0=\{0\}$. It is a routine calculation that $D_{1/2}$ and $x\gen x + u\gen u$ are incompatible symmetry algebras of our equation: $D_{1/2}$ gives $F=Kx^2\tau^{-1/3}$ with $\tau= 2x^2u_2-xu_1+u)$, whereas $x\gen x + u\gen u$ gives us the condition $xF_x+uF_u=u_2F_{u_2} + 3F$ which is not satisfied by $F=Kx^2\tau^{-1/3}$. Hence we conclude that $\mathsf{A}=D_{1/2}$ if $D_{1/2}\subset \mathsf{A}$.

\begin{lemma} The only extensions of
\[
\langle 2t\gen t + 2x\gen x,
-t^2\gen t-2tx\gen x - x\gen u, \gen t \rangle
\]
by solvable Lie algebras $\mathsf{A}$ such that $\mathsf{A}$ does not contain $D_{1/2}$ are as given in Theorem \ref{directsumsthm}:
\begin{align*}
&\langle 2t\gen t + 2x\gen x,
-t^2\gen t-2tx\gen x - x\gen u, \gen t \rangle\oplus\langle \gen u \rangle\\
&\langle 2t\gen t + 2x\gen x,
-t^2\gen t-2tx\gen x - x\gen u, \gen t \rangle\oplus\langle x\gen x + u\gen u\rangle\\
&\langle 2t\gen t + 2x\gen x,
-t^2\gen t-2tx\gen x - x\gen u, \gen t \rangle\oplus\langle \gen u, x\gen x + u\gen u\rangle.
\end{align*}
\end{lemma}

\smallskip\noindent{\bf Proof:} In this case we may decompose $\mathsf{A}$ as the direct sum $\mathsf{A}=\mathsf{k}_{\star}\dot{+}\mathsf{k}_0\dot{+}\mathsf{A}_0$ where now $\mathsf{A}_0$ is the abelian ideal containing operators commuting with $\Sl(2, \mathbb{R})$. We also note that $\mathsf{k}_0\dot{+}\mathsf{A}_0$ is a solvable Lie algebra and the action of $\Sl(2, \mathbb{R})$ on $\mathsf{k}_0\dot{+}\mathsf{A}_0$ is a direct sum action. Then we have from Theorem \ref{directsumsthm} that $\dim(\mathsf{k}_0\dot{+}\mathsf{A}_0)=1$ or $\dim(\mathsf{k}_0\dot{+}\mathsf{A}_0)=2$. If $\dim(\mathsf{k}_0\dot{+}\mathsf{A}_0)=1$, then $\mathsf{A}_0=\langle \gen u\rangle$ or $\mathsf{A}_0=\langle x\gen x + u\gen u\rangle$ and in both these cases $\mathsf{k}_0=\{0\}$. For $\dim(\mathsf{k}_0\dot{+}\mathsf{A}_0)=2$ we have $\mathsf{k}_0\dot{+}\mathsf{A}_0=\langle \gen u, x\gen x + u\gen u\rangle$ so that $\mathsf{A}_0=\langle \gen u\rangle$ and $\mathsf{k}_0=\langle x\gen x + u\gen u\rangle$.

\medskip\noindent For any $\mathsf{k}_J=\langle Q_1,\dots, Q_{2J+1}\rangle$ with $J>0$ we have $Q_{2J+1}=a\gen t + b(u)x^{-J+1}\gen x + c(u)x^{-J}\gen u$, and $a=0$ except possibly when $J=1$. For this case we have $Q_3=a\gen t + b(u)\gen x + c(u)x^{-1}\gen u$ and one can show that $b(u),\; c(u)$ are polynomials with ${\rm deg}\, b(u)\leq 2,\; {\rm deg}\, c(u)\leq 3$. Note that $\gen t=e_1$ so that writing $Q_3=ae_1+Q'_3$ we have $Q_2=ae_1 + Q'_2$ and $Q_1=ae_2 + Q'_1$ where $Q'_2=[e_2, Q_3]$ and $2Q'_1=[e_2, Q'_2]$. Since $e_1, e_2, e_3$ are already symmetries, we may reduce the problem of having $Q_1, Q_2, Q_3$ as symmetries to that of having $Q'_1, Q'_2, Q'_3$ as symmetries. This is just the special case of $J=1$ for $Q_{2J+1}=b(u)x^{-J+1}\gen x + c(u)x^{-J}\gen u$. Thus we consider only $\mathsf{k}_J=\langle Q_1,\dots, Q_{2J+1}\rangle$ for $J>0$ where $Q_{2J+1}=b(u)x^{-J+1}\gen x + c(u)x^{-J}\gen u$.

\medskip\noindent We also have
\[
Q_{2J}=2JtQ_{2J+1}-b'(u)x^{-J+2}\gen x + [b(u)-c'(u)]x^{-J+1}\gen u
\]
and we have either ${\rm rank}\,\mathsf{k}_J=1$ or ${\rm rank}\,\mathsf{k}_J=2$.

\medskip\noindent\underline{$\mathsf{A}_0=\langle \gen u\rangle,\;\; \mathsf{k}_0=\{0\},\;\; {\rm or}\, \mathsf{k}_0=\{x\gen x + u\gen u\}:$} Note that $[\mathsf{k}_J, \gen u]=\{0\}$ by Lemma \ref{usefullemma} so $b, c$ are constants and we obtain
\[
Q_{2J+1}=bx^{-J+1}\gen x + cx^{-J}\gen u,\quad Q_{2J}=2JtQ_{2J+1} + bx^{-J+1}\gen u
\]

\medskip\noindent If ${\rm rank}\, \mathsf{k}_J=1$  then $Q_{2J+1}\wedge Q_{2J}=0$ and this yields $b=0$ so that we may take $Q_{2J+1}=x^{-J}\gen u$ and $Q_{2J}=2JtQ_{2J+1}$. Putting these as symmetries into the equation for $G$, and noting that $G_t=0$, we obtain $Ju_1=0$ which is impossible for $J>0$. So we have no realization in this case.

\medskip\noindent If ${\rm rank}\, \mathsf{k}_J=2$ then $b\neq 0$ and we may take $b=1$ so that $Q_{2J+1}=x^{-J+1}\gen x + cx^{-J}\gen u$ and $Q_{2J}=2JtQ_{2J+1} + x^{-J+1}\gen u$. Note that we also have $ F=x^2f(\omega, \tau)$ with $\omega=xu_1-u,\; \tau=x^2u_2$. Since $\gen u$ is also a symmetry we must have $F_u=0$ so that $f_{\omega}=0$ giving $F=x^2f(\tau)$ and consequently $F_{u_1}=0$. We have two cases: $c=0$ and $c\neq 0$. If $c=0$ then $Q_{2J+1}=x^{-J+1}\gen x,\; Q_{2J}=2JtQ_{2J+1}+ x^{-J+1}\gen u$. Putting these two operators a symmetries into the equation for $F$, and noting that $F_t=F_u=F_{u_1}=0$ we obtain the equations
\begin{align*}
(J-1)[Jx^{-J-1}u_1-2x^{-J}u_2]F_{u_2}-3(J-1)x^{-J}F&=x^{-J+1}F_x\\
J(J-1)F_{u_2}=0.
\end{align*}
For $J\neq 1$ we clearly have $F_{u_2}=0$ from the second equation, giving $F=Kx^2$, and then the first equation gives $(3J-2)K=0$, and since $J$ is a half-integer we have $K=0$ so that $F=0$ which is a contradiction. For $J=1$ we have $Q_3=\gen x,\; Q_2=2t\gen x + \gen u$. Putting these into the equation for $G$, and noting that we have $G_t=G_u=0$ since both $\gen t$ and $\gen u$ are symmetries, we find that $2u_1=0$ which is a contradiction.

\medskip\noindent\underline{$\mathsf{A}_0=\langle x\gen x + u\gen u\rangle,\;\; \mathsf{k}_0=\{0\}:$} In this case we have $[Q_{2J+1}, x\gen x + u\gen u]=0$. With $Q_{2J+1}=b(u)x^{-J+1}\gen x + c(u)x^{-J}\gen u$ this yields $Q_{2J+1}=pu^{J}x^{-J+1}\gen x + qu^{J+1}x^{-J}\gen u$ where $p, q$ are constants. We know that $b(u)$ and $c(u)$ are polynomials, so $J$ must be an integer. Again, if ${\rm rank}\,\mathsf{k}_J=1$ then $p=0$ and then we may take $Q_{2J+1}=u^{J+1}x^{-J}\gen u$ which gives $Q_{2J}=2JtQ_{2J+1}$. Putting these into the equation for $G$, and noting that $G_t=0$, we obtain $Ju^{J+1}x^{-J}=0$ which is impossible since $J>0$.

For ${\rm rank}\,\mathsf{k}_J=2$ we have $p\neq 0$ so we may take $p=1$. In this case we find $Q_{2J+1}=u^{J}x^{-J+1}\gen x + qu^{J+1}x^{-J}\gen u$ and $Q_{2J}=2JtQ_{2J+1} -Ju^{J-1}x^{-J+2}\gen x -[q(J+1)-1]u^{J}x^{-J+1}\gen u$ and a calculation such as in the case of $Q_{2J+1}=u^mx^{-J+1}\gen x + u^{m+1}x^{-J}\gen u$ for integer $m>0$  gives $F=0$ (on using $x\gen x + u\gen u$ as a symmetry). This is a contradiction, so $\mathsf{k}_{\star}=\{0\}$ and we have only direct-sum extensions by solvable Lie algebras, as given in Theorem \ref{directsumsthm}.

\medskip\noindent In concluding this section, we note that in some of the above cases it is still possible to prove the non-existence of other extensions by calculating the maximal symmetry groups. However, there are cases where the structure of the equations leads to fearsome calculations. In these cases it is easier to use the present approach. We also wanted to have a unified approach as well as to illustrate how powerful this method can be, and to show that the calculations are in general less complicated than a direct calculation of the maximal symmetry algebra.

\section{Equations invariant under extensions of ${\rm sl}(2, \mathbb{R})$.} In this section we list the evolution equations which admit the various extensions by solvable Lie algebras $\mathsf{A}$ of $\Sl(2, \mathbb{R})$ as symmetry algebras.

\bigskip\noindent\underline{\bf Equations for extensions $\Sl(2, \mathbb{R})\uplus\mathsf{A}$ with $\dim\mathsf{A}=1$:}

\bigskip\noindent\underline{$\displaystyle \langle 2t\gen t + 2x\gen x,
-t^2\gen t-(x^2+2xt)\gen x, \gen t \rangle\oplus\langle \gen u\rangle$:}

\[
u_t=\frac{f(\tau)}{x^2u_1^4}u_3 + \frac{6f(\tau)}{x^3u_1^2}\left(\tau - \frac{1}{xu_1}\right)+\frac{g(\tau)}{x^2u_1} + u_1,
\]
with $\displaystyle \tau=\frac{u_2}{u_1^2}+\frac{2}{xu_1}.$

\medskip\noindent\underline{$\displaystyle\langle 2t\gen t + 2x\gen x,
-t^2\gen t-2xt\gen x - x\gen u, \gen t \rangle\oplus\langle \gen u\rangle$:}

\[
u_t=x^2f(\tau)u_3+x^{-1}[g(\tau)-x^2u_1^2],\;\; \tau=x^2u_2.
\]

\medskip\noindent\underline{$\displaystyle \langle 2t\gen t + 2x\gen x,
-t^2\gen t-2xt\gen x - x\gen u, \gen t \rangle\oplus\langle x\gen x +  u\gen u\rangle$:}

\[
u_t=x^4u_2f(\tau)u_3+x^{-1}[x^4u_2^2g(\tau) + u^2 -2xuu_1],\;\; \tau=\frac{xu_1-u}{x^2u_2}.
\]

\medskip\noindent\underline{$\displaystyle\langle 2x\gen x ,
-x^2\gen x, \gen x \rangle\oplus\langle \gen t\rangle$:}

\[
u_t=f(u)\left[\frac{u_3}{u_1^3}-\frac{3}{2}\frac{u_2^2}{u_1^4}\right] + g(u).
\]

\medskip\noindent\underline{$\displaystyle \langle 2x\gen x ,
-x^2\gen x, \gen x \rangle\oplus\langle \gen u\rangle$:}

\[
u_t=f(t)\left[\frac{u_3}{u_1^3}-\frac{3}{2}\frac{u_2^2}{u_1^4}\right] + g(t).
\]

\medskip\noindent\underline{$\displaystyle \langle 2x\gen x + 2u\gen u,
-x^2\gen x-2xu\gen u, \gen x \rangle\oplus\langle \gen t\rangle$:}

\[
u_t=f(\tau)u^3u_3+ug(\tau),\quad \tau=2uu_2-u_1^2.
\]

\medskip\noindent\underline{$\displaystyle \langle 2x\gen x + 2u\gen u,
-x^2\gen x-2xu\gen u, \gen x \rangle\oplus\langle u\gen u\rangle$:}

\[
u_t=\frac{f(t)u^3}{\tau^{3/2}}u_3+ug(t),\quad \tau=2uu_2-u_1^2.
\]

\medskip\noindent\underline{$\displaystyle\langle 2x\gen x + 2u\gen u,
-x^2\gen x-2xu\gen u, \gen x \rangle\oplus\langle tu\gen u\rangle$:}

\[
u_t=\frac{f(t)u^3}{\tau^{3/2}}u_3+ug(t) + \frac{u}{2t}\ln|\tau|,\quad \tau=2uu_2-u_1^2.
\]

\medskip\noindent\underline{$\displaystyle \langle 2x\gen x + 2u\gen u,
-(x^2-u^2)\gen x-2xu\gen u, \gen x \rangle\oplus\langle \gen t\rangle$:}

\[
u_t=u^3f(\tau)\left[\frac{u_3}{(1+u^2_1)^{3/2}}-\frac{3u_1u^2_2}{(1+u^2_1)^{5/2}}\right] + u\sqrt{1+u^2_1}h(\tau)
\]
where $\displaystyle \tau=\frac{1+u^2_1+uu_2}{(1+u^2_1)^{3/2}}$.

\medskip\noindent\underline{$\displaystyle\langle 2x\gen x + 2u\gen u,
-(x^2+u^2)\gen x-2xu\gen u, \gen x \rangle\oplus\langle \gen t\rangle$:}

\[
u_t=u^3f(\tau)\left[\frac{u_3}{|1-u^2_1|^{3/2}}-\frac{3u_1u^2_2}{|1-u^2_1|^{5/2}}\right] + u\sqrt{|1-u^2_1|}h(\tau)
\]
where $\displaystyle \tau=\frac{|1-u^2_1|+uu_2}{|1-u^2_1|^{3/2}}$. Note that here we must take the two cases $u^2_1>1$ and $u^2_1<1$ separately.

\bigskip\noindent\underline{\bf Equations for extensions $\Sl(2, \mathbb{R})\uplus\mathsf{A}$ with $\dim\mathsf{A}=2$:}

\bigskip\noindent\underline{$\displaystyle \langle 2t\gen t + 2x\gen x,
-t^2\gen t-(x^2+2xt)\gen x, \gen t \rangle\oplus\langle \gen u, u\gen u\rangle$:}

\[
u_t=\frac{K}{x^2u_1^4\tau^4}u_3 + \frac{6fK}{x^3u_1^2\tau^4}\left(\tau - \frac{1}{xu_1}\right)+\frac{L}{x^2u_1\tau^2} + u_1,
\]
with $K\neq 0$ and $\displaystyle \tau=\frac{u_2}{u_1^2}+\frac{2}{xu_1}.$

\medskip\noindent\underline{$\displaystyle \langle 2t\gen t + 2x\gen x,
-t^2\gen t-2xt\gen x - x\gen u, \gen t \rangle\oplus\langle \gen u, x\gen x + u\gen u\rangle$:}

\[
u_t=Kx^4u_2u_3-xu_1^2 + Lx^3u_2^2,\;\; K\neq 0.
\]

\medskip\noindent\underline{$\displaystyle\langle 2t\gen t + 2x\gen x,
-t^2\gen t-2xt\gen x - x\gen u, \gen t \rangle\uplus D_{1/2}$:}

\[
u_t=Kx^2\tau^{-1/3} + x^{-1}\left[\frac{3}{4}K(xu_1-u)\tau^{-1/3}+L(xu_1-u)^{3/2} + u^2 - 2xuu_1\right]
\]
where $\displaystyle \tau= 2x^2u_2 - xu_1 + u$ and $K\neq 0$.

\medskip\noindent\underline{$\displaystyle \langle 2x\gen x ,
-x^2\gen x, \gen x \rangle\oplus\langle \gen t, \gen u\rangle$:}

\[
u_t=K\left[\frac{u_3}{u_1^3}-\frac{3}{2}\frac{u_2^2}{u_1^4}\right] + L,\;\; K\neq 0.
\]

\medskip\noindent\underline{$\displaystyle \langle 2x\gen x ,
-x^2\gen x, \gen x \rangle\oplus\langle \gen t, t\gen t + u\gen u\rangle$:}

\[
u_t=Ku^2\left[\frac{u_3}{u_1^3}-\frac{3}{2}\frac{u_2^2}{u_1^4}\right] + L,\;\; K\neq 0.
\]

\medskip\noindent\underline{$\displaystyle \langle 2x\gen x ,
-x^2\gen x, \gen x \rangle\oplus\langle \gen u, t\gen t + u\gen u\rangle$:}

\[
u_t=Kt^2\left[\frac{u_3}{u_1^3}-\frac{3}{2}\frac{u_2^2}{u_1^4}\right] + L,\;\; K\neq 0.
\]

\medskip\noindent\underline{$\displaystyle\langle 2x\gen x + 2u\gen u,
-x^2\gen x-2xu\gen u, \gen x \rangle\oplus\langle \gen t, u\gen u\rangle$:}

\[
u_t=\frac{Ku^3}{\tau^{3/2}}u_3+Lu,\quad \tau=2uu_2-u_1^2,\;\; K\neq 0.
\]

\medskip\noindent\underline{$\displaystyle\langle 2x\gen x + 2u\gen u,
-x^2\gen x-2xu\gen u, \gen x \rangle\oplus\langle \gen t, t\gen t + qu\gen u\rangle$:}

\[
u_t=\frac{Ku^3}{\tau^{3/2+1/2q}}u_3 + \frac{Lu}{\tau^{1/2q}},\;\; q\neq 0,\;\; K\neq 0,
\]
where $\displaystyle \tau=2uu_2-u_1^2$. Note that $\displaystyle K=1,\; L=0,\; q=-\frac{1}{3}$ gives us the Harry Dym equation $\displaystyle u_t=u^3u_3$.

\bigskip\noindent´\underline{\bf Equations for extensions $\Sl(2, \mathbb{R})\uplus\mathsf{A}$ with $\dim\mathsf{A}=3$:} Here, there is only one three-dimensional extension of a semi-simple symmetry algebra:

\bigskip\noindent\underline{$\displaystyle \langle 2x\gen x, -x^2\gen x, \gen x\rangle\oplus\langle \gen t, \gen u, t\gen t + \left(qt+\frac{u}{3}\right)\gen u\rangle, \quad q\in \mathbb{R}$:}

\[
u_t=K\left(\frac{u_3}{u_1^3} - \frac{3}{2}\frac{u_2^2}{u_1^4}\right),\;\; K\neq 0.
\]

\section{Conclusion.} We have given a detailed account of the point symmetry classification of evolution equations of the type given in equation (\ref{eveqn}) and we have given an exhaustive list of all the types of equations which admit real semi-simple Lie algebras as symmetries, as well as all the possible extensions of these symmetry algebras by solvable Lie algebras. In doing so we have used a somewhat new approach to the problem of calculating the extensions of the semi-simple Lie algebras. This involves using elementary representation theory in order to find canonical forms for symmetry operator candidates which can then be tested as symmetry operators. Although this approach also involves detailed calculations, it avoids the heavy computation required if one is to try an calculate these extensions by the direct approach using the standard Lie algorithm. Indeed, our approach was necessitated by the almost impossible nature of these computations.

In the second paper (Part II), we will give a classification of all evolution equations (\ref{eveqn}) which admit solvable Lie symmetry algebras $\mathsf{A}$ and which are not linearized. We have found that there are 48 types of equations of the form (\ref{eveqn}) for $\dim\mathsf{A}=3$, 88 types for $\dim\mathsf{A}=4$ and there are 55 equations for $\dim\mathsf{A}=5$. At this stage, all non-linearities have been specified (up to a multiplicative constant). In this sense, we have been able to give a complete specification of these equations by point symmetries. Not all the types admitting $\dim\mathsf{A}=4$ admit a solvable $\mathsf{A}$ with $\dim\mathsf{A}=5$ and these types of equation then require higher-order symmetries in order to specify the non-linearities (up to constants). Thus, we see that our classification touches upon the problem of finding complete symmetry groups for an equation. We intend to revisit this question in the future.

When extending solvable Lie algebras one should note that there is also the possibility of having a {\it semi-simple extension}. For instance $\Sl(2, \mathbb{R})=\langle e_1, e_2, e_3\rangle$ with $[e_1, e_2]=2e_2, \; [e_1, e_3]=-2e_3,\; [e_2, e_3]=e_1$ is a semi-simple extension of the solvable Lie algebra $\langle e_1, e_2\rangle$, so that any direct-sum extension of $\Sl(2, \mathbb{R})$ is a semi-simple extension of a three- , four- or five-dimensional solvable Lie algebra. We shall discuss this phenomenon in Part II and list the appropriate semi-simple extensions.


\appendix
\section{Representations of ${\rm so}(3, \mathbb{R})$ and ${\rm sl}(2, \mathbb{R})$.}\label{irreps}

\medskip\noindent{$\displaystyle \Sl(2, \mathbb{K})$ with $\mathbb{K}=\mathbb{R}$ or $\mathbb{K}=\mathbb{C}$:} The irreducible finite-dimensional representations of $\Sl(2, \mathbb{K})$ are well-known (see for instance \cite{Varadarajan}). An irreducible representation space is defined by a half-integer $\displaystyle J=0,\frac{1}{2}, 1, \frac{3}{2}, 2,\dots$ and a vector space $\mathsf{k}_J=\langle Q_1,\dots, Q_{2J+1}\rangle$ and $\dim\mathsf{k}_J=2J+1$. Denoting by $e_k\cdot Q_l$ the representation of $e_k$ on $Q_l$, we have the relations

\[
Q_k=\frac{e_3^{k-1}}{(k-1)!}\cdot Q_1,\quad e_3\cdot Q_{2J+1}=0,
\]
as well as
\[
Q_{2J+1-k}=\frac{e_2^k}{k!}\cdot Q_{2J+1},\quad e_2\cdot Q_1=0,
\]
and
\[
e_1\cdot Q_k=2(J-k+1)Q_k.
\]
These relations hold for both real and complex vector spaces $\mathsf{k}_J$.

\medskip\noindent{$\displaystyle \So(3, \mathbb{R})$:} $\So(3, \mathbb{R})=\langle e_1, e_2, e_3\rangle$ is defined by the commutation relations $[e_1, e_2]=e_3,\; [e_2, e_3]=e_1,\; [e_3, e_1]=e_2$. In this case, we use the trick of complexifying. If $V$ is the {\em real} irreducible representation space of $\So(3, \mathbb{R})$, then define the {\em complex} vector space $W=V+iV$ as the space spanned by vectors $u+iv$ with $u, v\in V$. Next define $h=2ie_1,\; x=e_2+ie_3,\; y=-e_2+ie_3$. Then $\langle h, x, y\rangle$ is just $\Sl(2, \mathbb{C})$ and it acts {\em irreducibly} on $W$. So we have $W=\langle Q_1,\dots, Q_{2J+1}\rangle$ and $[h, Q_{2J-k+1}]=-2(J-k)Q_{2J-k+1}$ for $k=0, 1,\dots, 2J$. Further, $[y, Q_{2J+1}]=0$ and we also have $[h, Q_{2J+1}]=-2JQ_{2J+1}$. With  $Q_{2J+1}=X+iY$ for some $X, Y\in V$, we then find from $[h, Q_{2J+1}]=-2JQ_{2J+1}$ that $[e_1, X]=-JY,\; [e_1, Y]=JX$. Note also, that if $J$ is an integer, then there exists a vector $Z\in V$ with $[e_1, Z]=0$ (this follows from $[h, Q_{2J-k+1}]=-2(J-k)Q_{2J-k+1}$). Thus, in any irreducible representation of $\So(3, \mathbb{R})$ on a real Lie algebra $\mathsf{A}$ (whether abelian or not) we have $\dim \mathsf{A}=2J+1$ for some half-integer $J$,  non-zero vectors $X, Y$ with $[e_1, X]=\alpha Y, \; [e_1, Y]=-\alpha X$ for $\alpha>0$ for any $J>0$ and non-zero vectors $Z$ with $[e_1, Z]=0$ if $J$ is an integer.

Finally, we note that $\So(3, \mathbb{R})$ has no two-dimensional  irreducible representations: in fact if this were the case then there would be a representation by three real $2\times 2$ trace-free matrices of $\langle e_1, e_2, e_3\rangle$. These would then form a basis for $\Sl(2, \mathbb{R})$, contradicting the commutation relations for $\So(3, \mathbb{R})$.

\section{Proof of Theorem \ref{contactequiv}} In this appendix we give a proof of Theorem \ref{contactequiv}. To this end we begin with some informal preliminaries on contact structures.

We work on $k$-th order jet spaces $J^{k}(\mathbb{R}^n, \mathbb{R}^m)$, with local coordinates \newline  $(x,u,\underset{(1)}{u}, \underset{(2)}{u},\dots, \underset{(k)}{u})$ where $x=(x^1, \dots, x^n),\;\; u=(u^1,\dots, u^m)$ and where $\underset{(j)}{u}$ stands for the collection of all the $j$-th order partial derivatives of $(u^1,\dots, u^m)$ with respect to the $(x^1,\dots, x^n)$. On these spaces we introduce the {\sl contact one-forms}

\begin{align*}
\omega^l&=du^l-u^l_{\mu}dx^{\mu},\;\; \omega^l_{\mu_i}=du^l_{\mu_i}-u^l_{\mu_i\mu_j}dx^{\mu_j},\dots,\\
\omega^l_{\mu_1\dots\mu_k}&=du^l_{\mu_1\dots\mu_k}-u^l_{\mu_1\dots\mu_k\mu_{k+1}}dx^{\mu_{k+1}}
\end{align*}
for $l=1,\dots m$ and $\mu_i=1, \dots, n$, and we sum over repeated indices. The symbols $u^l_{\mu_i},\dots, u^l_{\mu_1\dots\mu_k}$ are then the partial derivatives of the functions $u^l$. These one-forms vanish on solutions of differential equations. They are known as the Cartan distribution (see \cite{VinogradovEtAl}). Note that a $k$-th order contact form

\[
\omega^l_{\mu_1\dots\mu_k}=du^l_{\mu_1\dots\mu_k}-u^l_{\mu_1\dots\mu_k\mu_{k+1}}dx^{\mu_{k+1}}
\]
is not a properly defined form on $J^{k}(\mathbb{R}^n, \mathbb{R}^m)$, because the $u^l_{\mu_1\dots\mu_k\mu_{k+1}}$ are coordinates on $J^{k+1}(\mathbb{R}^n, \mathbb{R}^m)$. To remedy this, we may work on the infinite jet bundle $J^{\infty}(\mathbb{R}^n, \mathbb{R}^m)$ (\cite{VinogradovEtAl}, \cite{Takens}, \cite{FatibeneFrancaviglia}), which may be thought of as a space of infinite sequences $(x, u, \underset{(1)}{u},\dots,\underset{(k)}{u},\dots)$, and we have projections $\pi^{\infty}_k: J^{\infty}(\mathbb{R}^n, \mathbb{R}^m)\to J^{k}(\mathbb{R}^n, \mathbb{R}^m)$ as well as projections $\pi^{k}_l: J^{k}(\mathbb{R}^n, \mathbb{R}^m)\to J^{l}(\mathbb{R}^n, \mathbb{R}^m)$ for $l\leq k$. In terms of the sequences $(x, u, \underset{(1)}{u},\dots,\underset{(k)}{u},\dots)$, we have $\pi^{\infty}_k(x, u, \underset{(1)}{u},\dots,\underset{(k)}{u},\dots)=(x, u, \underset{(1)}{u},\dots,\underset{(k)}{u})$ and $\pi^{k}_l(x, u, \underset{(1)}{u},\dots,\underset{(k)}{u})=(x, u, \underset{(1)}{u},\dots,\underset{(l)}{u})$. The space infinite jet space $J^{\infty}(\mathbb{R}^n, \mathbb{R}^m)$ is well-defined, as the inverse limit of the $k$-th order jet bundles $J^{k}(\mathbb{R}^n, \mathbb{R}^m)$ (see \cite{VinogradovEtAl}, \cite{Takens}, \cite{FatibeneFrancaviglia}). A function $F: J^{\infty}(\mathbb{R}^n, \mathbb{R}^m)\to \mathbb{R}$ is then defined to be smooth if $F$ is a smooth function of only $(x, u, \underset{(1)}{u},\dots,\underset{(k)}{u})$ for some natural number $k\in \mathbb{N}$. We work within this infinite jet bundle unless otherwise stated.

A {\sl tangent transformation} is then a transformation
\[
(x, u, \underset{(1)}{u},\dots, \underset{(k)}{u})\to (x', u', \underset{(1)}{u'},\dots, \underset{(k)}{u'})
\]
for any $k\geq 1$ such that the transformed contact forms

\begin{align*}
\omega'^l&=du'^l-u'^l_{\mu}dx'^{\mu},\;\; \omega'^l_{\mu_i}=du'^l_{\mu_i}-u'^l_{\mu_i\mu_j}dx'^{\mu_j},\dots,\\
\omega'^l_{\mu_1\dots\mu_k}&=du'^l_{\mu_1\dots\mu_k}-u'^l_{\mu_1\dots\mu_k\mu_{k+1}}dx'^{\mu_{k+1}}
\end{align*}
vanish when $\omega^l,\, \omega^l_{\mu_i},\dots, \omega^l_{\mu_1\dots\mu_k}$ vanish (see \cite{IbragimovAnderson}). Then the following result holds (\cite{IbragimovAnderson}):

\begin{thm} A tangent transformation is given by a prolongation of a point transformation
\[
(x,u)\to (x', u')
\]
if $u=(u^1,\dots, u^m)$ for $m\geq 2$. It is given by the prolongation of a contact transformation
\[
(x,u, \underset{(1)}{u})\to (x',u', \underset{(1)}{u'})
\]
if $m=1$, and then there is a (locally smooth) function $W(t,x,u, \underset{(1)}{u})$ such that
\[
x'^l=-W_{q^l},\;\; u'=W-pW_p-q^lW_{q^l},\;\; u'_{x'^l}=W_{x^l} + q^lW_u
\]
with $q^l=u_{x^l},\; l= 1,\dots, k$.
\end{thm}

Now let us consider smooth, real-valued functions $F(x,u,\underset{(1)}{u},\dots, \underset{(k)}{u})$. Then one can show (\cite{VinogradovEtAl}, \cite{FatibeneFrancaviglia}) that the exterior derivative $dF$ may be written as
\[
dF = DF + \sum_{j=1}^{k}\Lambda_l^{\mu_1\dots\mu_j}\omega^l_{\mu_1\dots\mu_j}
\]
for some functions $\displaystyle \Lambda_l^{\mu_1\dots\mu_j}$, when we work on the infinite jet bundle, where we sum over repeated indices. Here we have
\[
DF=D_{x^1}Fdx^1 + \dots D_{x^n}Fdx^n
\]
and $D_{x^i}$ is the operator of total differentiation with respect to $x^i$:
\[
D_{x^{\mu_i}}F=F_{x^i} + u^l_{\mu_i}F_{u^l} + u^l_{\mu_i\mu}F_{u^l_{\mu}}+\dots u^{l}_{\mu_i\mu_1\dots\mu_k}F_{u^{l}_{\mu_1\dots\mu_k}}.
\]
From this, it follows easily that if $D_{x^i}F=0$ then $F_{u^l}=F_{u^{l}_{\mu_1\dots\mu_j}}=0$ for $l=1,\dots, m$ and $j=1,\dots, k$, and then $DF=0$ implies $F=\;\text{constant}$, and so $DF=0$ if and only if $dF=0$. A further result in this direction is the following:

\begin{lemma}\label{invertibletransf} Suppose that
\[
(x,u,\underset{(1)}{u},\dots, \underset{(k)}{u})\to (x',u',\underset{(1)}{u'},\dots, \underset{(k)}{u'})
\]
defines an invertible tangent transformation, with $x'^1=X^1,\dots, x'^n=X^n$, then we have
\[
DX^1\wedge\dots \wedge DX^n\neq 0.
\]
\end{lemma}

\smallskip\noindent{\bf Proof:} There are two cases: if $u=(u^1,\dots, u^m)$ with $m\geq 2$, then the transformation is the prolongation of an invertible point transformation $(x, u)\to (X(x,u), U(x,u))$. Thus we have $dX^1\wedge\dots \wedge dX^n\wedge dU^1\wedge\dots \wedge dU^m\neq$ by invertibility.

Now suppose that $DX^1\wedge\dots \wedge DX^n=0$. We know from the previous remarks that $DX^i\neq 0$ for $i=1,\dots, n$. Hence $DX^1\in I(DX^2,\dots, DX^n)$ where $I(DX^2,\dots, DX^n)$ is the module generated by the forms $DX^2,\dots, DX^n$, that is, $I(DX^2,\dots, DX^n)$ consists of linear combinations of the form $\lambda_2DX^2+\dots+\lambda_nDX^n$ for some functions $\lambda_2,\dots, \lambda_n$. Now each $X^i$ is a function of $(x^1,\dots, x^n, u^1,\dots, u^m)$ and then we have
\[
dX^i=DX^i+\Lambda_l\omega^l.
\]
From this it now follows that $dX^1\in I(dX^2,\dots, dX^n, \omega^1,\dots, \omega^m)$. Further, since the transformation is a tangent transformation, we have  $dU^l-U^l_{i}dX^i=\lambda^l_j\omega^j$ for $l=1,\dots, m$,  so that $dU^l\in I(dX^1, dX^2,\dots, dX^n, \omega^1,\dots, \omega^m)$ and consequently $dU^l\in I(dX^2,\dots, dX^n, \omega^1,\dots, \omega^m)$ since we know that \newline $dX^1\in I(dX^2,\dots, dX^n, \omega^1,\dots, \omega^m)$. Hence each of the $m+n$ one-forms $dX^1,\dots, dX^n, dU^1,\dots, dU^m$ is a sum of $m+n-1$ one-forms, so that $dX^1\wedge\dots \wedge dX^n\wedge dU^1\wedge\dots\wedge dU^m=0$, which contradicts the invertibility of the transformation.

The case of just one function is treated similarly: in this case the tangent transformation is a prolongation of an invertible contact transformation
\[
(x^1,\dots, x^n, u, u_1,\dots u_n)\to (x'^1,\dots, x'^n, u', u'_1,\dots u'_n),
\]
where $u_i=u_{x^i}$, which is a particular case of a point transformation with $n+1$ functions $u, u_1,\dots, u_n$. Hence we must have $DX^1\wedge\dots\wedge DX^n\neq 0$.

\medskip We now come to the the proof of Theorem \ref{contactequiv}:

\begin{thm}
Any invertible contact transformation
\[
(t,x,u,p,q)\to (t', x', u', p', q')
\]
with $p=u_t,\;\; q=u_1$ and $p'=u'_{t'},\;\, q'=u'_{x'}$, preserving the form of an evolution equation of order $n$
\[
u_0=F(t,x,u,u_1,\dots, u_n),
\]
with $n\geq 2$, is such that $t'=T(t)$ with $\dot{T}(t)\neq 0$. Then the contact transformation has the form
\[
\begin{split}
t'&=T(t),\; x'=-W_{q}(t,x,u,q),\\
u'&= W(t,x,u,q) - qW_{q}(t,x,u,q),\\
p'&= -p\dot{T}(t) +pW_u(t,x,u,q) +W_t(t,x,u,q),\\
q'&=W_x(t,x,u,q) + qW_u(t,x,u,q)
\end{split}
\]
for some smooth function $W$.
\end{thm}

\smallskip\noindent{\bf Proof:} Our contact  transformation is

\begin{align*}
& t'=T(t, x, u, p, q),\; x'=X(t, x, u, p, q),\; u'=U(t, x, u, p, q),\\ & p'=P(t, x, u, p, q),\; q'=Q(t, x, u, p, q),
\end{align*}
and there is a function $W(t, x, u, p, q)$  so that
\[
T= - W_p,\;\; X= -W_q,\;\; U= W - pW_p - qW_q,\;\; P= W_t + pW_u,\;\; Q= W_x + qW_u
\]
(see \cite{IbragimovAnderson}). An evolution equation of order $n+1$ becomes in our notation
\[
p=F(t, x, u, q, q_1,\dots, q_n).
\]
where $q_1=D_xq, q_2=D^2_xq,\dots, q_n=D^n_xq$. We shall show that $D_tT\neq 0,\;\;D_xT=0$, from which it follows that $T=T(t)$.

We define the sequence of functions $Q^k,\;\; k\geq 0$, by $Q^0=Q$ and for $k\geq 1$ they are given by
\[
DT\wedge DQ^k=Q^{k+1}DT\wedge DX.
\]
The $Q^k$ are well-defined since we have $DT\wedge DX\neq 0$ by Lemma \ref{invertibletransf}. These functions are just the transformations of the functions $q_k=\partial^kq/\partial x^k$ defined by the contact conditions:
\[
dQ^k-Q^{k+1}dX - Q^k_0dT=0\quad  \text{mod}\;\; I
\]
where $Q^k_0=D_TQ^k,\;\; Q^{k+1}=D_XQ^k$  and where $I$ is the module of contact one-forms. From the fact that $dF=DF\;\; \text{mod}\;\; I$ we obtain $DQ^k-Q^{k+1}DX-Q^k_0DT=0$ on imposing the contact conditions. Then $DT\wedge DQ^k=Q^{k+1}DT\wedge DX$ follows easily.

Then we make the substitution $q'_{l}=Q^l$ in the evolution equation
\[
p'=F'(t',x',u',q',q'_1\dots, q'_n).
\]
The result is to be another evolution equation
\[
p=F(t,x,u,q,q_1\dots, q_n)
\]
so that the right-hand side contains the highest spatial derivative $q_n=D^n_xq$ and none of the derivatives $p_n=D^n_x p,\; p_0^{n}=D^n_tp,\; q_0^n=D^n_tq$ nor the mixed derivatives $q_{k,l}=D^k_tD^l_x q$ for $k\geq 1$, and $p_{k, l}=D^k_tD^l_x p$. Note that since $p=D_tu,\; q=D_x u$ we have $p_{k, l}=q_{k+1, l-1}$ and in particular $q_{1, 0}=q_0=p_{0, 1}=p_1$. Thus, we require that $Q^n_{p_0^{n}}=0$ and $Q^n_{p_n}=0$ as well as $Q^n_{q_n}\neq 0$.

We now show that $D_tT\neq 0$. For if $D_tT=0$ then $T=T(x)$. Then the relation $Q^{k+1}DT\wedge DX=DT\wedge DQ^k$ gives $Q^{k+1}T'(x)D_tXdt\wedge dx=D_tQ^kT'(x)dt\wedge dx$,
so that $Q^{k+1}=\frac{D_tQ^k}{D_tX}$. We have for $k=0$ that $Q^1=\frac{D_tQ}{D_tX}$ which shows that $Q^1=Q^1(t,x,u,p,q,p_0,q_0)$. Thus, by induction, we find that $Q^n$ is independent of the spatial derivatives of $p_n,\; q_n$, and hence we contradict the requirement that $Q^n_{p_n}\neq 0$. Consequently we must have $D_tT\neq 0$.

It follows from the fact that $D_tT\neq 0$ that $Q^1_{q_1}\neq 0$: in fact, we note that if $Q^k_{q_k}\neq 0$ then $DQ^k$ will depend linearly on $q_{k+1}$ and this appears only in the term $D_xQ^k$ as the coefficient of $Q^k_{q_k}$. So, differentiating $Q^{k+1}DT\wedge DX=DT\wedge DQ^k$ with respect to $q_{k+1}$ we find that $Q^{k+1}_{q_{k+1}}DT\wedge DX=Q^k_{q_k}D_tTdt\wedge dx$. Since we have $Q^n_{q_n}\neq 0,\;\; D_tT\neq 0,\; DT\wedge DX\neq 0$, it follows, by induction, that $Q^1_{q_1}\neq 0$.

Our next step is to show that $D_xT=0$ and to this end we assume that $D_xT\neq 0$. Then $Q^1_{p_0}=0$. In fact, if $Q^k_{p^k_0}\neq 0$ then $DQ^k$ will be linear in $p^{k+1}_0$, and this will appear in the coefficient of $Q^k_{p^k_0}$ in $D_tQ^k$. So, differentiating $Q^{k+1}_{q_{k+1}}DT\wedge DX=Q^k_{q_k}D_tTdt\wedge dx$ with respect to $p^{k+1}_0$ we find that $Q^{k+1}_{p^{k+1}_0}DT\wedge DX=-Q^k_{p^k_0}D_xTdt\wedge dx$. Thus, if $D_xT\neq 0$ then $Q^1_{p_0}\neq 0$ implies, by induction, that $Q^n_{p^n_0}\neq 0$, and this contradicts our requirement $Q^n_{p^n_0}=0$. Hence $Q^1_{p_0}=0$.

Another consequence of $D_xT\neq 0$ is that $Q_{q_0}=Q_{p_1}=0$. For suppose $Q^1_{q_0}\neq 0$. We know that $Q^k_{p_0^k}=0$ so that $D_xQ^k$ is independent of $q^{k+1}_0=D_xp^k_0$, and therefore, from $Q^{k+1}DT\wedge DX=DT\wedge DQ^k$ it follows that $Q^{k+1}_{q^{k+1}_0}DT \wedge DX=-Q^k_ {q^k_0}D_xTdt\wedge dx$ for $k\geq 1$. So, if $Q^1_{q_0}\neq 0$, then $Q^n_{q^n_0}\neq 0$, as follows from induction. Since we require $Q^n_{q^n_0}= 0$, we must have $Q^1_{q_0}=0$, because we also assume $D_xT\neq 0$. Note that $q_0=D_tq=D_tD_xu=D_xD_tu=D_xp=p_1$, and so we also have $Q^1_{p_1}=0$.

From all this we find that $Q^1=Q^1(t, x, u, p, q, q_1)$ and it then follows that $Q^n$ will contain the $n$-th order derivative $q_{n-1, 1}=D^{n-1}_tD_xq$ if $D_xT\neq 0$. In fact, putting $q_{k-1, 1}=D^{k-1}_tD_xq$ for $k\geq 1$, we have from $Q^{k+1}DT\wedge DX=DT\wedge DQ^k$ that
\[
Q^{k+1}_{q_{k, 1}}DT\wedge DX=-Q^{k}_{q_{k-1, 1}}D_xTdt\wedge dx
\]
for $k\geq 1$, since $q_{k, 1}$ is a derivative of order $k+1$ which can only occur in $D_tQ^k$. Now $q_{0, 1}=q_1$ and we know that $Q^1_{q_1}\neq 0$, so it follows by induction that $Q^n_{q_{n-1, 1}}\neq 0$ if $D_xT\neq 0$. However, this is a contradiction since we must have $Q^n_{q_{n-1, 1}}=0$ for our transformations to transform any given evolution equation into an evolution equation. Hence we conclude that $D_xT=0$ and it then follows easily that $T=T(t)$.

Finally, noting that
\[
T= - W_p,\;\; X= -W_q,\;\; U= W - pW_p - qW_q,\;\; P= W_t + pW_u,\;\; Q= W_x + qW_u
\]
for some function $W(t,x,u,p,q)$, we find, on integrating this system, that the contact transformation has the stated form. This proves the result.

\bigskip\noindent{\bf Acknowledgements.} V. Lahno thanks the Swedish Research
Council for financial support(grant number 624-2004-1073), and
Link\"oping University for hospitality. The research of F. G\"{u}ng\"{o}r
was supported by the Research Council of Turkey (T\"{U}B\.{I}TAK), grant number 107T846.

\bibliographystyle{unsrt}

\end{document}